\documentclass[aps,prc,twocolumn,superscriptaddress,showpacs,showkeys,
               floatfix,nofootinbib,altaffilletter]{revtex4}

\usepackage{xspace}
\usepackage{hyperref}
\usepackage{longtable}

\newcommand{\FlukaLong}{{\scshape Fluka2008}\xspace}

\newcommand{\UrqmdLong}{{\scshape Urqmd1.3.1}\xspace}

\newcommand{\GheishaLong}{{\scshape Gheisha2002}\xspace}

\newcommand{\Gheisha}{{\scshape Gheisha}\xspace}
\newcommand{\Corsika}{{\scshape Corsika}\xspace}
\newcommand{\Venus}{{\scshape Venus}\xspace}
\newcommand{\VenusLong}{{\scshape Venus4.12}\xspace}

\newcommand{\GiBUULong}{{\scshape GiBUU1.3.0}\xspace}
\usepackage[latin2]{inputenc}
\usepackage{indentfirst}
\usepackage{enumerate}

\usepackage{amsmath}
\usepackage{amssymb}
\usepackage[english]{babel}
\usepackage{url}

\newif\ifpdf

\ifx\pdfoutput\undefined

\pdffalse

\else

\pdfoutput=1

\pdftrue

\fi

\ifpdf

\usepackage[pdftex]{graphicx}

\else

\usepackage{graphicx}

\fi

\usepackage{stmaryrd}

\begin{document}
\DeclareGraphicsExtensions{.pdf,.png,.eps,.jpg,.ps}
\cleardoublepage

\pagenumbering{roman}

\title{\large \bf Measurements of
                  Cross Sections and Charged Pion Spectra  \\
                  in proton--carbon Interactions at 31~GeV/\textit{c}}

\begin{abstract}
Interaction cross sections and charged pion spectra
in p+C interactions at 31~GeV/\textit{c} were measured
with the large-acceptance NA61/SHINE spectrometer at the CERN SPS.
These data are required to improve predictions of the
neutrino flux for the T2K long-baseline neutrino oscillation 
experiment in Japan.
A set of data collected during the first NA61/SHINE run 
in 2007 with an isotropic graphite target with a thickness of 4\% 
of a nuclear interaction length was used for the analysis.
The measured p+C inelastic and production cross sections are
257.2 $\pm$ 1.9 $\pm$ 8.9 and 229.3 $\pm$ 1.9 $\pm$ 9.0~mb, respectively.
Inclusive production cross sections for negatively and positively charged pions
are presented as functions of laboratory momentum in ten intervals
of the laboratory polar angle covering the range from 0 up to 420~mrad.
The spectra are compared with predictions of several hadron production models.
\end{abstract}

\clearpage

 \author{N.~Abgrall}\affiliation{University of Geneva, Geneva, Switzerland}
 \author{A.~Aduszkiewicz}\affiliation{Faculty of Physics, University of Warsaw, Warsaw, Poland}
 \author{B.~Andrieu}\affiliation{LPNHE, University of Paris VI and VII, Paris, France}
 \author{T.~Anticic}\affiliation{Rudjer Boskovic Institute, Zagreb, Croatia}
 \author{N.~Antoniou}\affiliation{University of Athens, Athens, Greece}
 \author{J.~Argyriades}\affiliation{University of Geneva, Geneva, Switzerland}
 \author{A.~G.~Asryan}\affiliation{St. Petersburg State University, St. Petersburg, Russia}
 \author{B.~Baatar}\affiliation{Joint Institute for Nuclear Research, Dubna, Russia}
 \author{A.~Blondel}\affiliation{University of Geneva, Geneva, Switzerland}
 \author{J.~Blumer}\affiliation{Karlsruhe Institute of Technology, Karlsruhe, Germany}
% ...

 \collaboration{\bf The NA61/SHINE Collaboration}
 \noaffiliation

 \date{\today}

 \maketitle
 \clearpage

 \pagenumbering{arabic}

 \section{Introduction}

 The NA61/SHINE (SPS heavy ion and neutrino experiment) experiment
 at the CERN Super Proton Synchrotron (SPS) pursues a rich physics program in various
 fields~\cite{proposala,add1,proposalb,Status_Report_2008}.
 First, precise hadron production measurements are performed for
 improving calculations of the neutrino flux in the T2K neutrino
 oscillation experiment~\cite{T2K},  as well as
 for more reliable simulations of cosmic-ray air showers in the Pierre Auger and KASCADE
 experiments~\cite{Auger,KASCADE}.
 Second, p+p, p+Pb and nucleus+nucleus collisions will be studied extensively
 at SPS energies. A collision energy and system size scan
 will be performed with
 the aim to study properties of the onset of deconfinement and
 searching for the critical point of strongly interacting matter.
 This paper presents the first NA61/SHINE results on charged
 pion spectra in p+C interactions at 31~GeV/\textit{c}, which are needed
 for an accurate neutrino flux prediction in the T2K experiment.
 The results are based on the data collected during
 the first run in 2007. NA61/SHINE is a large-acceptance
 hadron spectrometer~\cite{proposala}. Its main detector components,
 software, calibration, and analysis methods were inherited
 from the NA49 experiment~\cite{NA49,pC_NA49}.

\begin{figure}[tb]
 \begin{center}
 \includegraphics[width=0.76\linewidth]{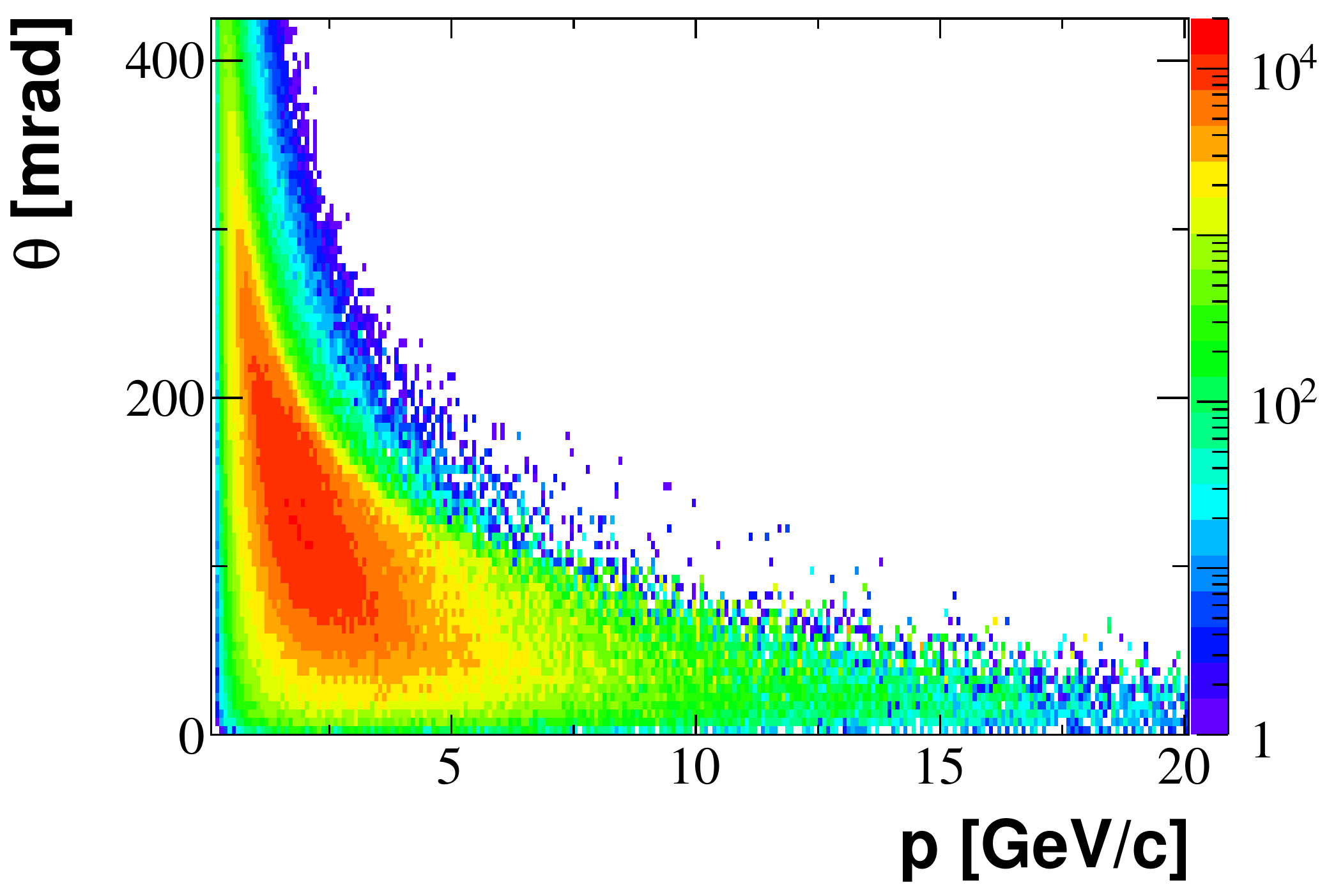}
 \end{center}
\vspace*{-0.3cm}
   \caption{(Color online)
 The prediction from the T2K beam simulation:
 the $\{ p,~\theta \}$ distribution for positively charged pions
 weighted by the probability that their decay produces a muon neutrino 
 passing through the SK detector.}
 \label{fig:p-theta-pi+SK}
 \end{figure}

 T2K is a long-baseline neutrino experiment in Japan, which uses
 a high-intensity neutrino beam produced at J-PARC.\footnote{The Japan Proton
 Accelerator Research Complex 
 organized 
 jointly by JAEA and KEK in Tokai, Japan.}
 It
 aims to precisely measure the $\nu_\mu \to \nu_e$ appearance
 and $\nu_\mu$ disappearance~\cite{T2K}.
 In order to generate the neutrino beam a high-intensity
 30~GeV (kinetic energy) proton beam impinging on a 90-cm-long graphite target is used,
 where $\pi$ and K mesons decaying into (anti)neutrinos are produced.
 The neutrino fluxes and spectra are then measured both at the near-detector
 complex, 280~m from the target, and by the Super-Kamiokande (SK) detector
 located 295~km away from the neutrino source and 2.5\textdegree~off-axis.
 Neutrino oscillations are probed by comparing the neutrino flux
 measured at SK to the predicted one. 
 In order to predict the flux at SK one uses the near-detector measurements
 and extrapolates them to SK with the help of Monte Carlo  simulations.
 Up to now, these Monte Carlo predictions
 have been based on hadron production models only.
 For more precise predictions,
 which would allow the reduction of systematic uncertainties to the level needed
 for the T2K physics goals,
 measurements of pion and kaon production off carbon targets 
 are essential~\cite{proposala,add1,proposalb}.
 The purpose of the NA61/SHINE measurements for T2K is to provide 
 this information
 at exactly the proton extraction energy of the J-PARC Main Ring
 synchrotron, namely, 30~GeV kinetic energy (approximately 31~GeV/\textit{c} momentum).
 Presently, the T2K neutrino beam-line is set up to focus positively charged
 hadrons, in such a way that it produces a $\nu_\mu$ beam.
 Spectra of positively charged pions presented in this paper constitute
 directly an essential ingredient in the neutrino flux calculation.
 The kinematic region of interest for positively charged pions
 whose daughter muon neutrinos pass through the SK detector
 is shown in Fig.~\ref{fig:p-theta-pi+SK}, for the kinematic
 variables $p$ -- the momentum of a given particle and
 $\theta$ -- its polar angle in the laboratory frame.

 In addition to providing reference
 data for T2K, precise results on particle production in proton-carbon
 interactions furnish important input to improve hadronic generators needed
 for the interpretation of air showers initiated by ultra-high-energy cosmic
 particles 
 (see, e.g.,~\cite{Heck:2003br, Drescher:2003gh, Meurer:2005dt, Maris:2009uc}).

 The NA61/SHINE data
 will also allow testing and improvement of existing
 hadron production
 models in an intermediate energy region that is not well
 constrained by measurements at present.

 Experimental data on proton-nucleus interactions in the region of momentum
 of a few tens
 of GeV/\textit{c} are rather scarce. Incident proton momenta from 3 up to 12 GeV/\textit{c}
 were explored by the HARP Collaboration~\cite{HARP} for a large variety of nuclear
 targets, including carbon~\cite{pC12_HARP}.
 Almost 40 years ago an experiment was performed at CERN at 24~GeV/\textit{c}~\cite{24GeV}.
 Several targets were used but the momentum and angular range
 of produced particles was limited
 to angles from 17 to 127~mrad and momenta from 4 to 18~GeV/\textit{c}.
 Recently the MIPP particle production experiment at Fermilab presented
 preliminary results from its first data collected at 120~GeV/\textit{c}~\cite{MIPP}.
 Precise measurements of pion production in proton-carbon
 interactions at 158~GeV/\textit{c} are available from the NA49~Collaboration~\cite{pC_NA49}.

\begin{figure*}[!ht]
\begin{center}
\includegraphics[width=0.9\textwidth]{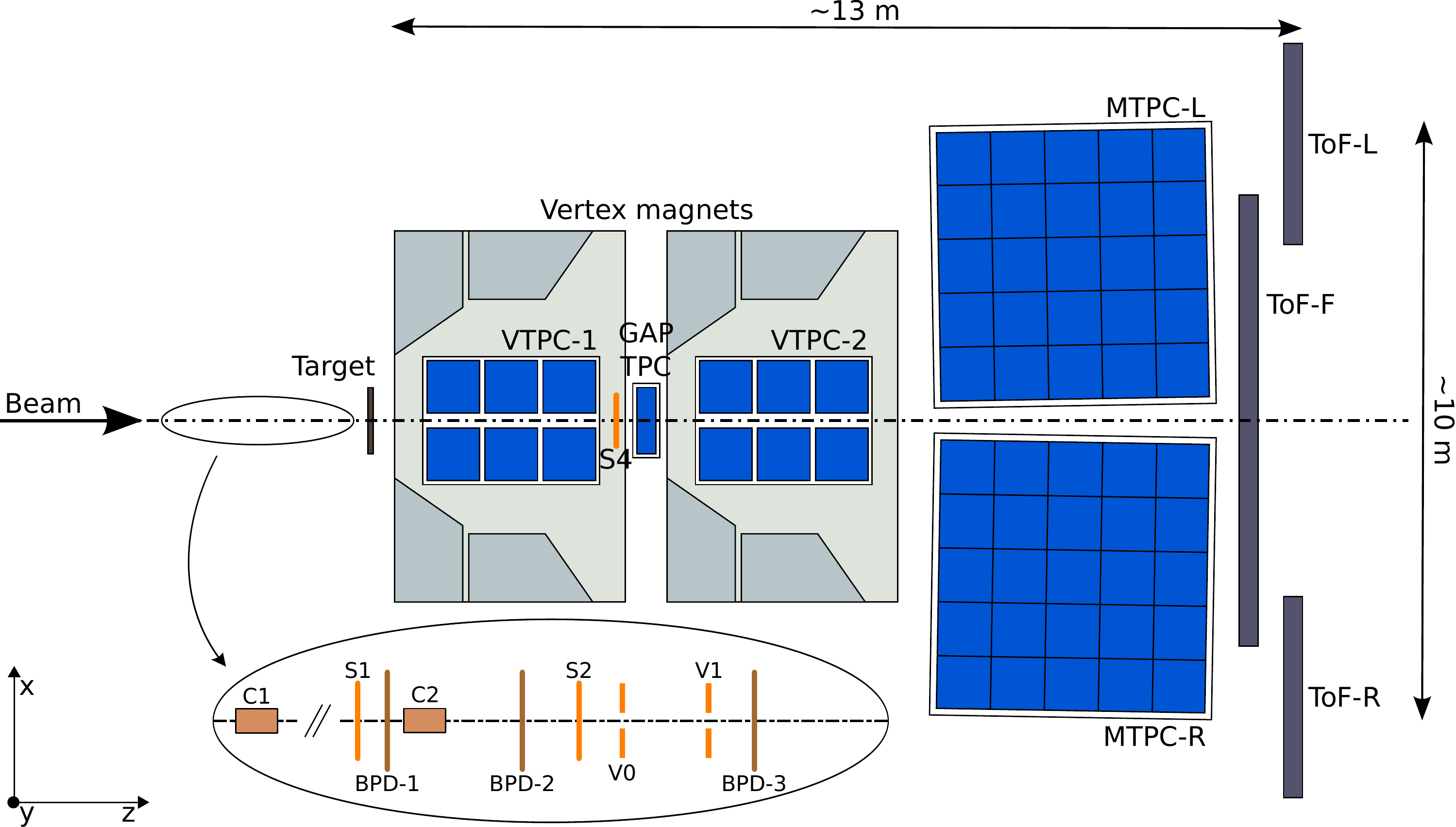}
\end{center}
  \caption{(Color online)
The layout of the NA61/SHINE experiment at the CERN SPS (top view, not to scale).
The chosen right-handed coordinate system is shown on the plot.
The incoming beam direction is along the $z$ axis.
The magnetic field bends charged particle trajectories
in the $x$-$z$ (horizontal) plane.
The drift direction in the TPCs is along the $y$ (vertical) axis.
}
  \label{fig:detector}
\end{figure*}

 The paper is organized as follows.
 In Sec.~II the NA61/SHINE experimental set-up is described.
 Details on the beam, trigger, and event selection are given in Sec.~III.
 Data reconstruction, simulation, and detector performance are described in
 Sec.~IV. Analysis techniques and final results are presented in Secs.~V and~VI,
 respectively. These results are compared with
 hadron production
 models in Sec.~VII.
 A summary in Sec.~VIII closes the paper.

\section{The NA61/SHINE set-up}
\label{Sec:set-up}

The NA61/SHINE experiment is a large-acceptance hadron spectrometer in
the North Area H2
beam-line of the CERN SPS.
The schematic layout is shown in Fig.~\ref{fig:detector} 
together with the overall dimensions.

The main components of the current detector were constructed and used by the
NA49 Collaboration~\cite{NA49}.
A set of scintillation and Cherenkov counters as well as beam position
detectors (BPDs) upstream of the spectrometer provide timing reference,
identification, and position measurements of the incoming beam particles. 
Details on this system are presented in
Sec.~\ref{Sec:trigger}.
The main tracking devices of the spectrometer
are large-volume time projection chambers (TPCs).
Two of them, the vertex TPCs (VTPC-1 and VTPC-2 in Fig.~\ref{fig:detector}),
are located in
a free gap of 100~cm between the upper and lower coils of the
two superconducting dipole magnets.
Their maximum combined bending power is 9~T$\cdot$m.
In order to optimize the acceptance of the detector at 31~GeV/\textit{c}
beam momentum, the magnetic field used during the 2007 data-taking period
was set to a bending power of 1.14~T$\cdot$m. 
Two large-volume main TPCs (MTPC-L and \mbox{MTPC-R}) are positioned downstream of
the magnets symmetrically to the beam line. The
TPCs are filled with  Ar:CO$_2$ gas mixtures in proportions 90:10 for VTPCs
and 95:5 for MTPCs.
The particle identification capability of the TPCs
based on measurements of the specific energy loss $dE/dx$
is augmented by time-of-flight measurements ($tof$) using
time-of-flight (ToF) detectors.
The ToF-L and \mbox{ToF-R} arrays of scintillator pixels
have a time resolution of better than 90~ps~\cite{NA49}.
Before the 2007 run the experiment was upgraded with a new
forward time-of-flight detector (ToF-F) in order to extend the acceptance.
The ToF-F consists of 64 scintillator bars with 
photomultiplier (PMT) readout at both ends 
resulting in a time resolution of about 115~ps.
The target under study is installed 80~cm in front of the VTPC-1.
The results presented here were obtained with an
isotropic graphite target of
dimensions 2.5(W)$\times$2.5(H)$\times$2(L)~cm and
with a density of $\rho = 1.84$~g/cm$^3$.
The target thickness along the beam is equivalent to about 4\% of
a nuclear interaction length~($\lambda_{\mathrm{I}}$).

\section{Beam, trigger and data samples}
\label{Sec:trigger}

\begin{figure}[htbp]
\begin{center}
\ifpdf
\includegraphics[width=0.6\linewidth]{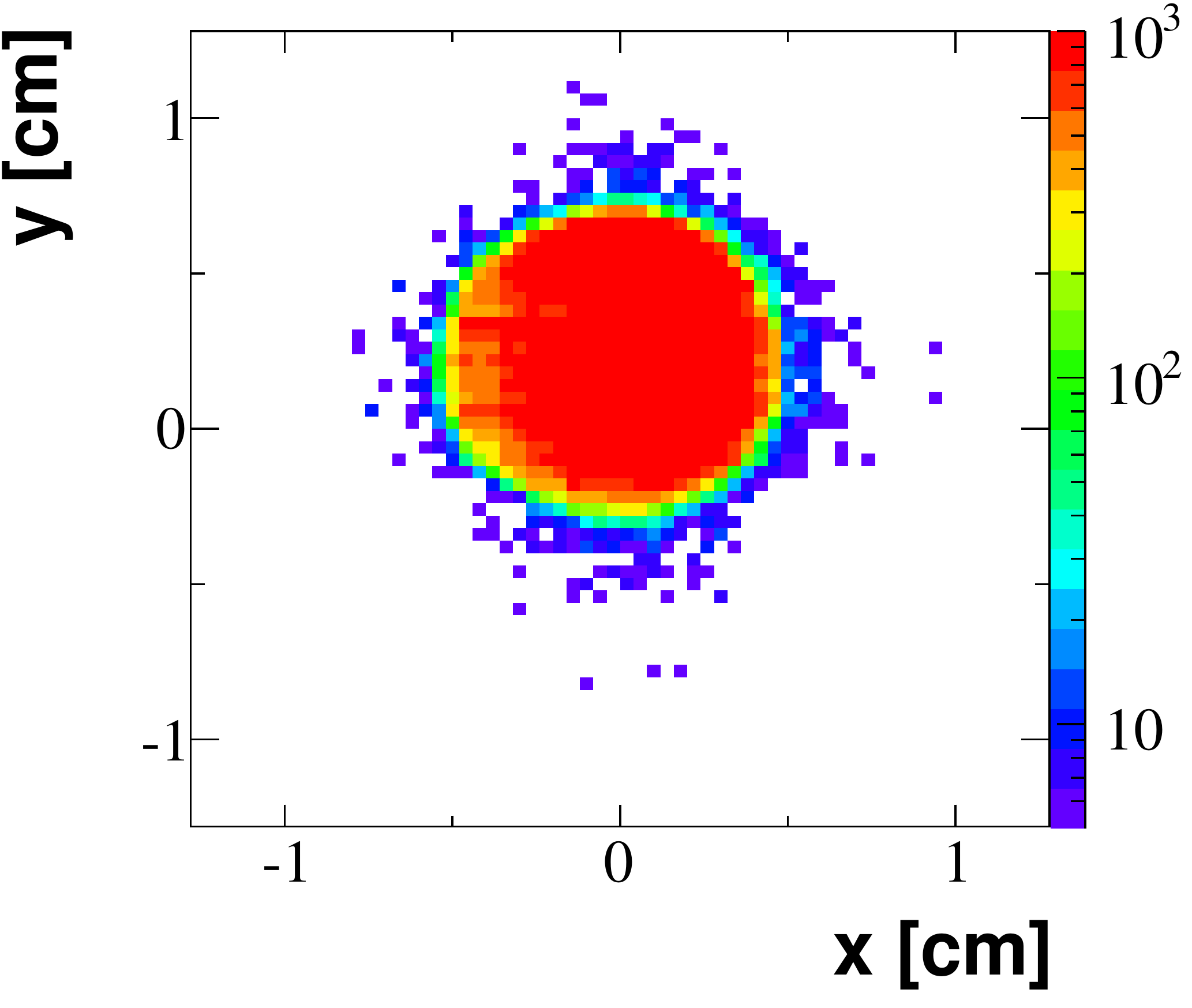}
\includegraphics[width=1.\linewidth]{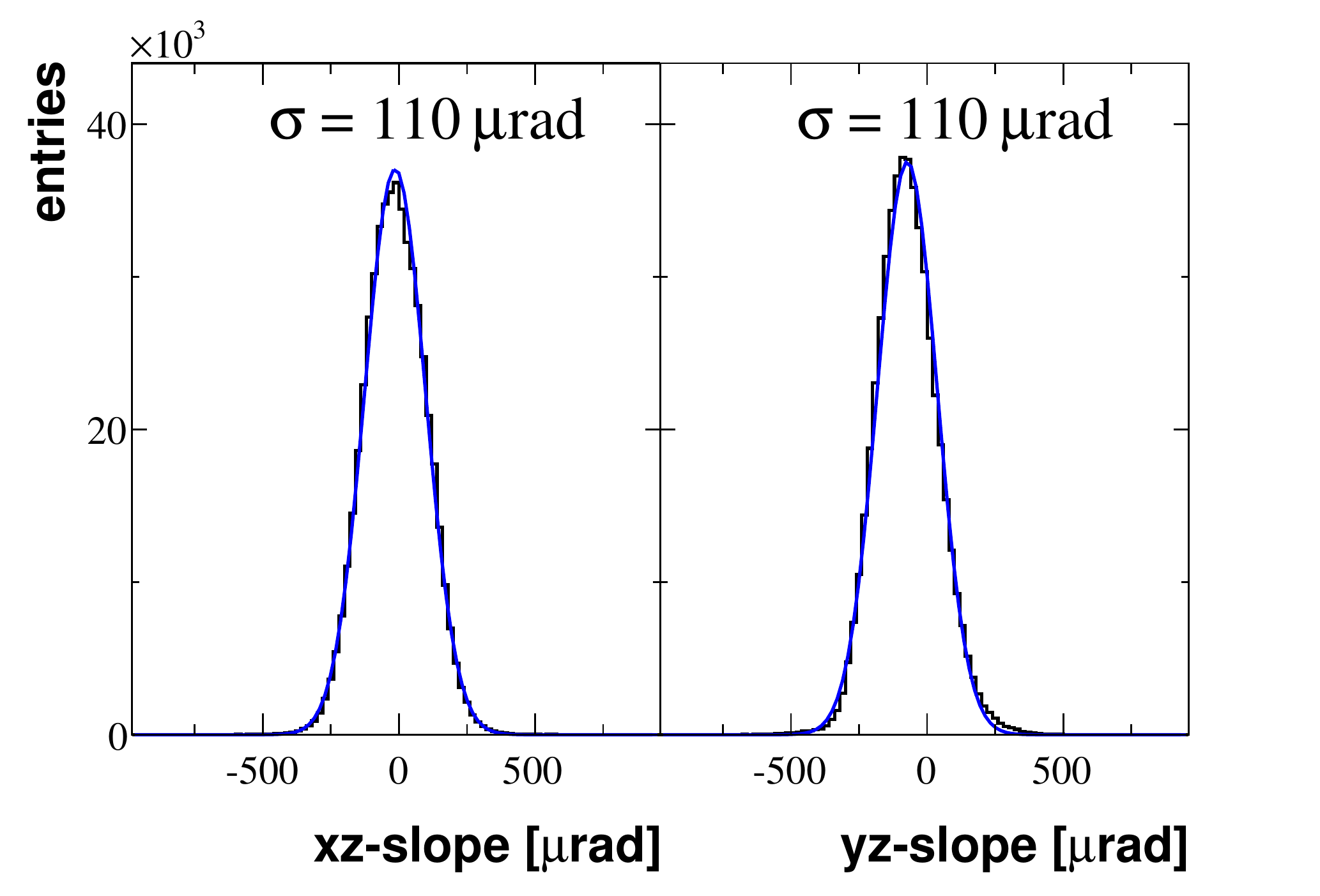}
\else
\includegraphics[scale=0.20]{cBeamPro2}
\includegraphics[scale=0.32]{fitDiv2}
\fi
\end{center}
\caption{(Color online)
{\it Top:}
The beam spot as measured by BPD-3 after the $\overline{\textrm{V1}}$ cut
described in the text.
{\it Bottom:}
The beam divergence in $x$ and $y$.} \label{profile}
\end{figure}

A 31~GeV/\textit{c} secondary hadron beam is produced from 400~GeV protons extracted
from the SPS in slow extraction mode.
The beam is transported along the H2 beam-line toward the experiment.
Collimators in the beam line are adjusted to get an average beam particle rate of 15~kHz.
The setup of beam detectors is illustrated in the inset
on Fig.~\ref{fig:detector}.
Protons from the secondary hadron beam are identified by two
Cherenkov counters, a CEDAR~\cite{CEDAR}
and a threshold counter, labeled C1 and
C2, respectively.
The CEDAR counter, using a six-fold
coincidence, provides positive identification of protons, while the
threshold Cherenkov counter, operated at pressure lower than the proton
threshold, is used in anti-coincidence in the trigger logic.
The fraction of protons in the beam was about 14\%.
A selection based on signals from Cherenkov counters allowed
the identification of
beam protons with a purity of about 99\%.
A consistent value for the purity was found by bending the beam into 
the TPCs with the full magnetic field and using the $dE/dx$ identification method.

Two scintillation counters S1 and S2 provide
beam definition, together with the two veto
counters V0 and V1 with a 1-cm-diameter hole,
which are collimating the beam on the target.
The S1 counter provides also the timing (start time for all counters). 
Beam protons are then selected by the coincidence
$\textrm{S1}\cdot\textrm{S2}\cdot\overline{\textrm{V0}}\cdot
\overline{\textrm{V1}}\cdot\textrm{C1} \cdot\overline{\textrm{C2}}$. 
The trajectory of individual beam particles is measured
in a telescope of beam position detectors along the beam line 
(BPD-1, -2 and -3 in Fig.~\ref{fig:detector}). 
These counters are small (3$\times$3~cm)
proportional chambers with cathode strip readout,
providing a resolution of about 200~$\mu$m in two orthogonal
directions; see~\cite{NA49} for more details. 
The beam profile and divergence obtained from the
BPD measurements are presented in Fig.~\ref{profile}.

\begin{figure}[htbp]
\begin{center}
\ifpdf
\includegraphics[width=0.85\linewidth]{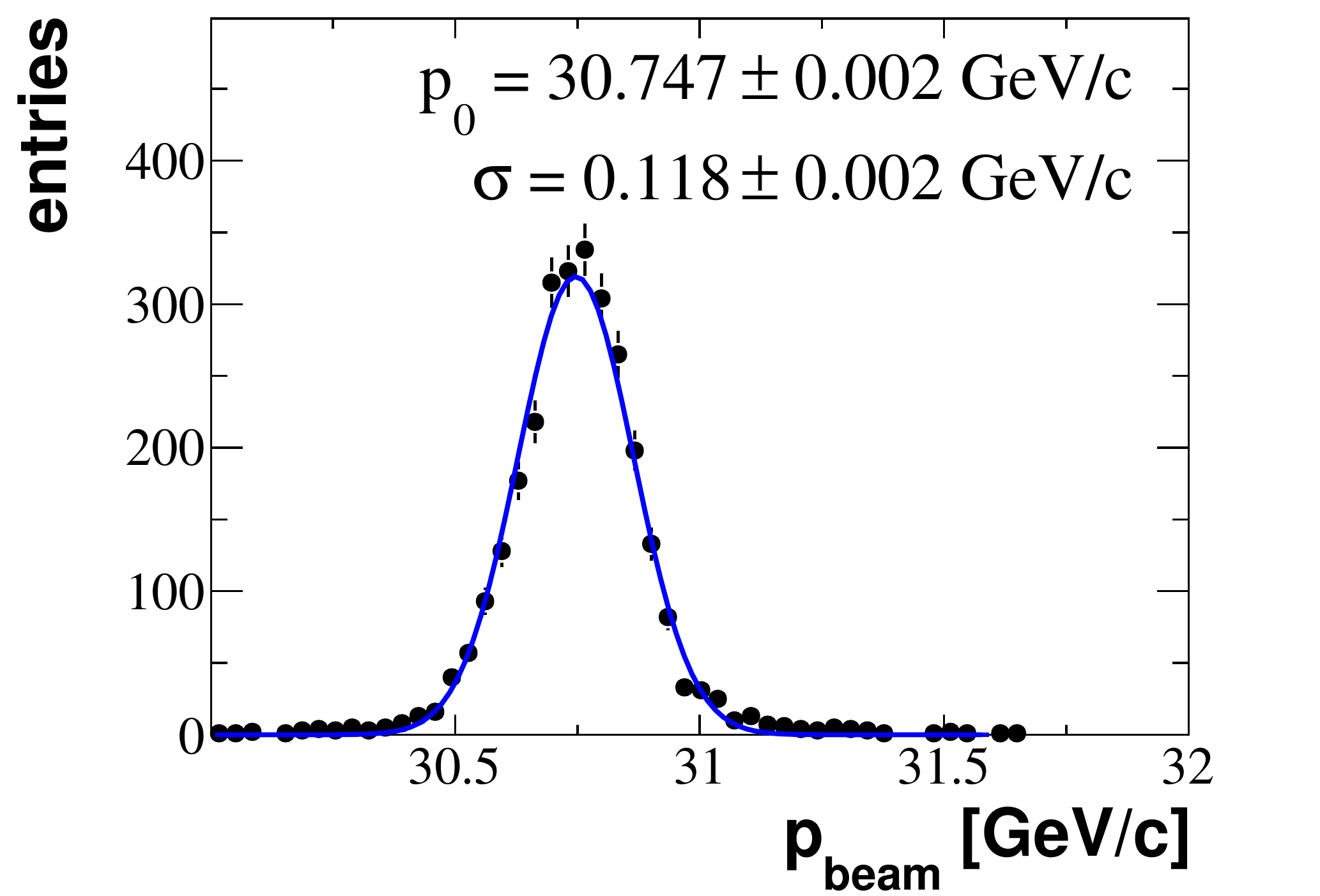}
\else
\includegraphics[width=0.85\linewidth]{beam_momentum}
\fi
\end{center}
\caption{(Color online) 
  The beam momentum distribution measured by the reconstruction of
  beam particles in the TPCs. Only statistical errors are shown.
}
 \label{beam_momentum}
\end{figure}

The beam momentum was measured directly 
in a dedicated run by bending
the incoming beam particles into the TPCs with the full magnetic field.
The measured beam momentum distribution is shown in Fig.~\ref{beam_momentum}.
The mean value of 30.75~GeV/\textit{c} agrees with the set value of
30.92~GeV/\textit{c} within the available precision of setting the beam magnet currents
($\approx$0.5\%) in the H2 beam-line.

Interactions in the target are selected by an anti-coincidence of the incoming beam protons with
a small, 2-cm-diameter, scintillation counter (S4) placed on the beam trajectory
between the two vertex magnets (see Fig.~\ref{fig:detector}). This
minimum bias trigger is based on the disappearance of the incident
proton. A measurement of the interaction cross section
is discussed in Sec.~\ref{Sec:Claudia}.

The results presented in this paper are based on the analysis of $667 \times 10^3$
proton interaction triggers recorded with the carbon target inserted 
and $46 \times 10^3$ proton interaction triggers recorded with the carbon target 
removed. Additionally $47 \times 10^3$ events were recorded with beam proton triggers 
and 33 $\times 10^3$ with beam triggers without any beam particle identification.

\begin{figure*}[htbp]
\begin{center}
\includegraphics[width=0.70\linewidth,angle=-90]{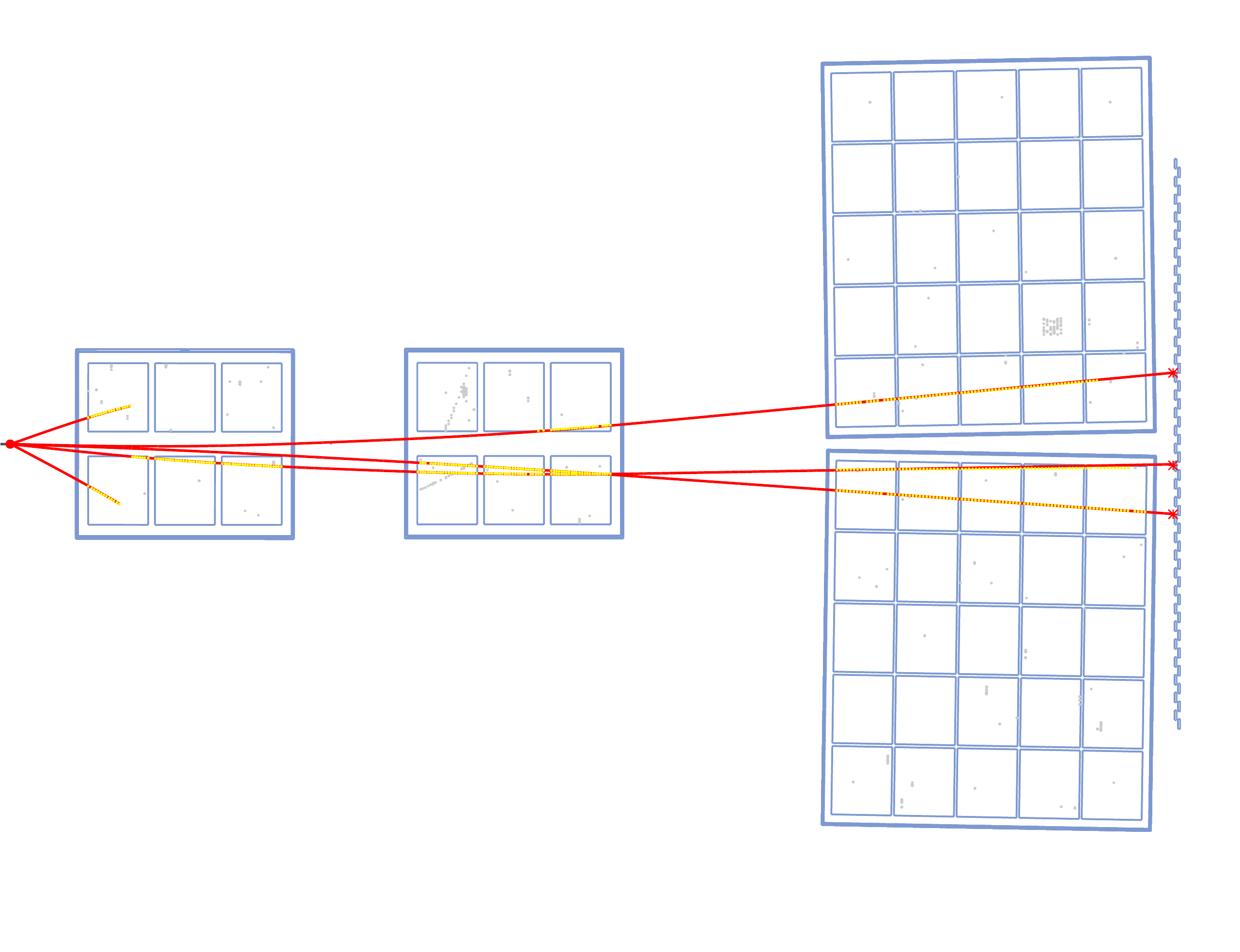}
\end{center}
\vspace*{-1cm}
  \caption{(Color online) 
Topview (projection to the $x$-$z$ plane) of a p+C interaction 
as detected by the four NA61 TPCs and
the ToF-F detector
(for details see Sec.~\ref{Sec:set-up} and Fig.~\ref{fig:detector}). 
The red lines correspond to the trajectories
of tracks produced in the interaction and
reconstructed using the TPC clusters indicated by yellow points.
Stars correspond to the reconstructed ToF-F hits. 
}
\label{fig:real_data_event}
\end{figure*}

\begin{figure}[h]
\begin{center}
\includegraphics[width=0.85\linewidth]{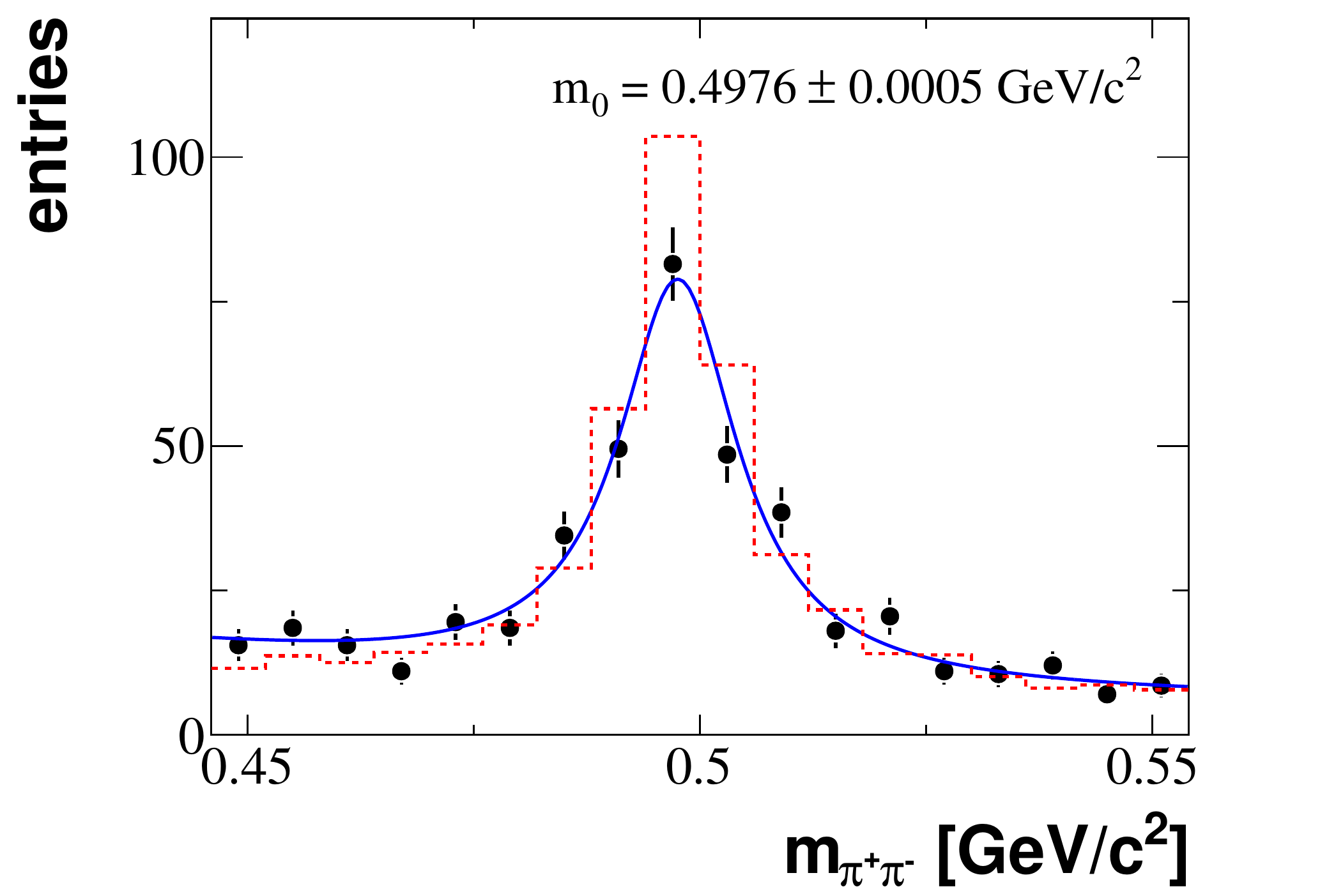}
\end{center}
  \caption{(Color online)
  Invariant mass distribution of reconstructed $K^0_S$ candidates.
  The fitted mean value of the peak is given in the legend. 
  The dashed histogram shows the simulated distribution, which
  is normalized to the data on the right side of the peak.
}
\label{fig:K0_study}
\end{figure}

\section{Data reconstruction,  simulation and detector performance}
\subsection{Calibration}
\label{calibration}
The calibration procedure of the 2007 NA61 data was largely based on the
approach
developed for the NA49 experiment~\cite{NA49}
and consists of several consecutive steps resulting
in optimized parameters for

\begin{enumerate}[(i)]
\setlength{\itemsep}{1pt}
\item detector geometry, drift velocity, and residual corrections,
\item magnetic field,
\item time-of-flight measurements, and
\item specific energy loss measurements.
\end{enumerate}
Each step
involved reconstruction of the data required to optimize a
given set of calibration constants followed by
verification  procedures.
Details of the procedure and quality assessment are presented in
Ref.~\cite{Status_Report_2008}. The quality of detector calibration in
quantities relevant for this paper is illustrated in the following subsections.

\begin{figure}[h!]
\begin{center}
\includegraphics[width=0.85\linewidth]{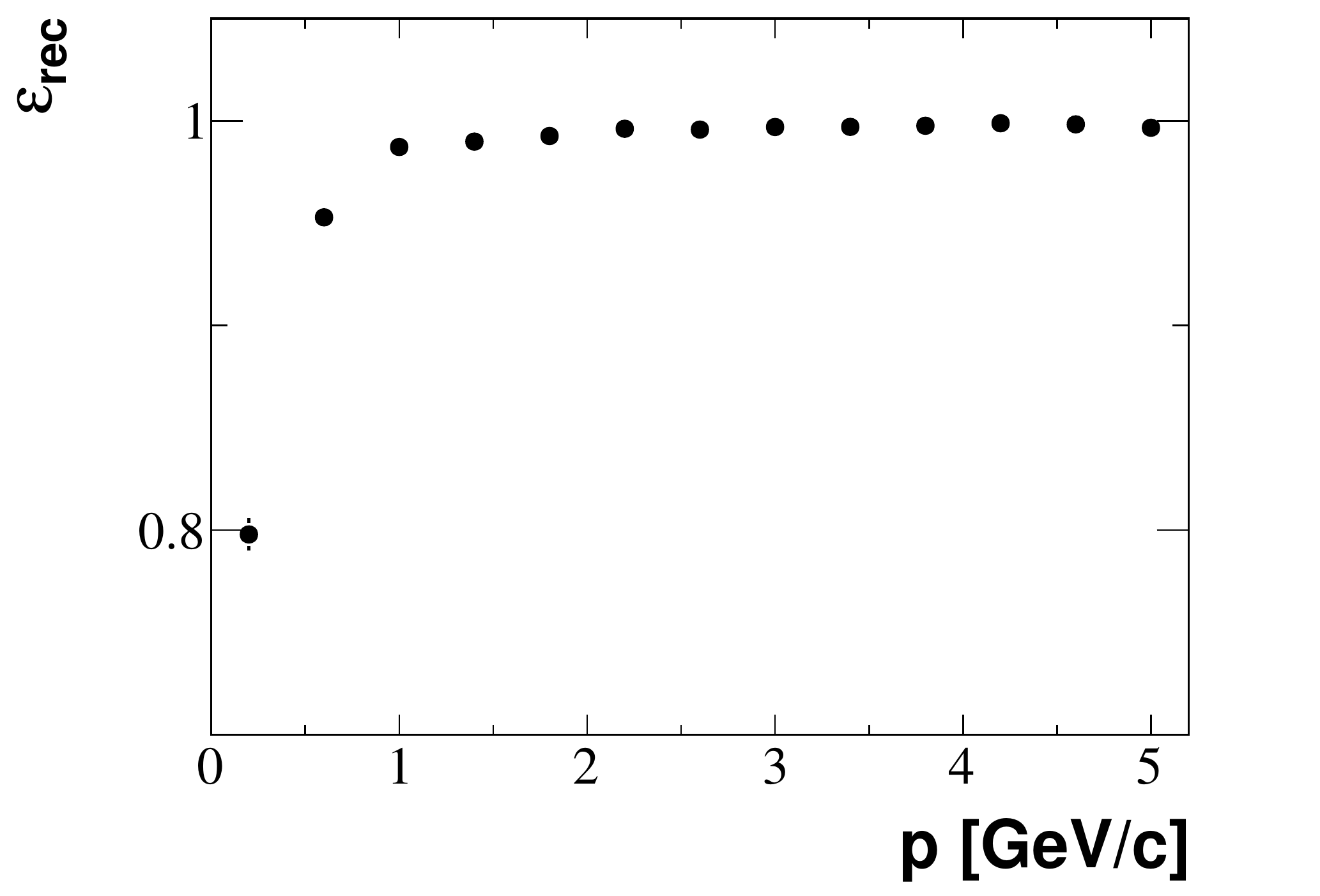}
\end{center}
  \caption{(Color online) 
   Track reconstruction efficiency for negatively charged
   particles as a function of momentum in the
   polar angle interval [100,140]~mrad.
}
\label{fig:track_eff}
\end{figure}

\subsection{Reconstruction}
Reconstruction algorithms used for the analysis described here
are based on those developed by
the NA49 Collaboration~\cite{NA49}.
The main steps of the
reconstruction procedure are

\begin{enumerate}[(i)]
\setlength{\itemsep}{1pt}
\item cluster finding in the TPC raw data, and calculation of a cluster
center of gravity and total charge,
\item reconstruction of local track segments in each TPC
separately,
\item matching of track segments from different TPCs
into global tracks,
\item track fitting through the magnetic field and
determination of track parameters at the first measured TPC cluster,
\item determination of the interaction vertex
as the intersection point of the incoming beam particle
with the middle target plane,
\item refitting the particle trajectory using the interaction vertex as an
additional point and determining the particle momentum at the interaction vertex,  and
\item matching of ToF-F hits with the TPC tracks.
\end{enumerate}

An example of a reconstructed event is shown
in Fig.~\ref{fig:real_data_event}.
One can observe different track topologies, including long tracks
hitting the ToF-F.

\begin{figure}[h!]
\includegraphics[width=0.85\linewidth]{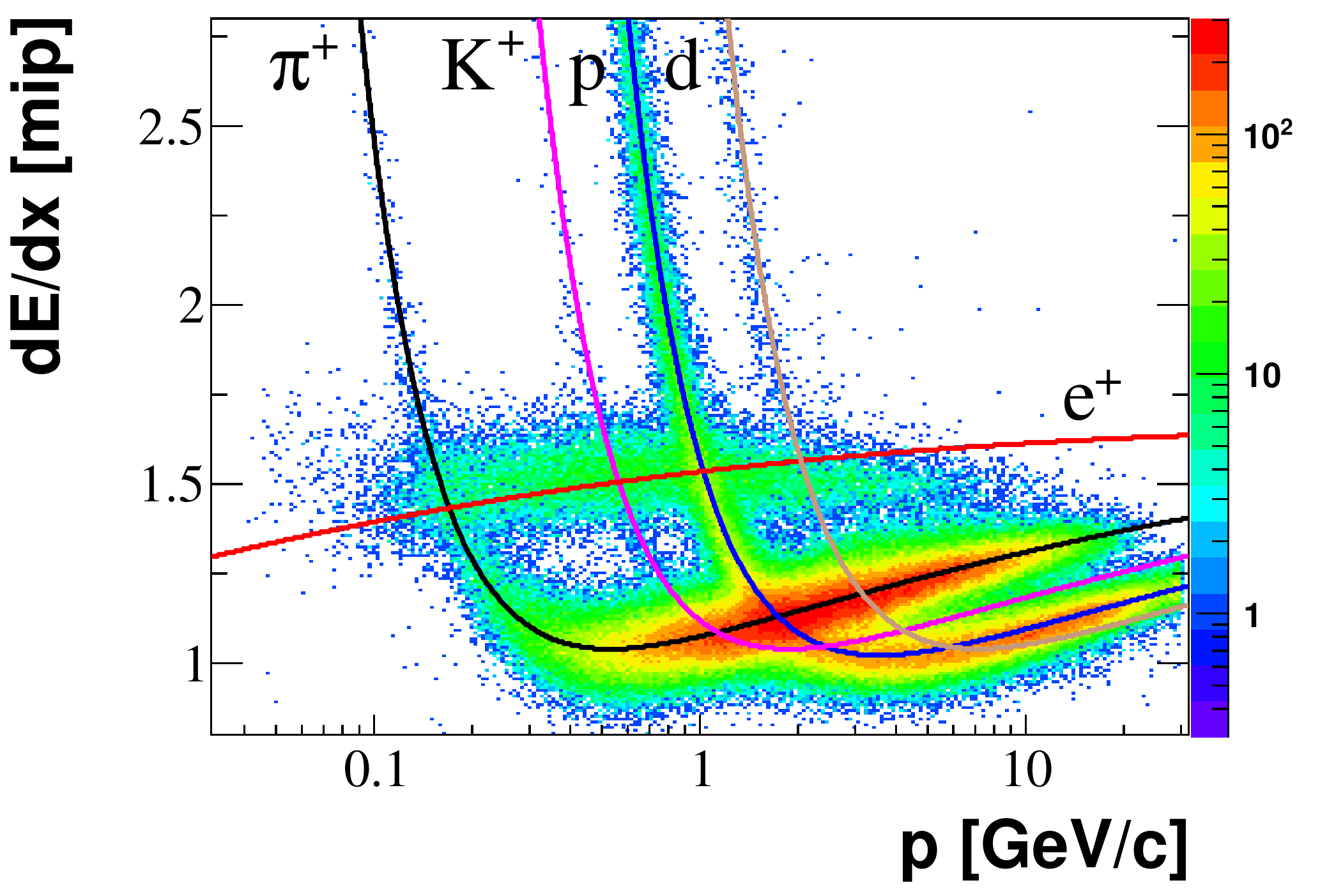}
\includegraphics[width=0.85\linewidth]{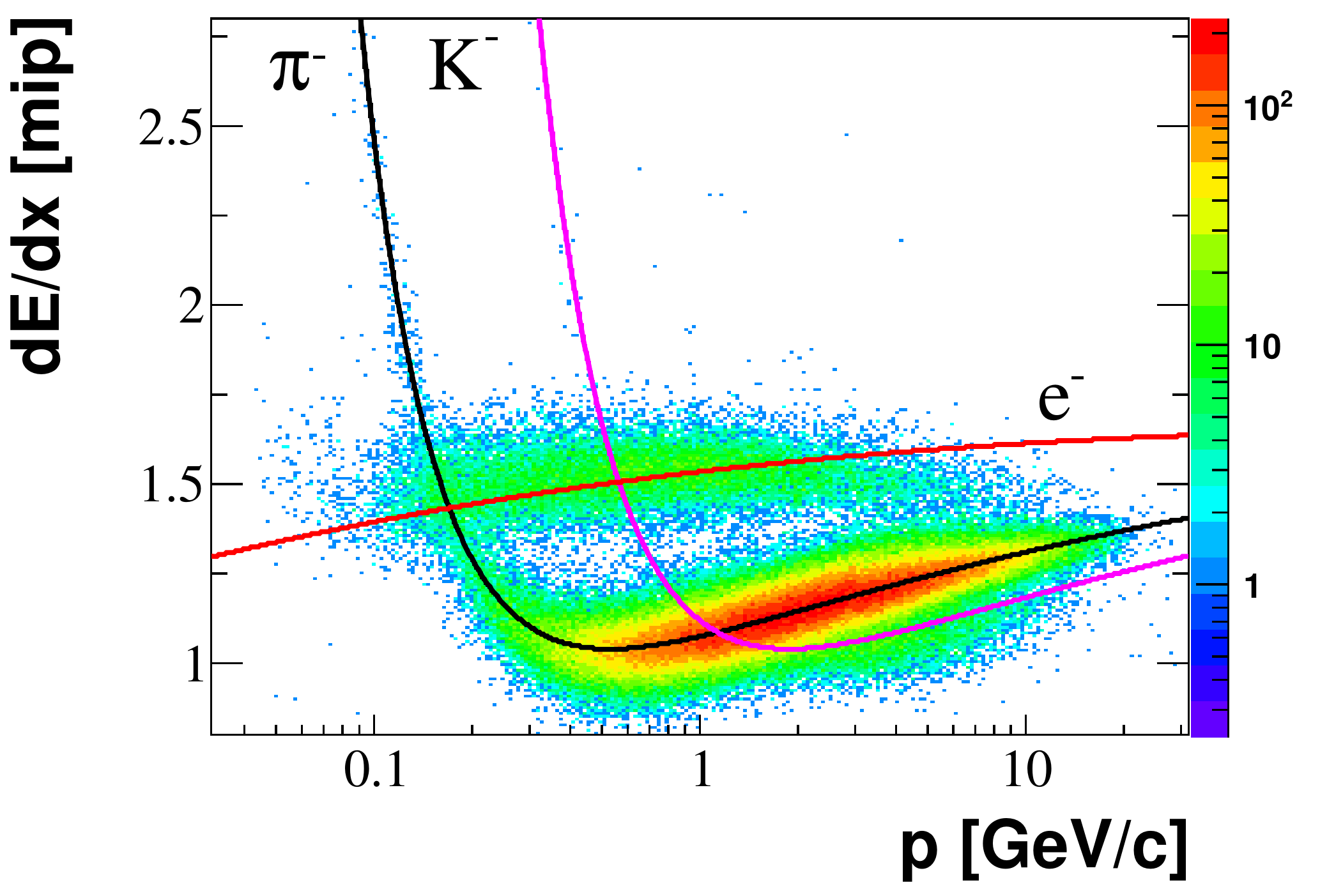}
\caption{(Color online) 
Specific energy loss in the TPCs for positively ($top$) and negatively ($bottom$) charged particles as a function of momentum. Curves show parametrizations of the mean $dE/dx$ calculated for different particle species.
}
\label{dedx_pos_neg}
\end{figure}

\subsection{Monte Carlo simulation}
\label{Sec:MC}
Simulation of the NA61 detector response used to correct
the raw data consists of the following steps 
(see~\cite{Nicolas} for more details):

\begin{enumerate}[(i)]
\setlength{\itemsep}{1pt}

\item
generation of p+C interactions at 31~GeV/\textit{c} using the
\VenusLong~\cite{Venus} model,

\item
propagation of outgoing particles through the detector material
using the GEANT 3.21 package~\cite{Geant3}, which
takes into account the magnetic field as well as relevant physics processes,
such as particle interactions and decays,

\item
simulation of the detector response using dedicated
NA61 packages which introduce distortions corresponding to
all corrections applied to the real data (see Sec.~\ref{calibration}),

\item
storage of the
simulated events in a file which has the same
format as the raw data,

\item reconstruction of the simulated events with the same reconstruction
chain as used for the real data, and

\item
matching of the simulated and reconstructed tracks.

\end{enumerate}

Finally, the ratio of the number of reconstructed (matched) tracks 
in a given ($p$,~$\theta$) bin to the number of generated tracks
of the specific particle type in this bin is used as the global
correction factor (Sec.~\ref{Sec:Tomek}). 
Alternatively, flat-phase-space Monte Carlo simulation was used to calculate 
the same ratio 
for a given type of primary particle and similar results were obtained 
within errors for the geometrical acceptance and reconstruction efficiency 
corrections. The simulation based on the VENUS4.12 model, which
has realistic ratios between the different particle species,
was used to obtain the corrections due to the admixture 
of secondary particles e.g., from weak decays of
other particles and from secondary interactions. 
The systematic uncertainties due to the dependence of the corrections 
on the model are discussed in Sec.~\ref{Sec:Syst}.

\subsection{Detector performance}

The quality of measurements was studied
by reconstructing the momentum of beam particles (see Sec.~\ref{Sec:trigger})
and masses of $K^0_S$ and $\Lambda$ particles from their
$V^0$ decay topology.
As an example the invariant mass distribution of $K^0_S$
candidates
is plotted in
Fig.~\ref{fig:K0_study}.
The  peak position (0.4976 $\pm$ 0.0005~GeV/\textit{c}$^2$)
agrees  with the known $K^0_S$ mass.
The widths of the $K^0_S$ and $\Lambda$
invariant mass peaks are
reasonably
reproduced by the Monte Carlo simulation, which confirms
its validity for the studies presented below.

\begin{figure}[h]
\includegraphics[width=0.85\linewidth]{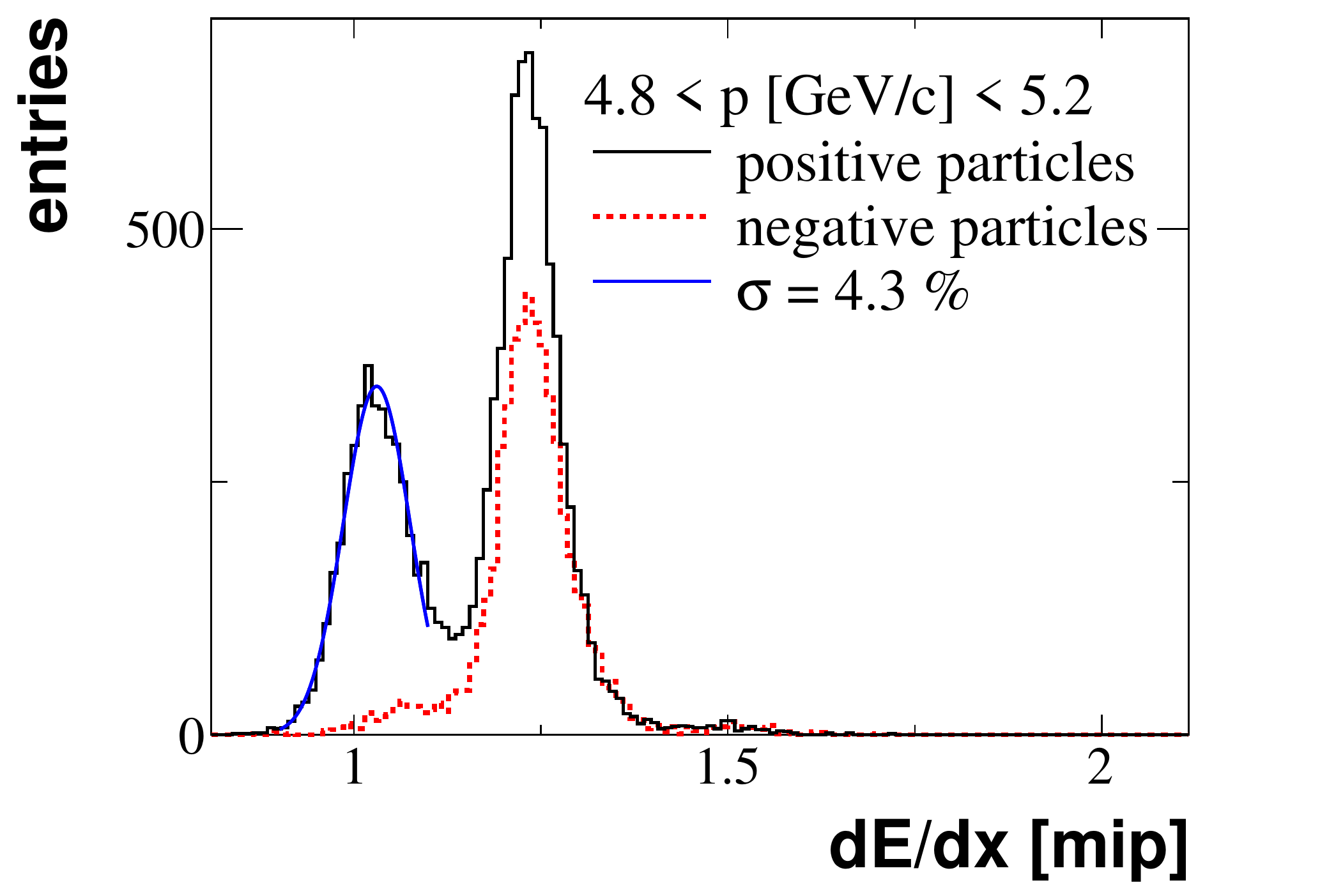}
\includegraphics[width=0.85\linewidth]{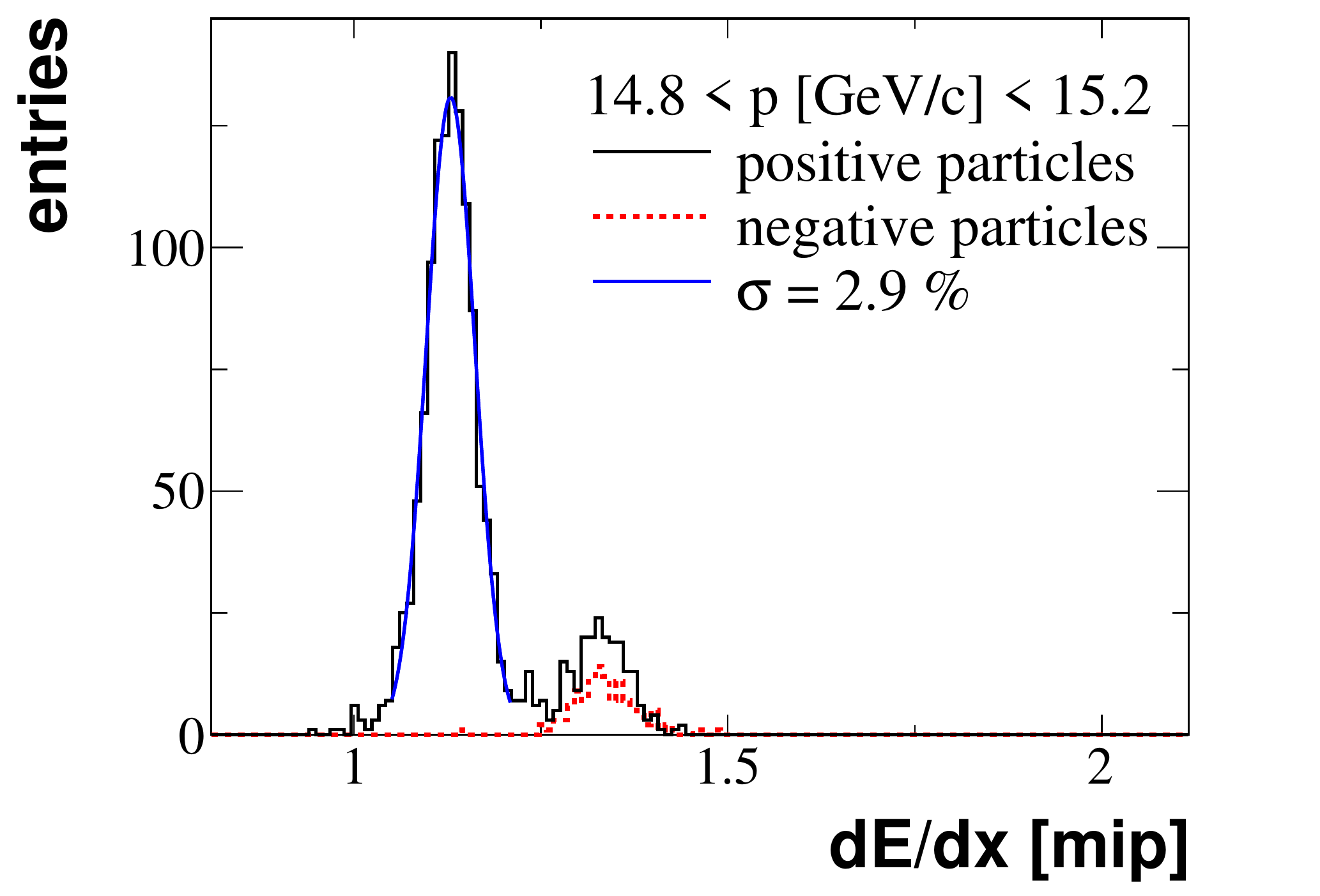}
\caption{(Color online) 
  The $dE/dx$ distributions for positively (full line)
  and negatively (dashed line)
  charged particles at 5 and 15~GeV/\textit{c} particle momenta; the momentum bin
  width is 0.4~GeV/\textit{c}.
}
\label{dedx_spectra_pos_neg}
\end{figure}

\begin{figure}[h]
\begin{center}
\includegraphics[width=0.8\linewidth]{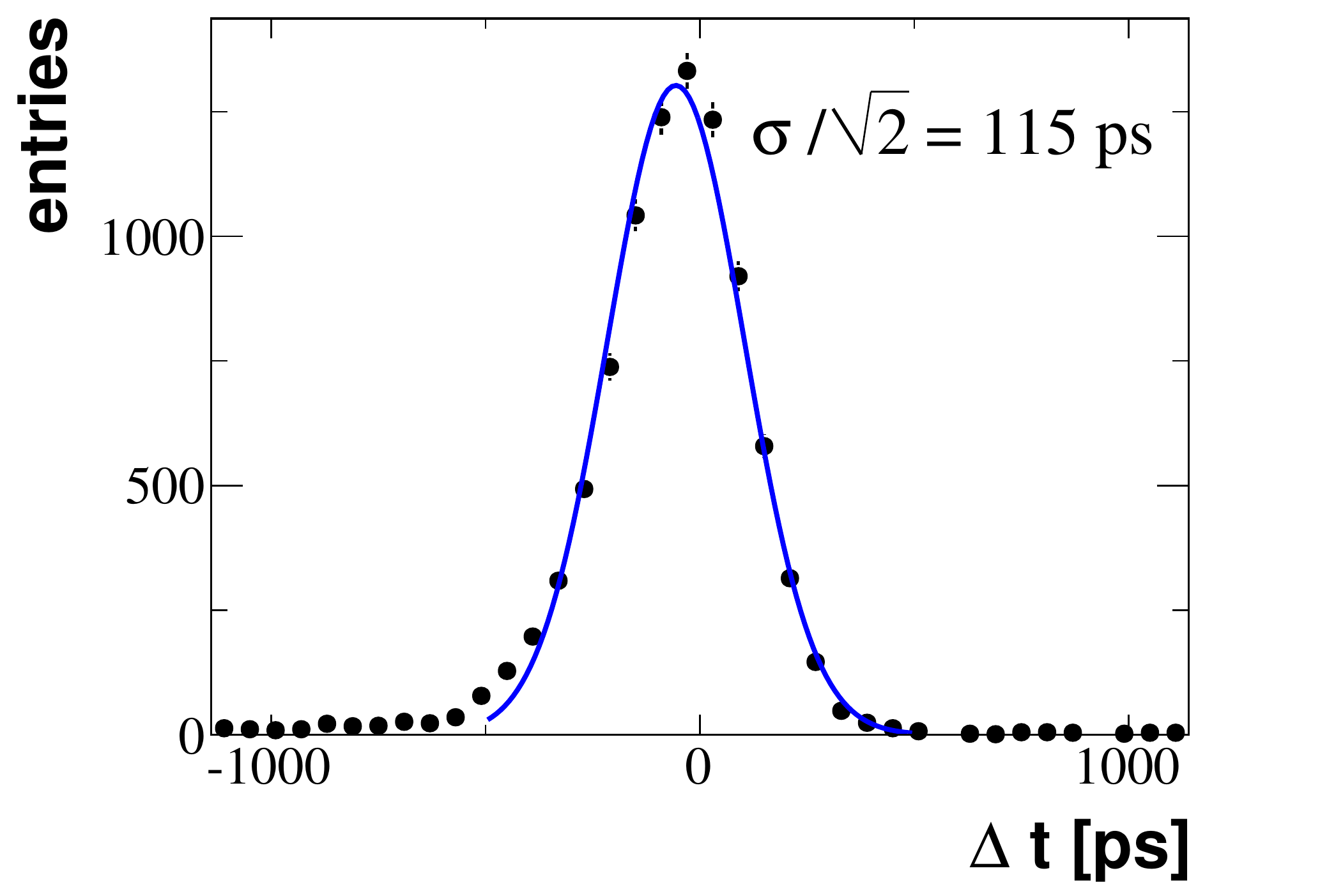}
\includegraphics[width=0.8\linewidth]{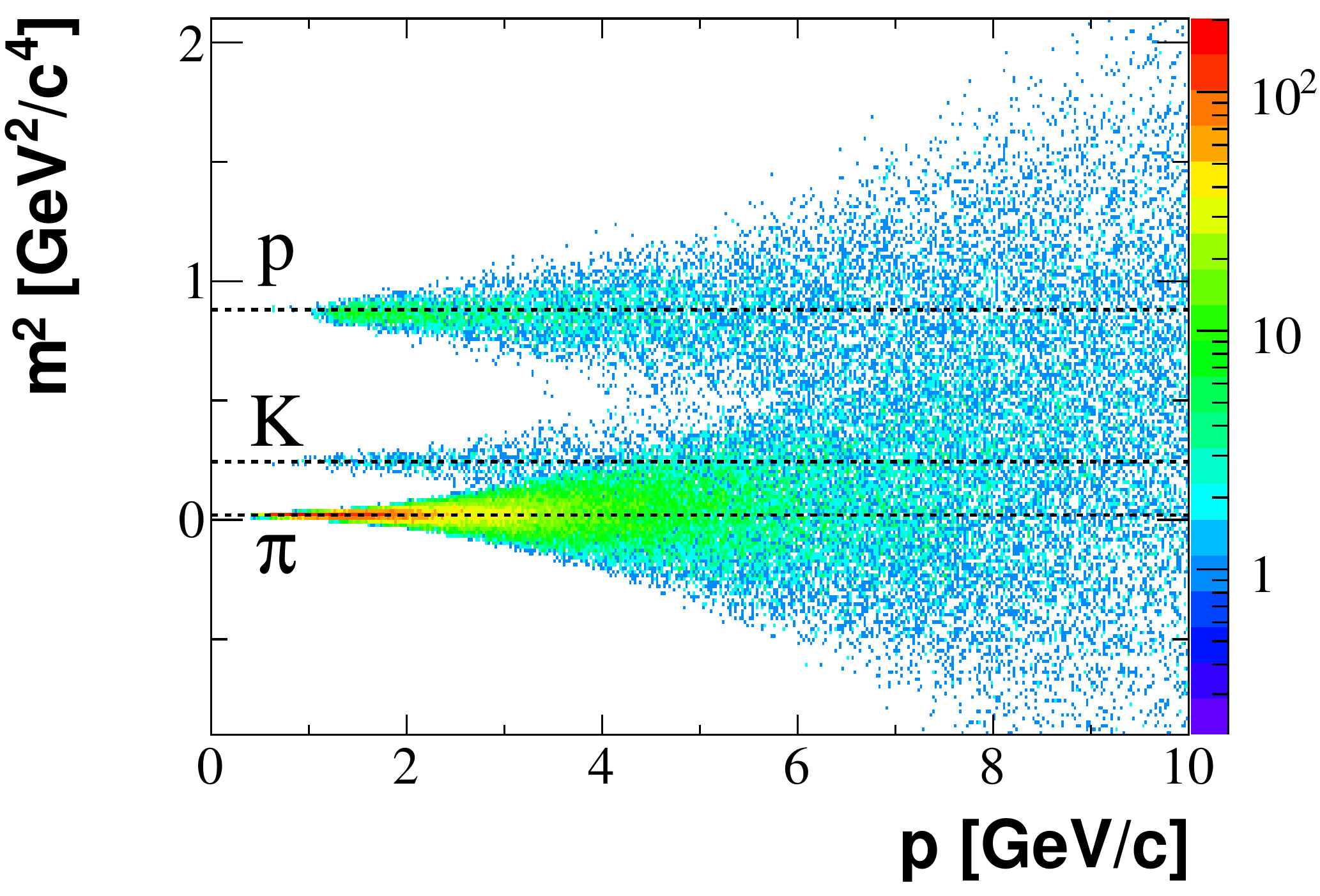}
\end{center}
  \caption{(Color online) 
   {\it Top:} Distribution of the difference between a particle's time-of-flight
   measured independently by two overlapping scintillator bars of the ToF-F
   detector. The width of the distribution is about 160~ps, indicating
   a $tof$ resolution of about 115~ps for a single measurement.
  {\it Bottom:} Mass squared, derived from the ToF-F measurement and the fitted 
   path length and momentum, versus momentum $p$.
  The lines show the expected mass squared values for different particles.
}
\label{fig:ftof_resolution}
\end{figure}

\begin{figure*}[tb]
\begin{center}
\includegraphics [width=0.31\linewidth]{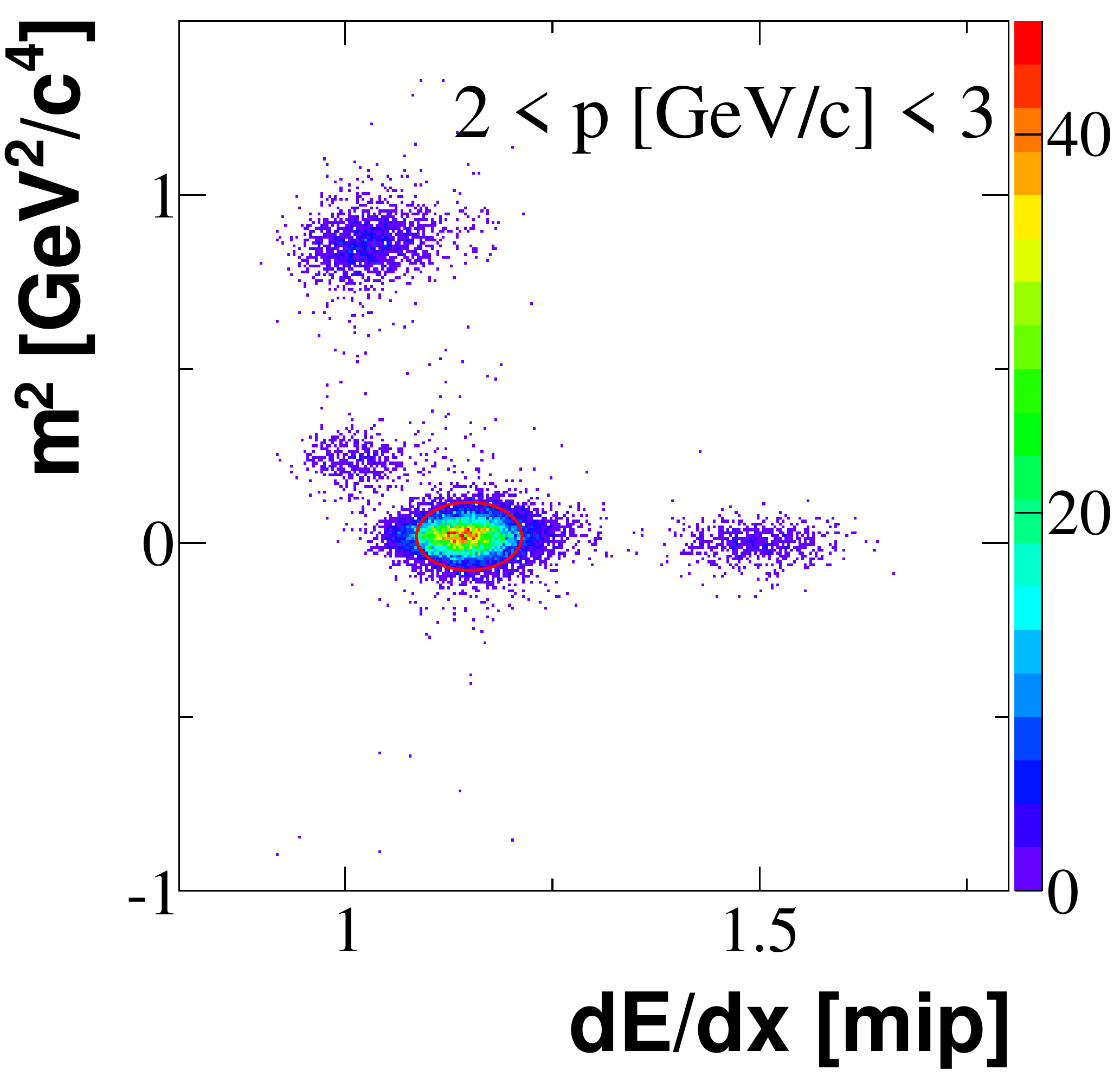}
\includegraphics [width=0.31\linewidth]{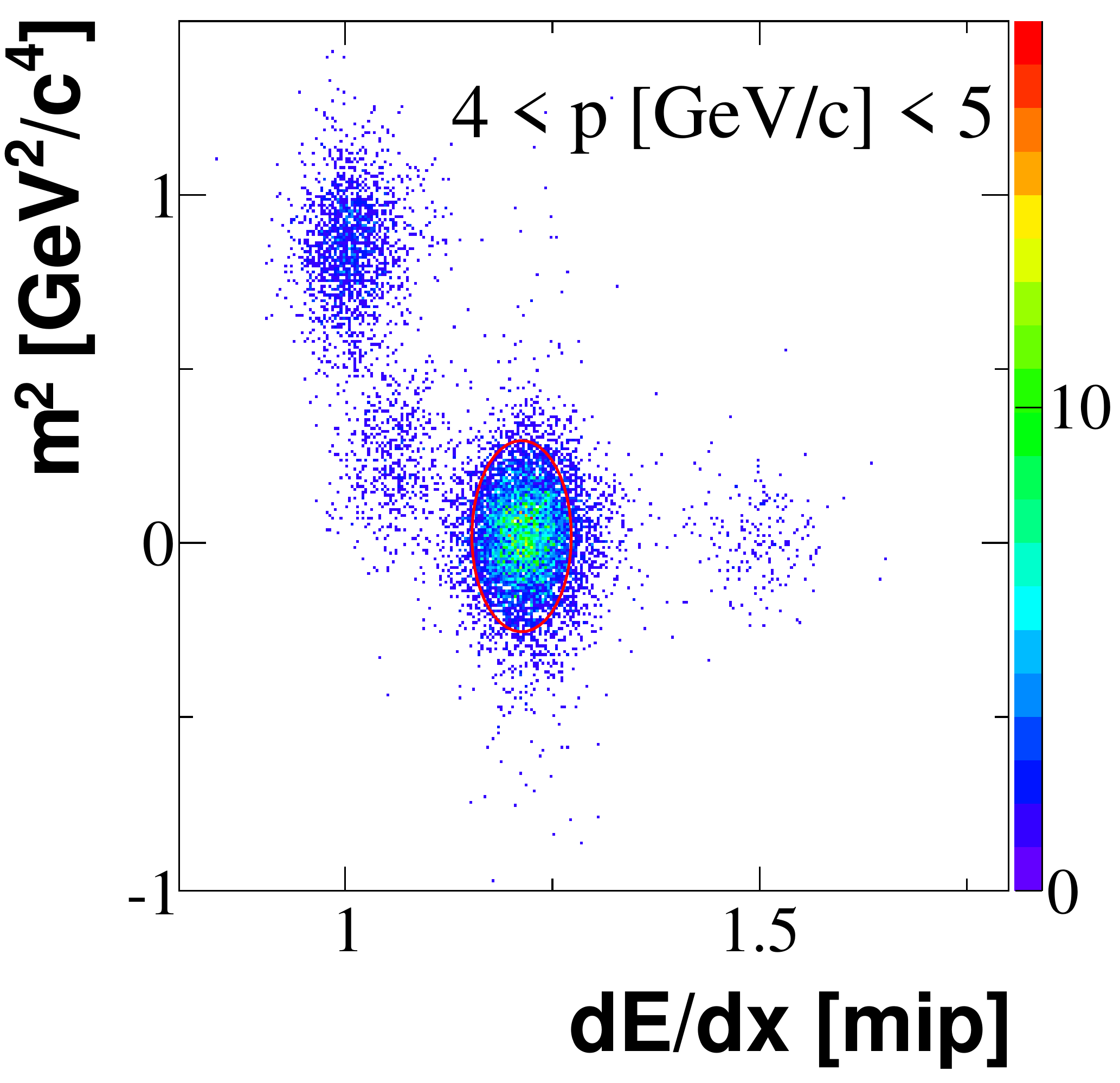}
\includegraphics [width=0.31\linewidth]{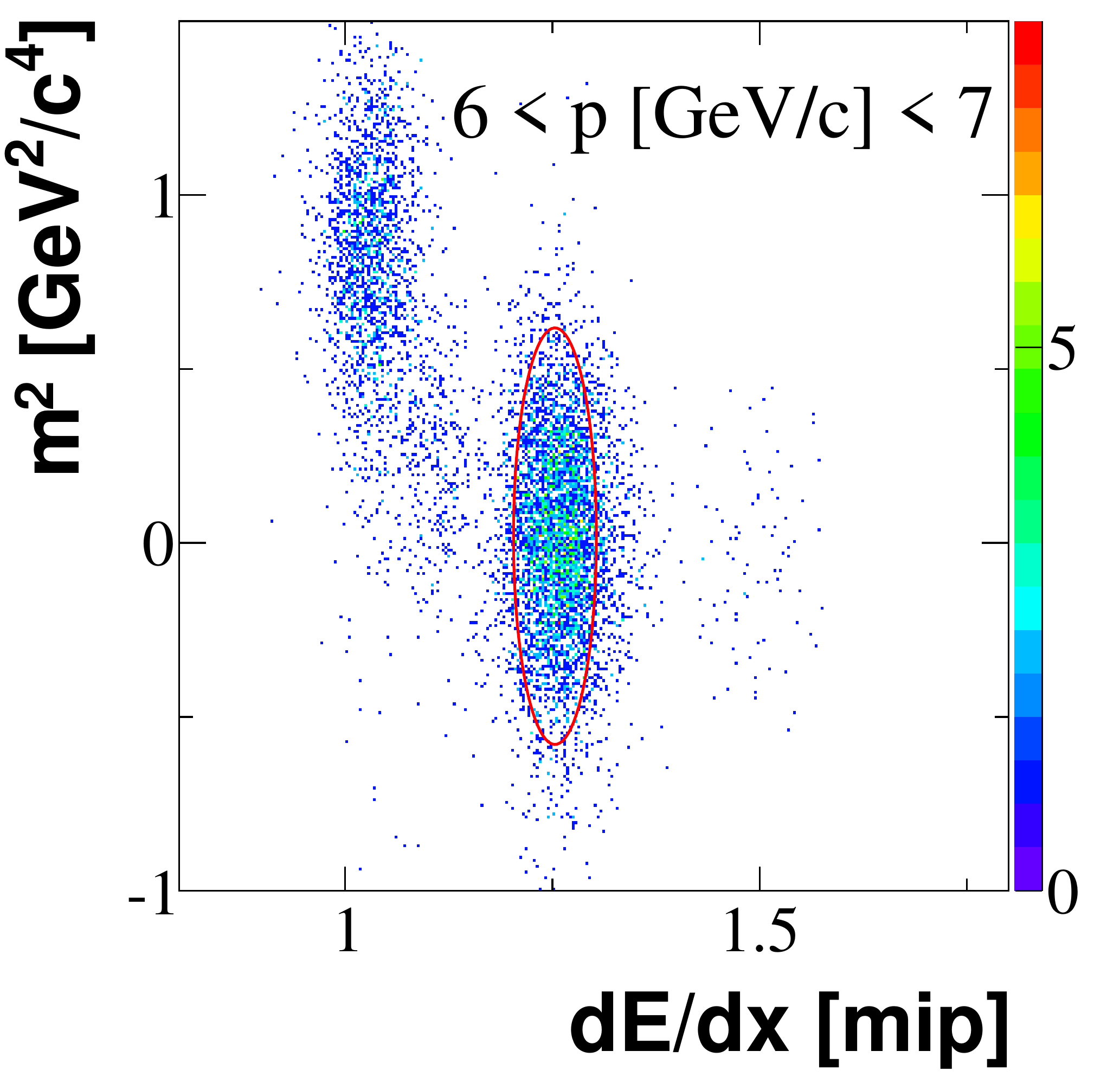}
\caption{(Color online) 
  Examples of two-dimensional $m^2$--$dE/dx$ plots for
  positively charged particles in three
  momentum intervals indicated in the panels. 
  $2\sigma$ contours around fitted pion peaks are shown.
  The left and middle plots correspond to the $dE/dx$ cross-over
  region while the right plot is at such a high momentum that the ToF-F
  resolution becomes a limiting factor. The combination of both
  measurements provides close to 100\% purity in the pion
  selection over the whole momentum range.
}
\label{fig:tof-dedx}
\end{center}
\end{figure*}

The track reconstruction efficiency and resolution of
kinematic quantities were calculated by matching of
simulated and reconstructed tracks.
As an example,
the reconstruction efficiency
as a function
of momentum for negatively charged tracks in the polar
angle interval [100,140]~mrad is shown in
Fig.~\ref{fig:track_eff}.
The momentum resolution $\sigma(p)/p^2$, averaged
over different track topologies in the detector, was estimated to be 
about 2$\times$10$^{-3}$,
7$\times$10$^{-3}$, and 3$\times$10$^{-2}$~(GeV/\textit{c})$^{-1}$ at
$p>5$\,GeV/\textit{c}, $p=2$\,GeV/\textit{c}, and $p=1$\,GeV/\textit{c}, respectively.
The results depend somewhat on particle production properties, and
the quoted numbers refer to negatively
charged particles (more than 90\% of them are pions)
produced by \Venus and passing the track selection cuts
described in Sec.~\ref{Sec:cuts}.

Particle identification was performed based on
the energy loss measurements in the TPCs and on the time-of-flight
information from the ToF-F detector.
The calibrated $dE/dx$ distributions as a function of particle momentum 
for positively and negatively
charged particles are presented in Fig.~\ref{dedx_pos_neg}.
The Bethe-Bloch parametrization of the mean energy loss, scaled to
the experimental data~(see Sec.~\ref{Sec:Magda}), is
shown by the curves for positrons (electrons),
pions, kaons, protons, and deuterons.
The typical achieved $dE/dx$ resolution is 3\%-5\%; see Fig.~\ref{dedx_spectra_pos_neg}.

The intrinsic $tof$ resolution of the ToF-F detector is
about 115~ps as derived from measurements of $tof$ for particles traversing
the overlap region of two adjacent scintillator bars
(see Fig.~\ref{fig:ftof_resolution}).
Figure~\ref{fig:tof-dedx} shows distributions of the mass squared
$m^2$ (derived from the $tof$, fitted momentum, and path length)
versus $dE/dx$ (measured in the TPCs)
in selected intervals of particle momentum.
The combined $m^2$ and $dE/dx$ measurements allow us to extract yields of
identified particles even in the cross-over region of the Bethe-Bloch
curves (1--4~GeV/\textit{c} momentum range).

The geometrical acceptance of the ToF-F detector is limited to
particle momenta above about 0.8~GeV/\textit{c}.
However, in the low-momentum region (less than about 1~GeV/\textit{c})
the $dE/dx$ information alone is sufficient to distinguish pions
from electrons/positrons, kaons, and protons;
see
Fig.~\ref{dedx_pos_neg}.

\begin{figure}[h!]
\begin{center}
\includegraphics[width=0.85\linewidth]{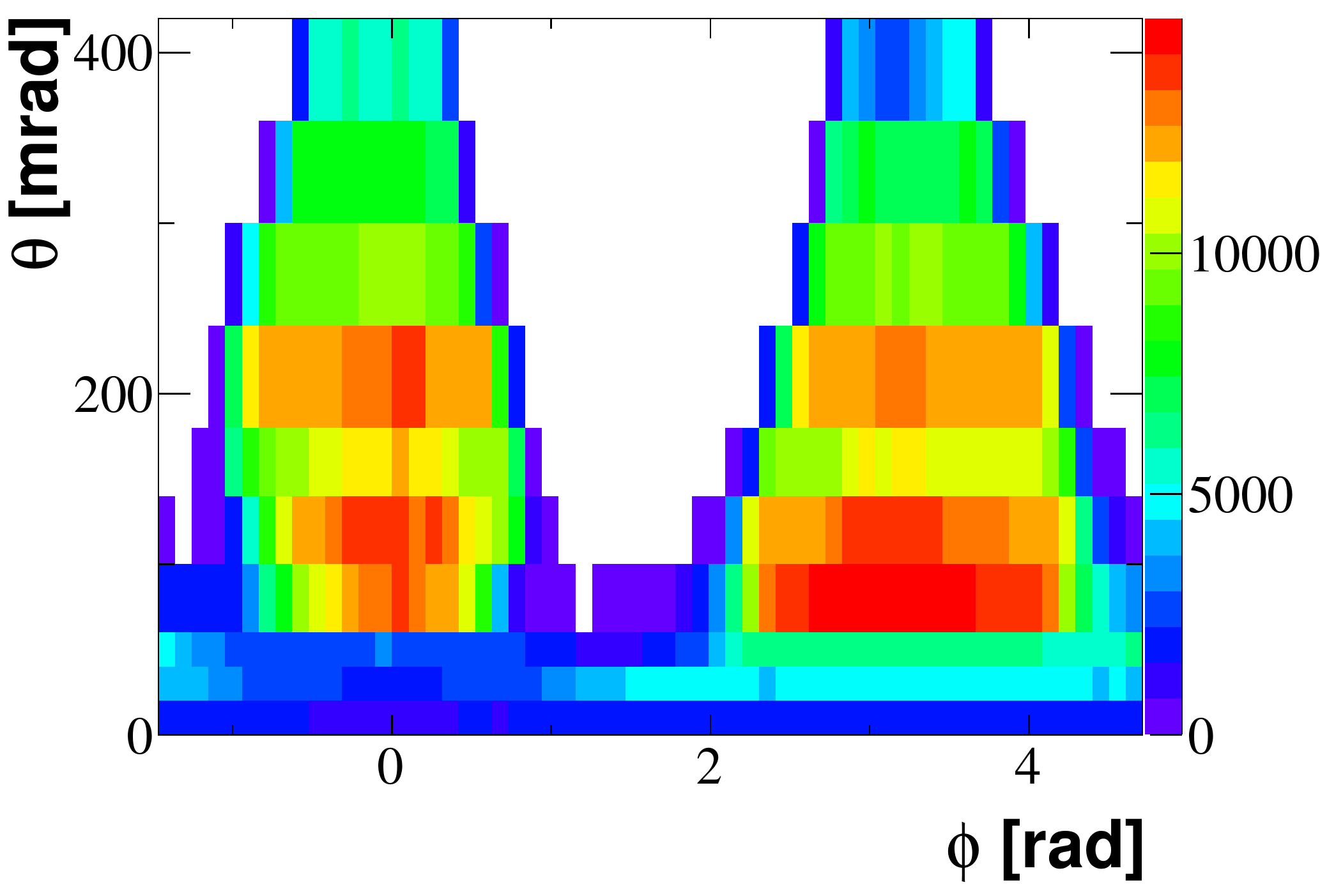}
\end{center}
  \caption{(Color online) 
    Polar angle ($\theta$) vs azimuthal angle ($\phi$) distribution
    for reconstructed negatively charged
    particles in the momentum interval 
    $ 0.5 < p~\mbox{[GeV/\textit{c}]} < 5.0 $.
}
\label{fig:track_phi}
\end{figure}

\begin{figure}[h!]
\begin{center}
\includegraphics[width=0.85\linewidth]{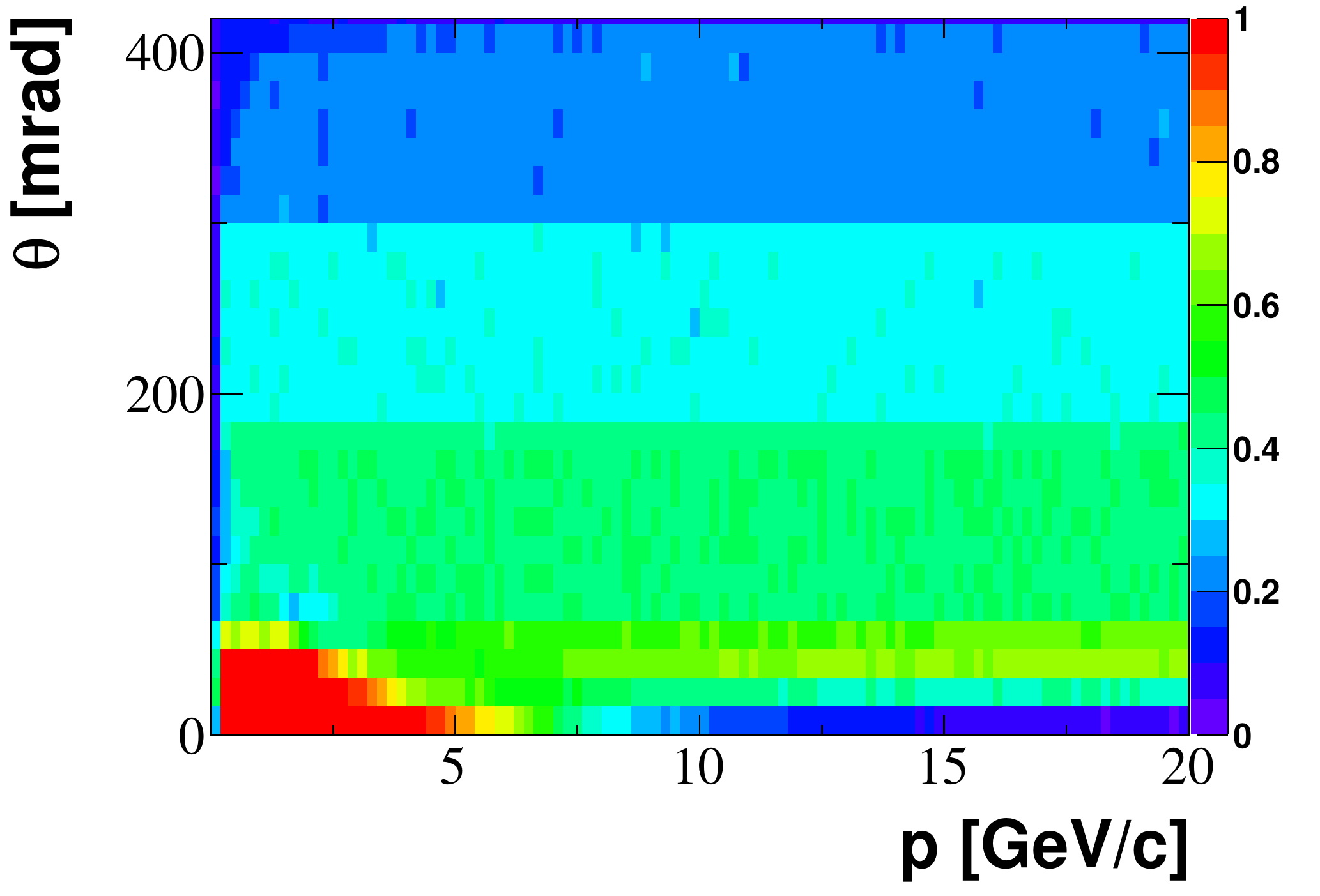}
\includegraphics[width=0.85\linewidth]{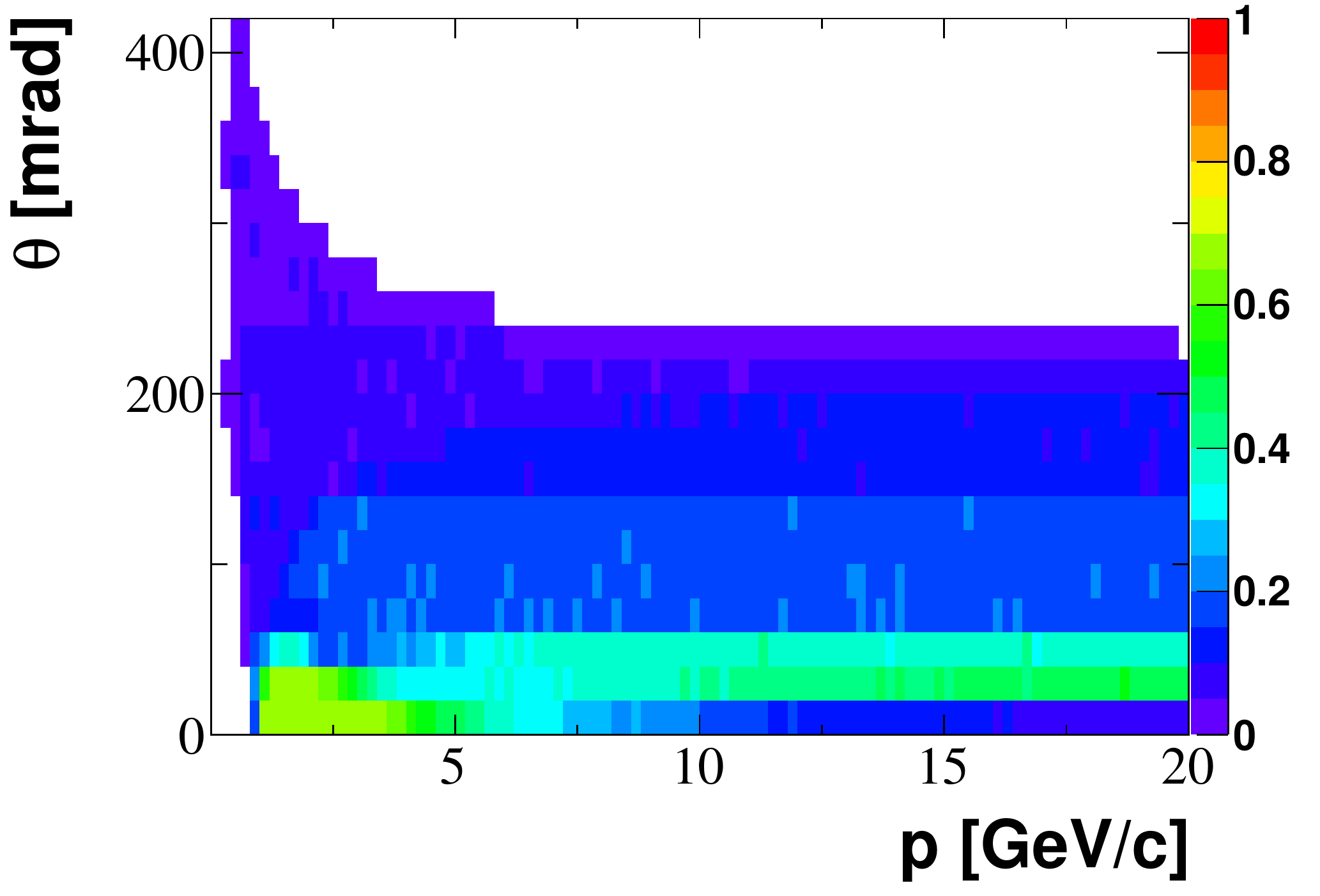}
\caption{(Color online) 
  Fraction of accepted particles as a function of momentum
  and polar angle, after the track acceptance cuts (see Sec.~\ref{Sec:cuts}) 
  for negatively charged tracks ($top$),
  and after an additional ToF-F acceptance cut (see Sec.~\ref{Sec:Sebastien}) 
  for positively charged tracks ($bottom$).
  The first polar angle bin, [0,20]~mrad, is fully covered by accepted particles
  up to 7.6~GeV/\textit{c}. 
}
\label{fig:geometrical}
\end{center}
\end{figure}

About  47\% of all charged particles produced at the
interaction vertex pass the standard TPC track selection
cuts (see Sec.\ref{Sec:cuts}).
In particular,
the uninstrumented region 
around
the beam-line in VTPC-1 and VTPC-2
(see Fig.~\ref{fig:detector}) and the magnet apertures lead to a limited
acceptance in azimuthal angle. 
This is illustrated in Fig.~\ref{fig:track_phi}, where a distribution of all accepted
negatively charged particles is shown as a function of azimuthal and
polar angles.
The fraction of charged particles accepted in the TPCs
and in the TPCs plus ToF-F is
plotted as a function of momentum and polar angle
in Fig.~\ref{fig:geometrical}.
As the azimuthal acceptance varied with the polar angle $\theta$ 
an individual wedge in azimuthal angle $\phi$ was adjusted 
for each particular $\theta$ bin.
Comparing plots in Fig.~\ref{fig:geometrical} with
Fig.~\ref{fig:p-theta-pi+SK},
one concludes that the NA61/SHINE acceptance fully covers the phase space region
of interest for T2K.

\section{Analysis techniques}
\label{sec:analysis}

This section presents the procedures used for data analysis.
Crucial for this analysis is the identification of the pions produced.
Depending on the momentum interval, different approaches were
adopted, which led also to different track selection criteria.
The task is facilitated for the negatively charged pions by the
observation that more than 90\% of primary negatively charged particles
produced in p+C interactions at this energy are $\pi^-$, and thus the analysis
of $\pi^-$ spectra can also be carried out without additional particle
identification.

This section starts from a presentation of 
event and track selection cuts. 
Next, the analysis of interaction cross sections is
described~\cite{Claudia}.
Finally, three analysis methods applied to obtain pion spectra are
introduced.
These are

\begin{enumerate}[(i)]
\setlength{\itemsep}{1pt}

\item analysis of $\pi^-$ mesons via measurements of
negatively charged particles
({\it $h^-$ analysis}~\cite{Tomek}); see Sec.~\ref{Sec:Tomek};
\item analysis of $\pi^+$ and $\pi^-$ mesons
identified via $dE/dx$ measurements
in the TPCs ({\it $dE/dx$ analysis at low momentum}~\cite{Magda}); see Sec.~\ref{Sec:Magda}; and
\item analysis of $\pi^+$ and $\pi^-$ mesons
identified via time-of-flight and $dE/dx$
measurements in the ToF-F and TPCs, respectively
({\it $tof$--$dE/dx$ analysis}~\cite{Sebastien}); see Sec.~\ref{Sec:Sebastien}.
\end{enumerate}

Each analysis yields fully corrected pion spectra with
independently calculated statistical and systematic errors.
The spectra  were compared
in overlapping phase-space domains to check their consistency.
Complementary domains were combined to reach maximum acceptance.

The final results  refer to pions
(denoted as \textit{primary pions})
produced in p+C interactions at 31~GeV/\textit{c}
by strong interaction processes and in the electromagnetic decays of
produced hadrons.

\subsection{Event and track selection}
\label{Sec:cuts}

This section presents event and track selection criteria
common for all analysis methods. Selection criteria
specific to a given method are described in the corresponding
subsection below.

The analysis was based on a sample of $521 \times 10^3$ events selected 
from the total sample of $667 \times 10^3$ registered and reconstructed 
proton interaction triggers recorded with the carbon target inserted.
The selected events are required to have signals in each plane of all three
BPD detectors with properly reconstructed beam tracks which include 
measured points on both planes of \mbox{BPD-3}.
This criterion essentially removes
contamination by interactions upstream of the target.
For the event sample with the target removed, the selection reduces the
number of events from $46  \times 10^3$ to $17  \times 10^3$.

In order to select well-measured tracks in the TPCs as well as to reduce the contamination
of tracks from secondary interactions and weak decays the following
track selection criteria were applied:

\begin{enumerate}[(i)]
\setlength{\itemsep}{1pt}

\item  the track momentum fit at the
interaction vertex should have converged,
\item  the total number of reconstructed points on a track should
be at least~30,
\item the sum of the number of reconstructed points in VTPC-1 and VTPC-2 should
be at least~12,
\item the ratio of the total number of reconstructed points to
the maximum possible number of points derived from the track trajectory
with respect to the detector geometry
should be larger than 0.5,
\item the distance between a track extrapolated to the target plane
and the interaction point (impact parameter) 
should be smaller than 4~cm in both
transverse directions separately, and
\item the track azimuthal angle should be within an
azimuthal angle wedge around the horizontal plane;
the wedge size depends on polar angle and
the smallest one ($\pm20\textdegree$ for $h^-$ and $dE/dx$ analyses
and $\pm10\textdegree$  for the $tof$--$dE/dx$ analysis)
was used for the largest
polar angles; see Fig.~\ref{fig:track_phi}.
\end{enumerate}

\subsection{Cross section measurements}
\label{Sec:Claudia}

For normalization and cross section measurements we adopted 
the same procedure as the one developed by 
the NA49 Collaboration~\cite{norm_NA49}. 
The minimum bias trigger on proton interactions, 
described in Sec.~\ref{Sec:trigger}, 
allows us to define a ``trigger'' cross section 
which is used both for the normalization of the differential inclusive 
pion distributions and for the determination of the inelastic 
and production cross sections.

From the numbers of selected interactions, fulfilling the
requirements on BPD signals and reconstruction of the
proton beam particles as detailed in Sec.~\ref{Sec:cuts}, 
we compute an interaction probability of
(6.022 $\pm$ 0.034)\% with the carbon
target inserted and of (0.709 $\pm$ 0.007)\% with
the carbon target removed. These measurements lead to an interaction
probability of (5.351 $\pm$ 0.035)\%
in the carbon target, taking into account
the reduction of the beam intensity in the material along its trajectory.
The corresponding ``trigger'' cross section  is
(298.1 $\pm$ 1.9 $\pm$ 7.3)~mb, after correcting
for the exponential beam attenuation in the target.
The systematic error on the ``trigger'' cross section was conservatively 
evaluated by comparing this value with the one obtained without 
any event selection criteria.

Two classes of processes are considered, inelastic and production
interactions. The production processes are defined as those
in which new hadrons are produced.
The inelastic processes include 
in addition interactions which result only in disintegration
of the target nucleus (quasi-elastic interactions).

The inelastic cross section $\sigma_{inel}$ was defined as the sum of all
processes due to strong p+C interactions except
coherent nuclear elastic scattering. Thus it includes interactions
with production of new hadrons (production processes) and
quasi--elastic interactions which lead only to break up of
the carbon nucleus.
The inelastic cross section  was
derived from the ``trigger'' cross section by applying two corrections:

\begin{enumerate}[(i)]
\setlength{\itemsep}{1pt}

\item
subtraction of the contribution from coherent elastic scattering
(47.2~$\pm$~0.2~$\pm$~5.0~mb),
i.e., removal of those events in which the incoming beam
particle undergoes a large-angle coherent
elastic scatter on the carbon nuclei and misses
the S4 counter, and
\item
addition of the lost inelastic events because of emitted
charged particles hitting the S4 counter
(5.7~$\pm$~0.2~$\pm$~0.5~mb for protons and
 0.57~$\pm$~0.02~$\pm$~0.35~mb for pions and kaons).
\end{enumerate}

The corrections were calculated  based on the GEANT4~\cite{GEANT4}  
simulation (with the QGSP\_BERT physics list) of the beam line setup using
the measured profile and divergence of the proton beam.
The GEANT4 angular distributions for coherent
elastic scattering and quasi-elastic scattering, as well as
the total inelastic cross section, were cross-checked against
available experimental measurements. The largest discrepancy
of 7.5\% was found for the coherent elastic events in which the scattered
proton is outside of the S4 acceptance. This discrepancy
was taken into account in the calculation of the systematic
error, which includes also uncertainties in other simulation
parameters, namely, beam position and divergence, as well as the size of the S4 counter.
The total systematic error on the inelastic cross section amounts to 8.9~mb.

The production cross section is determined 
from the inelastic cross section 
by subtracting the cross section of quasi-elastic p+C interactions
at 31~GeV/\textit{c} which amounts to $27.9\pm1.5~(sys)$~mb according to Glauber
model calculations~\cite{Glauber}.

Details of the cross section analysis procedure can be found in Ref.~\cite{Claudia}.

\subsection{The {\it $h^-$} analysis}
\label{Sec:Tomek}

More than 90\% of primary negatively charged particles produced
in p+C interactions at 31~GeV/\textit{c} are negatively charged pions.
Thus $\pi^-$ meson spectra can be obtained by subtracting
the estimated non-pion contribution from the spectra of negatively charged
particles  and additional particle identification
is not required.
Note that this method is not applicable to the analysis of
$\pi^+$ meson spectra because of the much larger and unknown contribution of protons and
$K^+$ mesons to all positively charged particles.

\begin{figure}[!h]
\begin{center}
\includegraphics[width=0.85\linewidth]{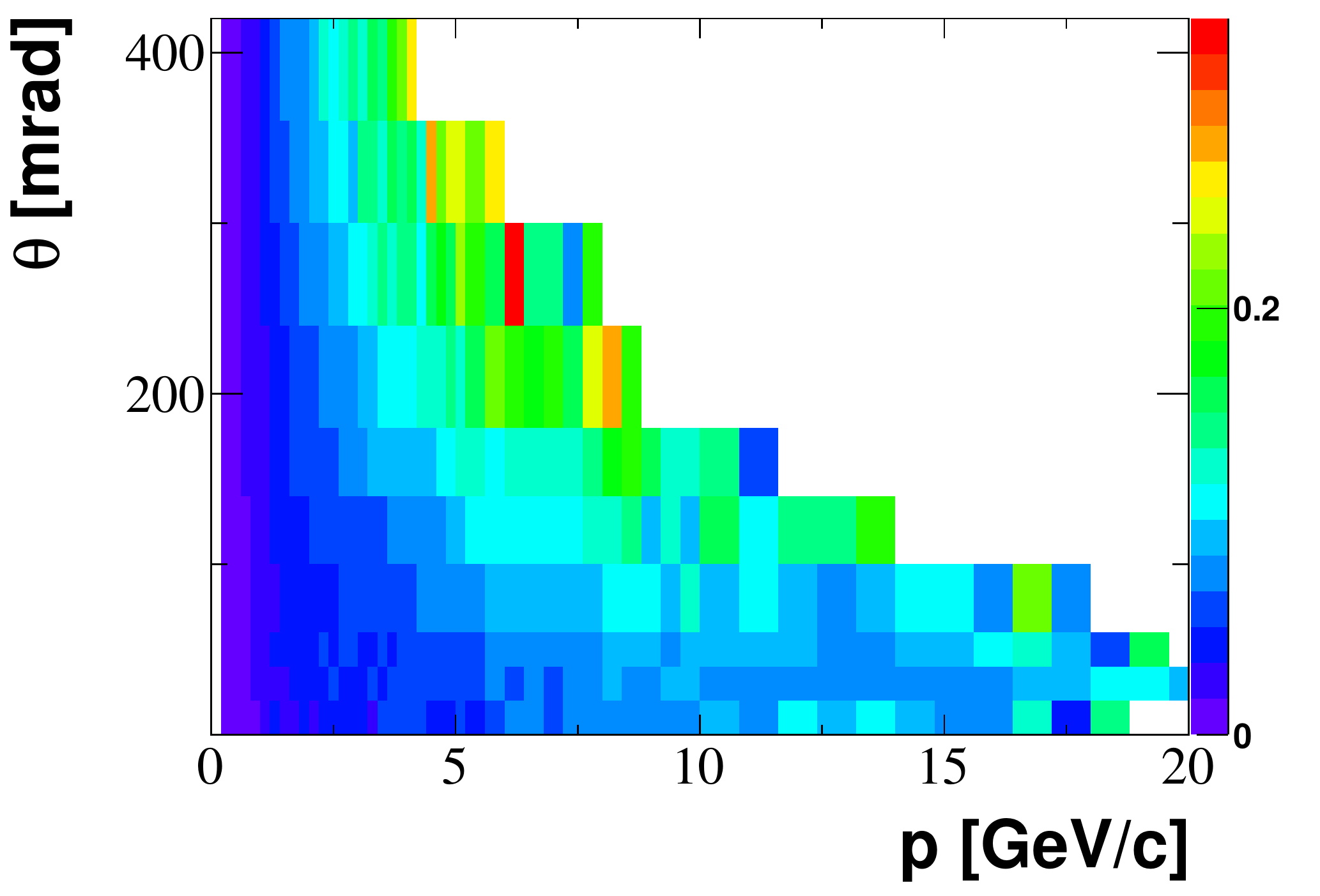}
\end{center}
  \caption{(Color online) 
    The relative contribution of accepted primary
    $K^-$ and $\overline{p}$
    to accepted primary negatively charged pions
    in different ($p$,~$\theta$) intervals calculated within the \Venus model.
}
\label{fig:non-pions}
\end{figure}

First, spectra of all accepted negatively charged particles
were obtained in ($p$,~$\theta$) bins.
The event and track selection criteria presented in Sec.~\ref{Sec:cuts}
were applied.

Second,
the Monte Carlo simulation described in Sec.~\ref{Sec:MC}
was used to calculate corrections for the contribution of electrons and
primary $K^-$ and $\overline{p}$ 
as well as secondary particles from weak decays (feed down), secondary interactions,
and photon conversions in the target and the detector material. Furthermore, the corrections
include track reconstruction efficiency and resolution as well as
losses  due to the
limited geometrical acceptance of the detector.
Bin-by-bin correction factors were calculated as
the ratio of all generated primary $\pi^-$ mesons to all reconstructed and
accepted negatively charged particles in a given bin.

\begin{figure}[!h]
\begin{center}
\includegraphics[width=0.85\linewidth]{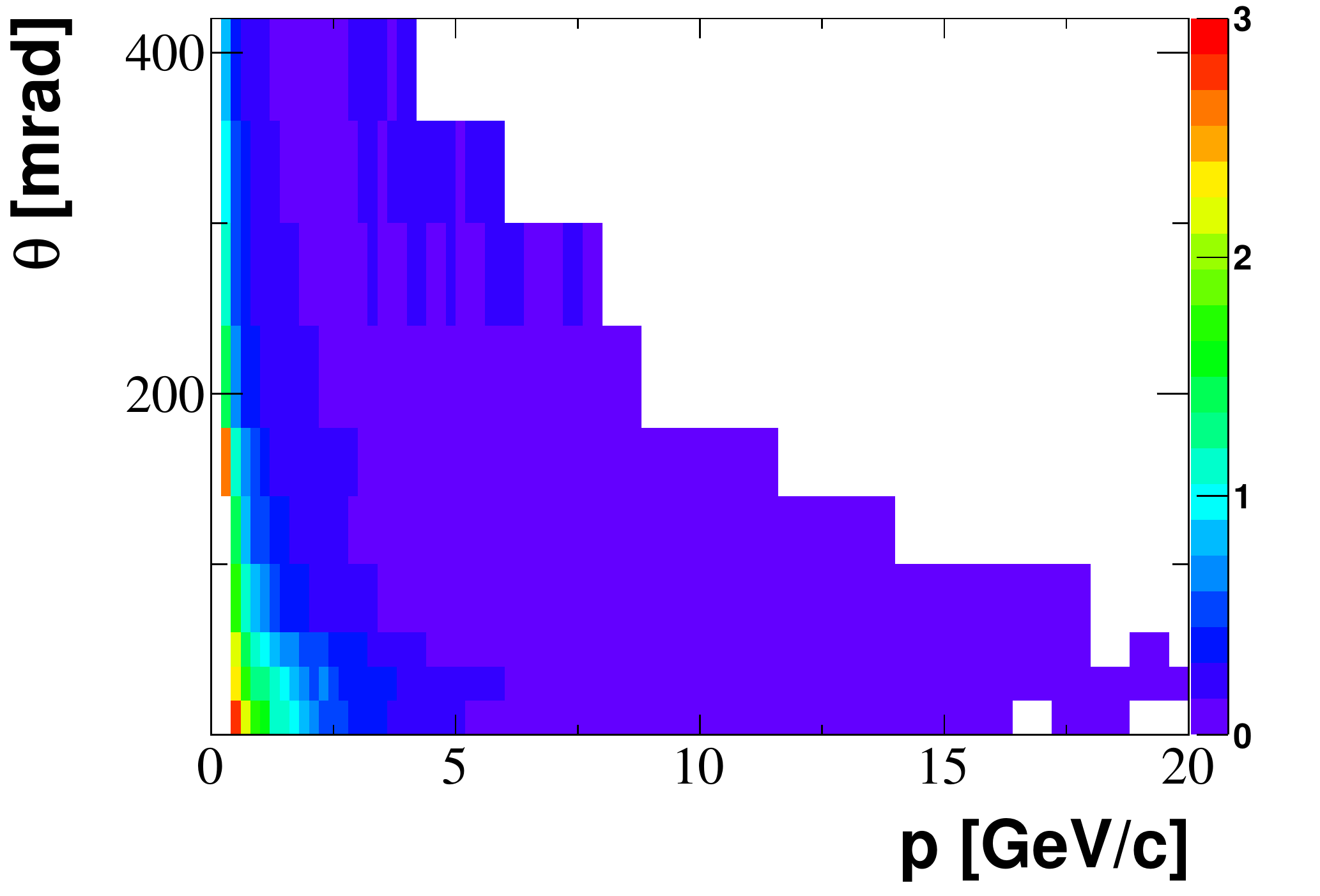}
\end{center}
  \caption{(Color online) 
    The relative contribution of accepted electrons and non-primary
    negatively charged hadrons
    to  accepted primary  negatively charged pions in different 
    ($p$,~$\theta$) intervals
    calculated within the \Venus model.
    The low-momentum region where the ratio is larger than 3 is
    excluded.
}
\label{fig:non-primary}
\end{figure}

Different contributions to the correction factors were studied
separately. The losses due to the
track reconstruction efficiency and the
limited geometrical detector acceptance
are shown in Figs.~\ref{fig:track_eff} and~\ref{fig:geometrical},
respectively.
The relative contribution of accepted primary $K^-$ and $\overline{p}$  to
accepted primary negatively charged pions is presented in Fig.~\ref{fig:non-pions}.
Finally, the relative contribution of accepted electrons and non-primary 
negatively charged hadrons to
accepted primary negatively charged pions is given in Fig.~\ref{fig:non-primary}.
The correction for secondary pions and electrons is largest for small momenta.
The former are mainly due to strange particle decays close to the primary vertex;
the latter originate mostly from conversion of $\pi^0$ decay photons.
Uncertainty 
due
to this correction is discussed in Sec.~\ref{Sec:Syst}.

\begin{figure}[!hb]
\begin{center}
\includegraphics[width=0.85\linewidth]{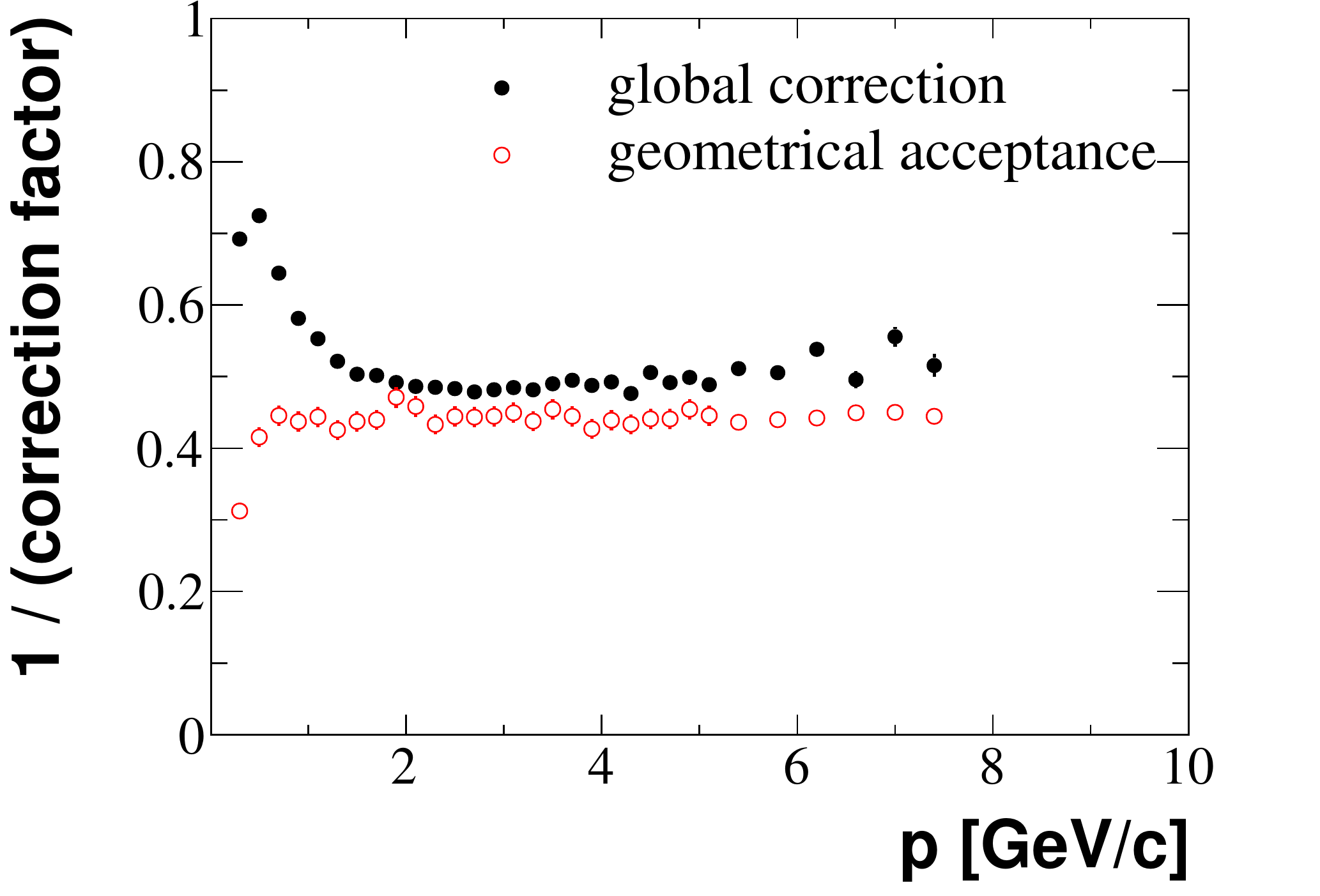}
\includegraphics[width=0.85\linewidth]{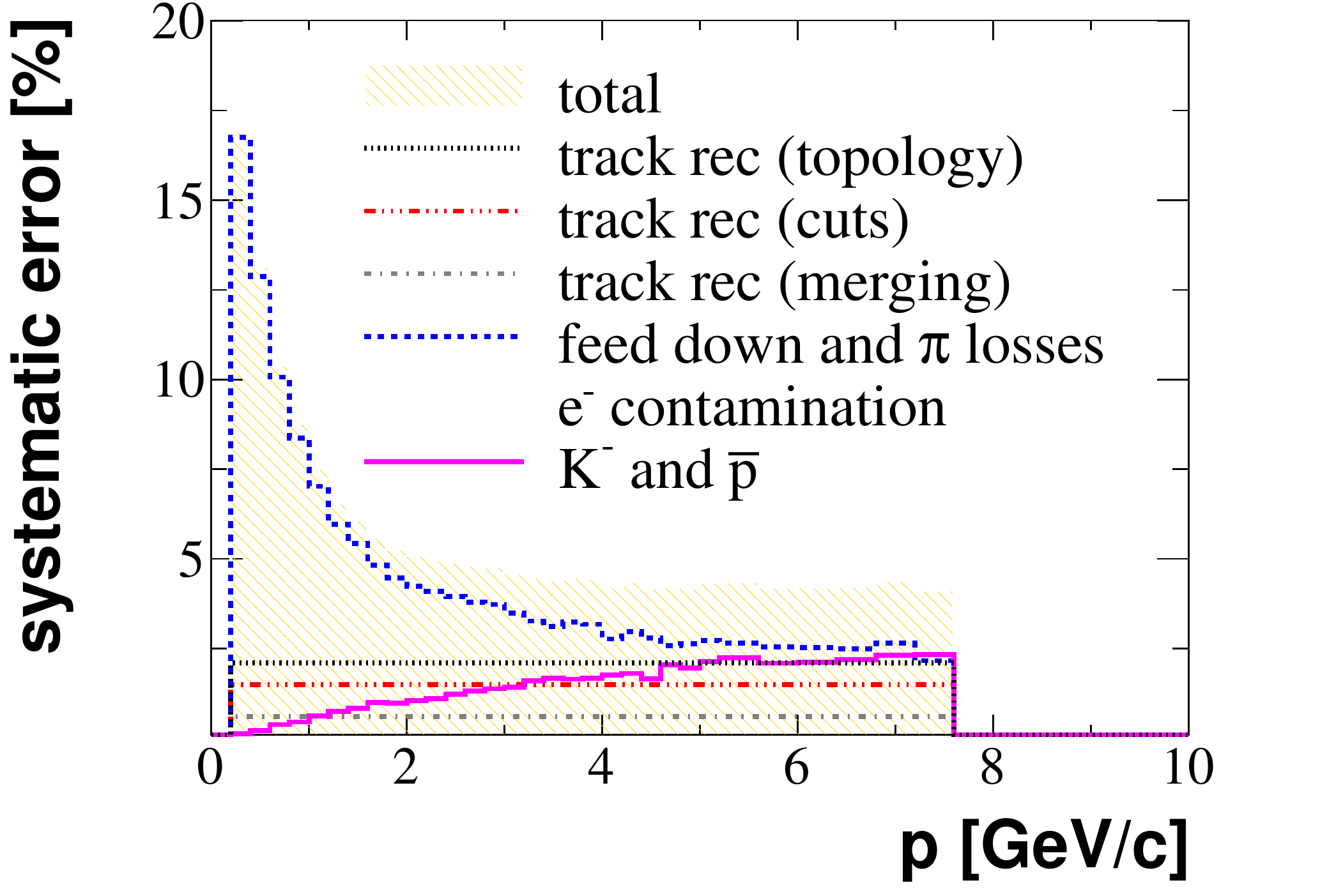}
\caption{(Color online) 
   Example of momentum dependence of the 
   inverse correction factor and geometrical acceptance~({\it top}) 
   and systematic errors~({\it bottom}) 
   for the  {\it $h^-$} analysis  
   for negatively charged pions 
   in the polar angle interval [140,180]~mrad.
   For details concerning systematic uncertainties, see
   Sec.~\ref{Sec:Syst}.
}
\label{fig:hminus-cor}
\end{center}
\end{figure}

The inverse correction factor and the geometrical acceptance
versus momentum for the polar angle interval
[140,180]~mrad are shown in Fig.~\ref{fig:hminus-cor} as an example.

\subsection{The $dE/dx$ analysis at low momentum}
\label{Sec:Magda}

The analysis of charged pion  production at low momentum
was done  by means of particle identification via
measurements of specific energy loss in the TPCs.
Measurements of $tof$ are not available for
low-momentum particles since they do not
reach the ToF-F detectors.
A reliable  identification of $\pi^{+}$ mesons was not possible 
at momenta above 1~GeV/\textit{c} where the Bethe-Bloch (BB) curves
for pions, kaons, and protons cross each other (see Fig.~\ref{dedx_pos_neg}). 
On the other hand, for $\pi^{-}$ mesons, where the contribution of $K^-$ and antiprotons 
is almost negligible, the $dE/dx$ analysis could be extended 
in momentum up to 3~GeV/\textit{c} allowing consistency checks 
with the other analysis methods in the region of overlap.

The procedure of particle identification, described below, is tailored 
to the region where a fast change of energy loss with momentum is observed.
In order to optimize the parametrization of the BB function, 
samples of $e^\pm$, $\pi^\pm$, $K^\pm$, $p$, and $d$ tracks with
reliable particle identification were chosen in the $\beta \gamma$ 
range from 0.2 up to 100. The dependence of the BB function on 
$\beta \gamma$ was then fitted to the data using the
Sternheimer and Peierls parametrization of Ref.~\cite{stern}.
This function was then used to calculate for every track 
of a given momentum the expected $(dE/dx)_{BB}$
values for all possible identification hypotheses 
for comparison with the measured mean $(dE/dx)_{data}$. 
A small (a few percent) dependence of the mean $(dE/dx)_{data}$ 
on the track polar angle had to be corrected for.

The identification procedure was performed in ($p$,~$\theta$) bins. 
Narrow momentum intervals (of 0.1~GeV/\textit{c} for $p < 1$~GeV/\textit{c} and 
0.2~GeV/\textit{c} for $1 < p < 3$~GeV/\textit{c}) were chosen to account for the strong 
dependence of $dE/dx$ on momentum.
The event and track selection criteria presented in Sec.~\ref{Sec:cuts}
were applied. In each ($p$,~$\theta$) bin an unbinned 
maximum likelihood fit (for details see Ref.~\cite{marek})  
was performed  to extract yields
of $\pi^{+}$ and $\pi^{-}$ mesons. The probability density functions 
were assumed to be a sum of Gaussian functions for each particle species, 
centered on $(dE/dx)_{BB}$ with variances derived from data. 
The $dE/dx$ resolution is a function of 
the number of measured points and the particle momentum. 
In the $\pi^{+}$ analysis three independent abundances were fitted 
($\pi^{+}$, $K^{+}$ and proton)
while in the $\pi^{-}$ analysis we were left with only two independent 
abundances ($\pi^{-}$ and $K^{-}$). The $e^{+}$ and $e^{-}$ abundances
were determined from the total number of particles in the fit.

\begin{figure}[!hb]
\begin{center}
\includegraphics[width=0.85\linewidth]{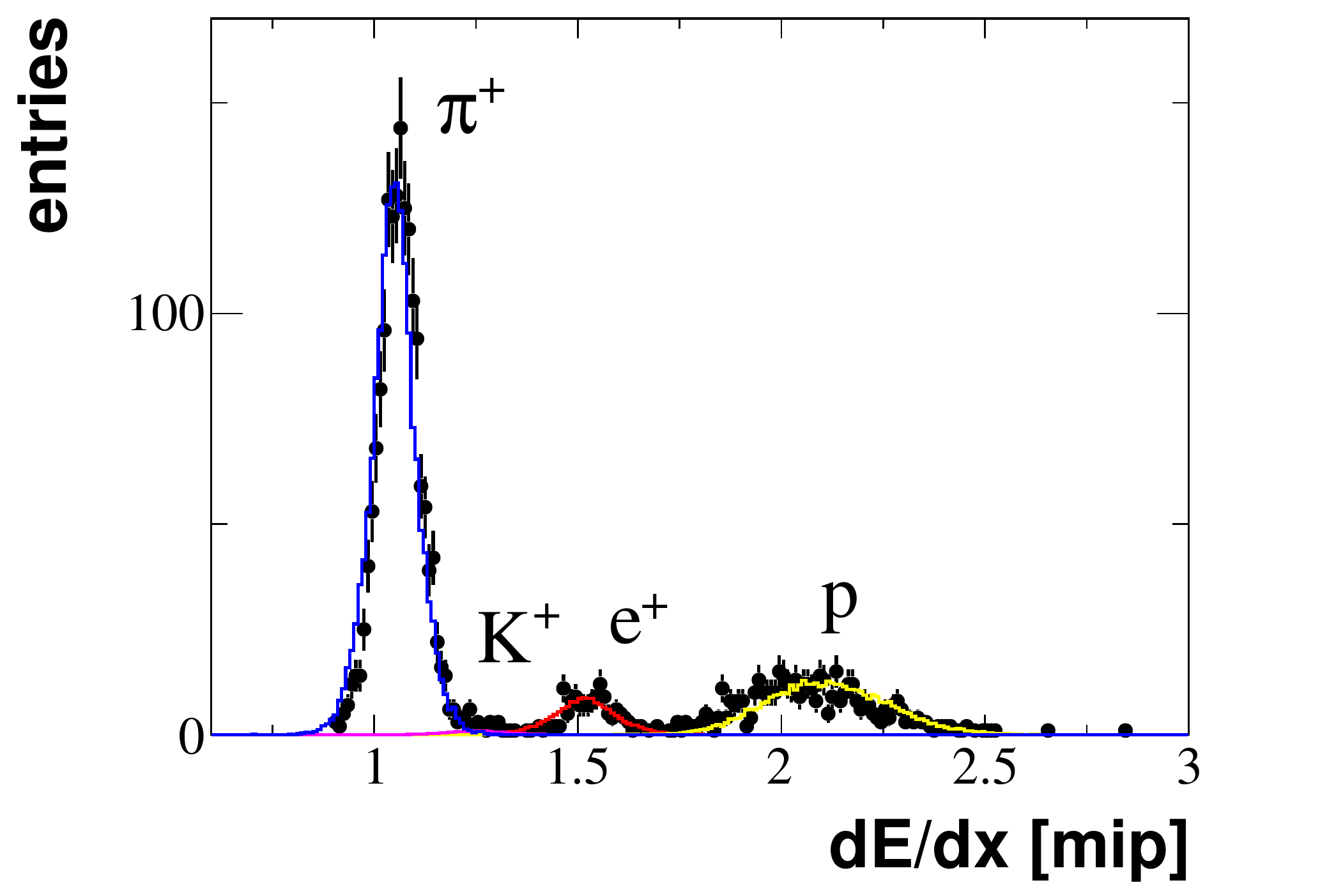}
\caption{(Color online)
   The $dE/dx$ distribution for positively charged
   particles in the momentum bin [0.7,0.8]~GeV/\textit{c} 
   and angular bin [180,240]~mrad compared
   with the distribution calculated using the fitted relative abundances. 
   A small systematic shift between the fit curves (Gaussian peaks centered on the
   Bethe-Bloch values) and data is observed.  
   This effect results in a major contribution to the total systematic error
   in the particle identification procedure.
}
\label{dedx_fit_example}
\end{center}
\end{figure}

As an example, the $dE/dx$ distribution for positively charged
particles in the momentum bin [0.7,0.8]~GeV/\textit{c} and angular bin [180,240]~mrad
is shown in Fig.~\ref{dedx_fit_example} compared with the distribution
obtained using the fitted parameters.

\begin{figure}[!hb]
\begin{center}
\includegraphics[width=0.87\linewidth]{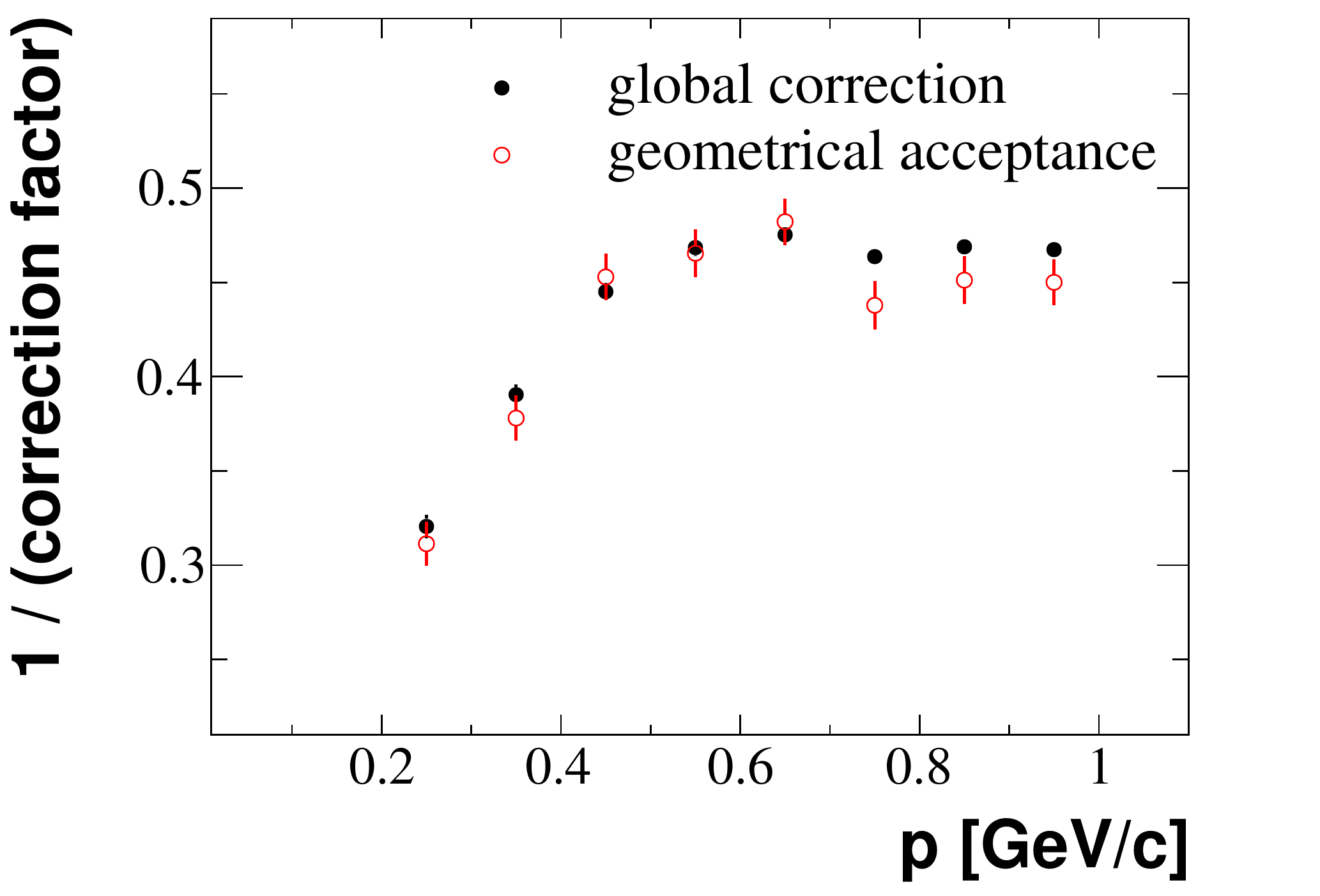}
\includegraphics[width=0.87\linewidth]{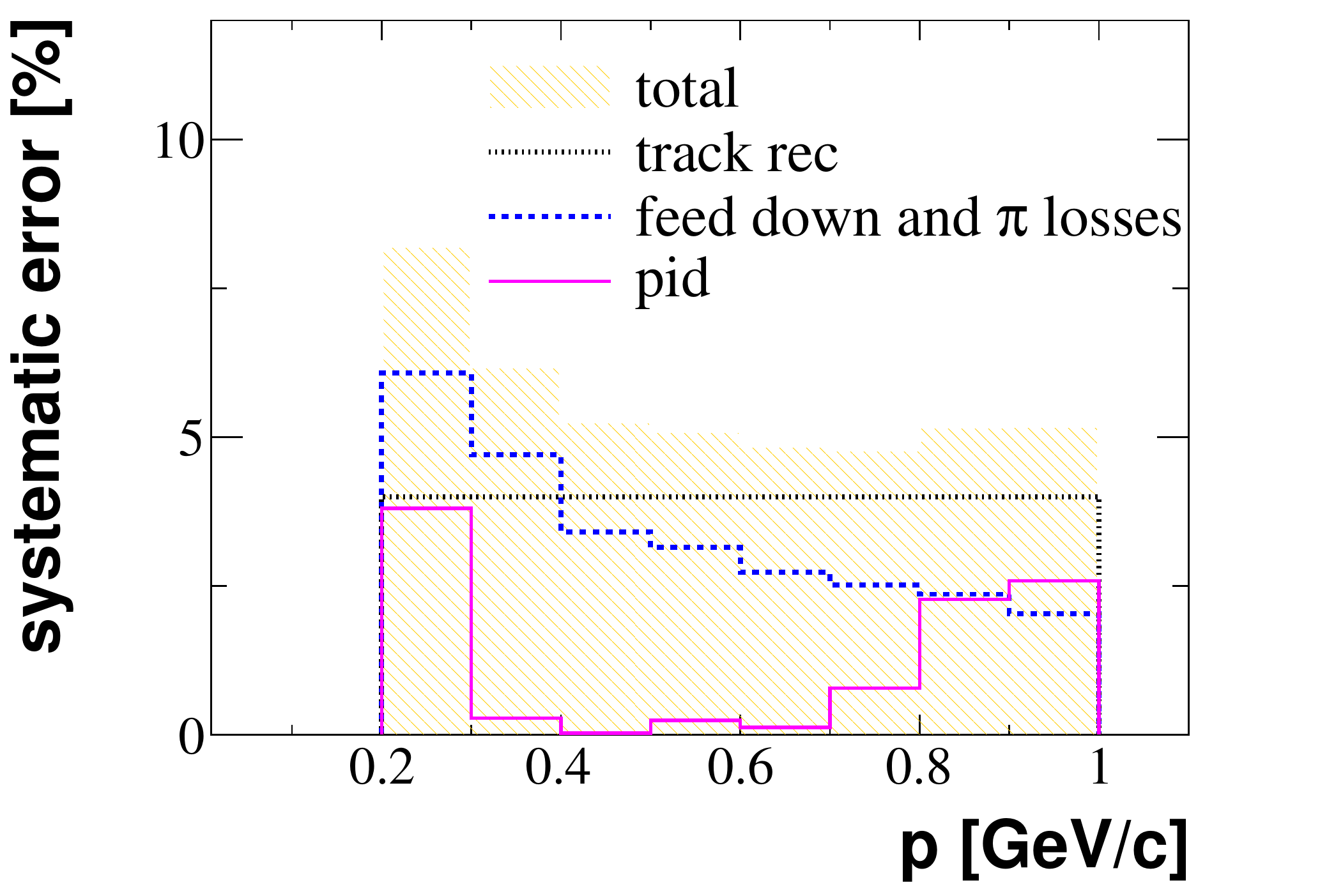}
\caption{(Color online)
   Example of the momentum dependence of the 
   inverse correction factor and geometrical acceptance~({\it top})
   and systematic errors~({\it bottom})
   for the $dE/dx$ analysis
   for positively charged pions 
   in the polar angle interval [140,180]~mrad.
   The small differences between geometrical acceptance and global correction
   are due to the fact that pion loss and feed down corrections, which are
   the most important ingredients of the global correction, cancel to a large
   extent.
   For details concerning systematic uncertainties see
   Sec.~\ref{Sec:Syst}.
}
\label{fig:dedx-cor}
\end{center}
\end{figure}

Finally, the Monte Carlo simulation described in Sec.~\ref{Sec:MC}
was used to calculate bin-by-bin corrections for
pions from weak decays and
interactions in the target and the detector material. The corrections
include also track reconstruction efficiency and resolution as well as
losses  due to pion decays and the
limited geometrical acceptance of the detector.
The inverse correction factor and geometrical acceptance
versus momentum for the polar angle interval
[140,180]~mrad is shown in Fig.~\ref{fig:dedx-cor} as an example.

\subsection{The $tof$--$dE/dx$ analysis}
\label{Sec:Sebastien}
High purity particle identification can be performed by combining
the $tof$ and $dE/dx$ information. Moreover, in the momentum range 
1--4~GeV/\textit{c}, where $dE/dx$ bands for different particle species overlap,
particle identification is in general only possible using the $tof$ method; see
Figs.~\ref{dedx_pos_neg} and~\ref{fig:tof-dedx} for illustration.
The ToF-F detector was designed to cover the necessary acceptance in
momentum and polar angle required by the T2K experiment.

\begin{figure}[!hb]
\begin{center}
\includegraphics[width=0.87\linewidth]{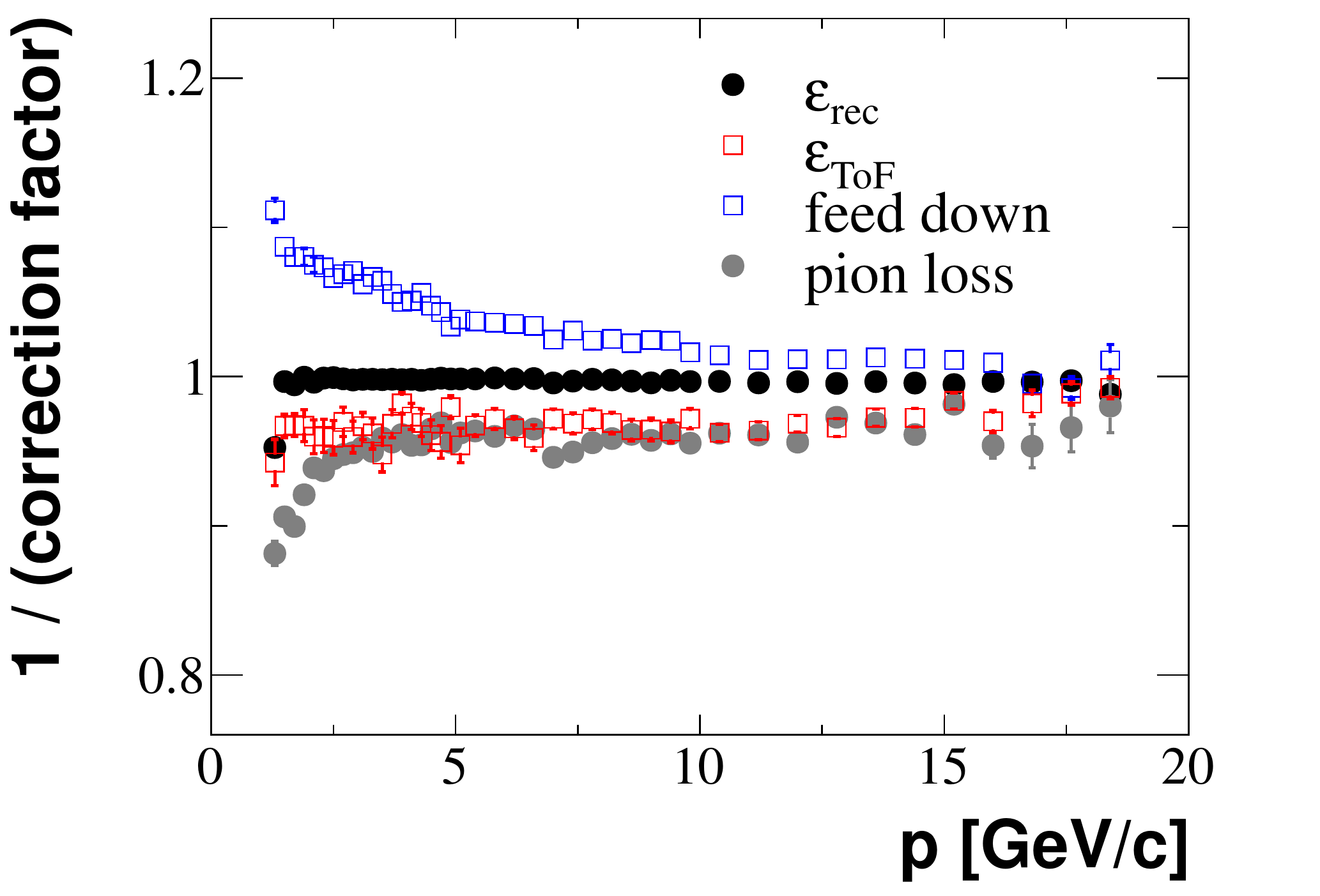}
\includegraphics[width=0.87\linewidth]{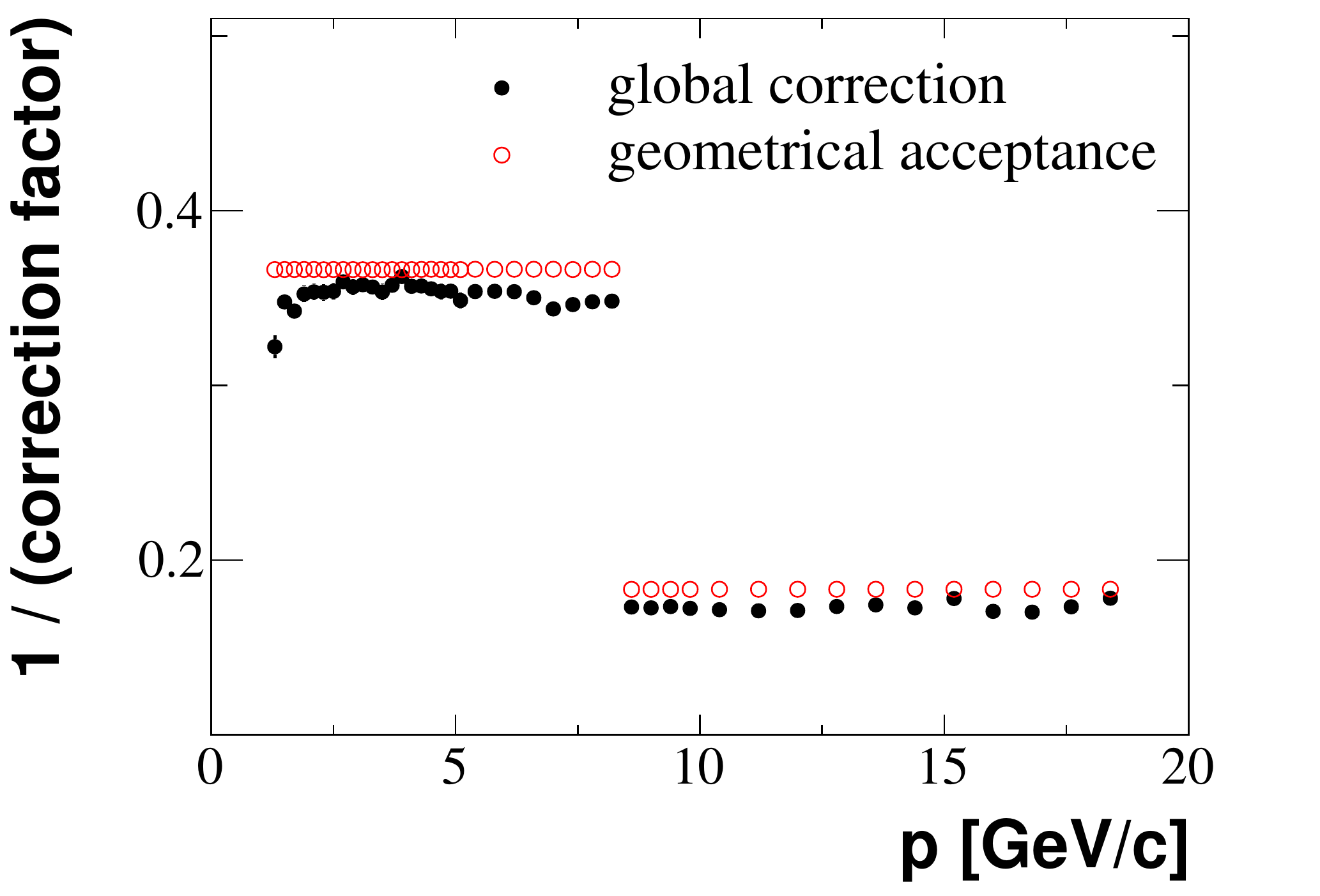}
\includegraphics[width=0.87\linewidth]{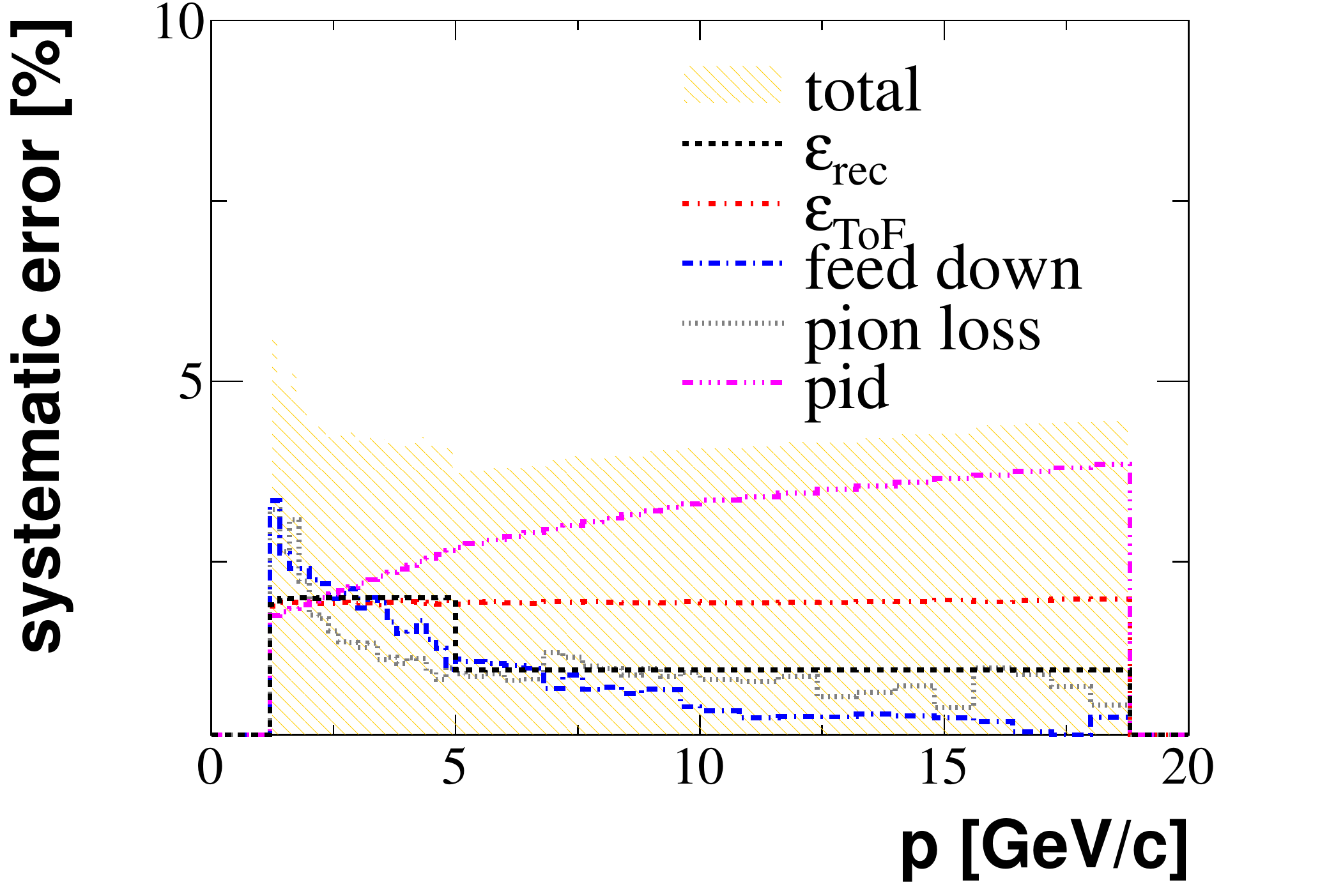}
\caption{(Color online) 
   Example of momentum dependence of the 
   inverse correction factors~({\it top}),
   total inverse correction factor and geometrical acceptance~({\it middle}),
   and systematic errors~({\it bottom}) 
   for the $tof$--$dE/dx$ analysis 
   for positively charged pions 
   in the polar angle interval [40,60]~mrad.
   $\epsilon_{rec}$ and $\epsilon_{ToF}$ are the efficiencies 
   of the reconstruction and of the ToF-F, respectively. 
   The feed-down correction accounts for pions from weak decays 
   which are reconstructed as primary particles, while the pion loss 
   accounts for pions lost due to decays or secondary interactions.
   For details concerning systematic uncertainties, see
   Sec.~\ref{Sec:Syst}.
   For the $tof$--$dE/dx$ analysis, within a given theta bin, in addition to
   the cut on the azimuthal angle, a momentum-dependent selection 
   on the track topology was applied. The purpose of this cut is to select 
   only regions with a flat acceptance along the whole momentum range.
}
\label{fig:tof-dedx-cor}
\end{center}
\end{figure}

First, a particle mass squared, $m^2$, was calculated
using the $tof$, momentum, and path length measurements.
The $m^2$-$dE/dx$ distributions of all accepted positively (and negatively)
charged particles were then obtained in each ($p$,~$\theta$) bin.
The event and track selection criteria presented in Sec.~\ref{Sec:cuts}
were applied.
In addition only tracks which gave a signal in the ToF-F detector were selected.
Two independent topologies of tracks, emitted to the left and to the right with respect 
to the incoming beam direction, have different acceptances.
The analysis was performed for each topology separately. When both
topologies are present the final result is the weighted average of the two.
Momentum-dependent azimuthal angle cuts were applied such that
in these angular intervals the detector acceptance was close
to 100\%. 
These cuts select tracks with a large number of measured
points as well, thus with a high reconstruction efficiency. 
The averaged individual as well as combined global correction 
factors are shown in Fig.~\ref{fig:tof-dedx-cor} for illustration.

The pion accumulations in these distributions were parametrized
by a product of Gaussian functions in $m^2$ and $dE/dx$.
In each ($p$,~$\theta$) bin the bin-by-bin
maximum likelihood method was applied to fit yields
of $\pi^+$ and $\pi^-$ mesons.
In the fit the signal shapes parametrized in the first step
were allowed to vary within narrow limits. The pion yields were
calculated summing all particles within 2$\sigma$ around the fitted pion
peak; see Fig.~\ref{fig:tof-dedx}. 
In addition the integral of the fitted function was computed and
used for the evaluation of the systematic uncertainty.

Finally,
the Monte Carlo simulation described in Sec.~\ref{Sec:MC}
was used to calculate corrections for
pions from weak decays and
interactions in the detector material and target,
track reconstruction efficiency and resolution,
and losses due to pion decays.
The ToF-F detection efficiency was estimated by requiring that a track
traversing the ToF-F wall generates a hit in the ToF-F. 
The ToF-F inefficiency is due also to double hits and readout
inefficiencies.
The bin-by-bin correction factors for different biases were calculated
separately and applied to the data.
The inverse correction factors versus momentum, 
total correction factor and geometrical acceptance for the polar angle interval
[40,60]~mrad are shown in Fig.~\ref{fig:tof-dedx-cor} as an example.

\subsection{Derivation of spectra}
\label{Sec:norma}

The  procedures presented in 
Secs.~\ref{Sec:Tomek}, \ref{Sec:Magda}, and \ref{Sec:Sebastien} were used
to analyze events with the carbon target inserted (I) and 
with the carbon target removed (R).
The corresponding
corrected numbers of $\pi^-$ and $\pi^+$ mesons in $p$ bins
and $\theta$ intervals are denoted as
$\Delta n^I_\alpha$ and
$\Delta n^R_\alpha$, where $\alpha$ stands for
$\pi^-$ and $\pi^+$.
Note that the same event and track selection criteria as well as
corrections discussed in Secs.~\ref{Sec:Tomek}, \ref{Sec:Magda},
and \ref{Sec:Sebastien} were used in the analysis of events
with the target inserted and removed.
The latter events allow us to correct the results for the contribution of
out-of-target interactions.

Then, the differential inclusive
cross section of $\pi^+$ and $\pi^-$ mesons is calculated as
\begin{equation}
    \label{eq:xsecmeas3}
\frac {d \sigma_{\alpha}}{d p} = \frac {\sigma_{trig}} {1-\epsilon}
\left( \frac {1} {N^I}  \frac { \Delta n^I_\alpha } { \Delta p } -
\frac {\epsilon} {N^R}  \frac { \Delta n^R_\alpha } { \Delta p }\right)~,
\end{equation}
where
\begin{enumerate}[(i)]
\setlength{\itemsep}{1pt}

  \item $\sigma_{trig} = (298.1 \pm 1.9 \pm 7.3)$~mb is
  the ``trigger'' cross section
  as given in Sec.~\ref{Sec:Claudia},

  \item $N^I$ and $N^R$ are the numbers of events selected (see Sec.~\ref{Sec:cuts}) 
         for the analysis
         of events with the target inserted and removed, respectively,

  \item $\Delta p$ is the bin size in momentum, and

  \item $\epsilon = 0.118 \pm 0.001$ is the ratio of the interaction probabilities 
  for operation with the target removed and inserted.

\end{enumerate}

The correction for the contribution of particles from out-of-target
events (the term 
$\epsilon / N^R \cdot \Delta n^R_\alpha / \Delta p$ in Eq.~\ref{eq:xsecmeas3}~)
amounts on average to about 7\% and 3\% in the first two polar angle intervals.
It is smaller than about 2\% for polar angle bins above 40~mrad.

The pion spectra normalized to the mean pion multiplicity in
production interactions was calculated as
\begin{equation}
    \label{eq:xsecprod}
\frac {d n_{\alpha}}{d p} = \frac{1}{\sigma_{prod}}
\cdot \frac {d \sigma_{\alpha}}{d p}~,
\end{equation}
where $\sigma_{prod}$ is the cross section for production processes.

\subsection{Statistical and systematic errors}
\label{Sec:Syst}

Statistical errors on the pion spectra include contributions
from the finite statistics of data and from the Monte Carlo simulation
used to obtain the correction factors. The Monte Carlo statistics was
about ten times larger than the data statistics and the total
statistical errors are dominated by the statistical uncertainty
of the data.

Systematic errors on the pion spectra were estimated by
varying track selection and identification criteria as well as
parameters used to calculate the corrections.
The following track selection criteria were varied: the minimum number of
points measured on the track, the azimuthal angle, and the 
impact parameter cuts.
It was
found that the influence of such changes is small as compared to the
statistical errors.
The accuracy of corrections for acceptance and reconstruction efficiency
was checked by comparison of the results obtained with independent track topologies
as well as using two
different algorithms for merging track segments from different TPCs
into global tracks (track merging algorithm).
The dominant contributions to the systematic error for all three analysis
methods come from the uncertainty in the correction
for secondary interactions and for weak decays of strange particles. We
assigned an uncertainty of 30\% of the correction for both sources.
 The systematic error due to the admixture of pions from the
decays of strange particles reconstructed at the primary vertex
depends on the knowledge of strange particle production. The estimate of
this error was based on the following.
\begin{enumerate}[(i)]
\setlength{\itemsep}{1pt}
\item
Comparison of the number of $V^0$ decays reconstructed in the data and in
the  \VenusLong model with default parameters.
\item
Variation of the strange particle yields in different Monte Carlo
generators; for example the  $K^-/\pi^-$  ratio in
p+C interactions at 31~GeV/\textit{c} from
\FlukaLong~\cite{Fluka}, \UrqmdLong~\cite{Urqmd}
and \GiBUULong~\cite{GIBUU} (with default value of physics parameters) 
is
3.73\%, 3.67\% and 3.35\%, respectively;
these values can be compared
with 4.06\% from the \VenusLong~\cite{Venus} generator used in this paper 
for calculation of corrections for pions from weak decays.
\item
Comparison of the \VenusLong predictions with the measured $K^-/\pi^-$ and 
$K^+/\pi^+$ ratios from p+B$_4$C interactions at 24~GeV/\textit{c} 
at large momenta and small angles.
\end{enumerate}
In the $h^-$ analysis a 20\% uncertainty in the correction for electron
admixture to the pion sample was assumed.
Examples of systematic errors for all three analyses are presented in the bottom 
panels of 
Figs.~\ref{fig:hminus-cor},~\ref{fig:dedx-cor}, and~\ref{fig:tof-dedx-cor}.
Losses of inelastic interactions because an emitted
particle hits the S4 counter were estimated
to be negligibly small, namely,
about 1\% or smaller in the relevant phase space for the analysis.
The total systematic error was calculated as a sum
of different contributions added in quadrature.
It does not include the overall uncertainty due to the
normalization procedure, namely, 2.5\% and 2.3\% for the normalization to
the inclusive cross section and mean pion multiplicity in production
events, respectively.
Statistical and systematic uncertainties were added in quadrature
in order to calculate
the total error.
The first quoted error always refers to the statistical and
the second to the systematic uncertainty.

\section{Results}
\label{Sec:results}

This section presents  results on inelastic and production
cross sections as well as on differential spectra of $\pi^+$ and $\pi^-$ mesons in
p+C interactions at 31~GeV/\textit{c}.

\subsection{Inelastic and production cross sections}
\label{Sec:totalxsection}

The corrections to the ``trigger'' cross section described in Sec.~\ref{Sec:Claudia}
result in a total inelastic cross section of
\begin{eqnarray*}
    \label{eq:finalsiginel}
\sigma_{inel} = 257.2 \pm 1.9 \pm 8.9~\mathrm{mb}.
\end{eqnarray*}

The production cross section was calculated from the inelastic
cross section by subtracting the quasi-elastic contribution.
The result is
\begin{eqnarray*}
    \label{eq:finalsigprod}
\sigma_{prod} = 229.3 \pm 1.9 \pm 9.0~\mathrm{mb}.
\end{eqnarray*}

\begin{figure}[!h]
\begin{center}
\includegraphics[width=0.85\linewidth]{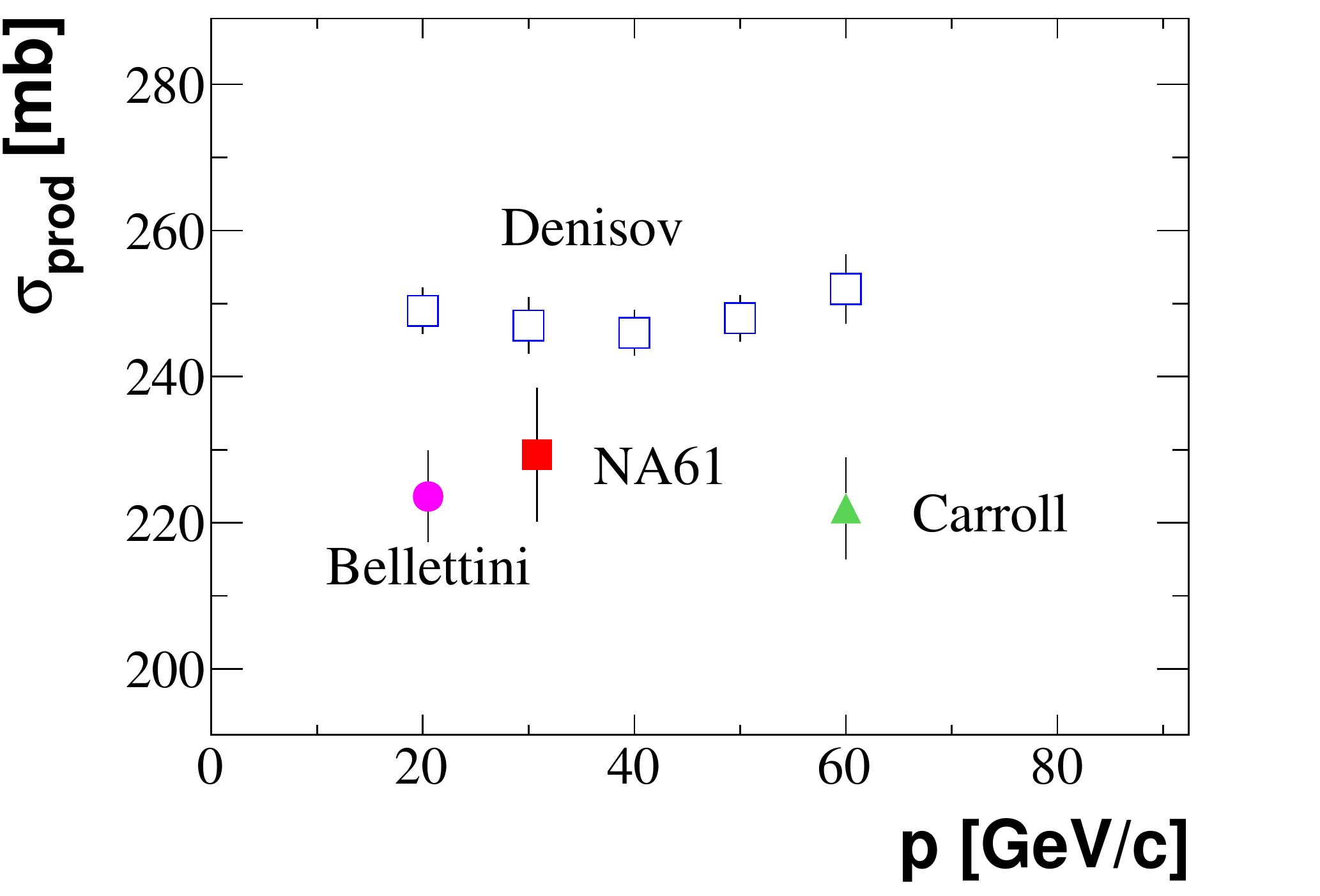}
\end{center}
 \caption{(Color online) 
 Beam momentum dependence of the production cross section
 for p+C interactions. The NA61/SHINE result (filled square)
 is compared with previous measurements:
 Bellettini \emph{et al.} (circle)~\cite{Bellettini},
 Carroll \emph{et al.} (triangle)~\cite{Carroll} and Denisov \emph{et al.} (open squares)~\cite{Denisov}.
 For the NA61/SHINE point, the error bar indicates statistical and systematic uncertainties
  added in quadrature. The result from Ref.~\cite{Bellettini} was recalculated by
 subtracting from the measured inelastic cross section
 a quasi-elastic contribution at 20~GeV/\textit{c} of $30.4 \pm 1.9~(sys)$~mb.
}

\label{fig:inelastic}
\end{figure}

The production cross section is compared to previous measurements
in Fig.~\ref{fig:inelastic}. The NA61/SHINE result at 31~GeV/\textit{c} is consistent with
the measurement of Ref.~\cite{Bellettini} at 20~GeV/\textit{c}, after
subtraction of the quasi-elastic contribution of $30.4\pm1.9~(sys)$~mb, 
and with that of Ref.~\cite{Carroll} at 60~GeV/\textit{c}.
The measurements presented in Ref.~\cite{Denisov} at different momenta 
are found to be systematically larger, as was already noted 
in Ref.~\cite{Carroll}, where the authors commented 
``We know of no reason for this discrepancy''.
Note that the data from Ref.~\cite{Denisov} are about 30~mb higher than other results; 
if interpreted as measurements of inelastic cross section, 
they would  agree with the other production cross section data, 
after the correction for the quasi-elastic contribution. 

\subsection{Spectra of $\pi^+$ and $\pi^-$ mesons}

\begin{figure*}[tb]
\begin{center}
\includegraphics[width=0.9\linewidth]{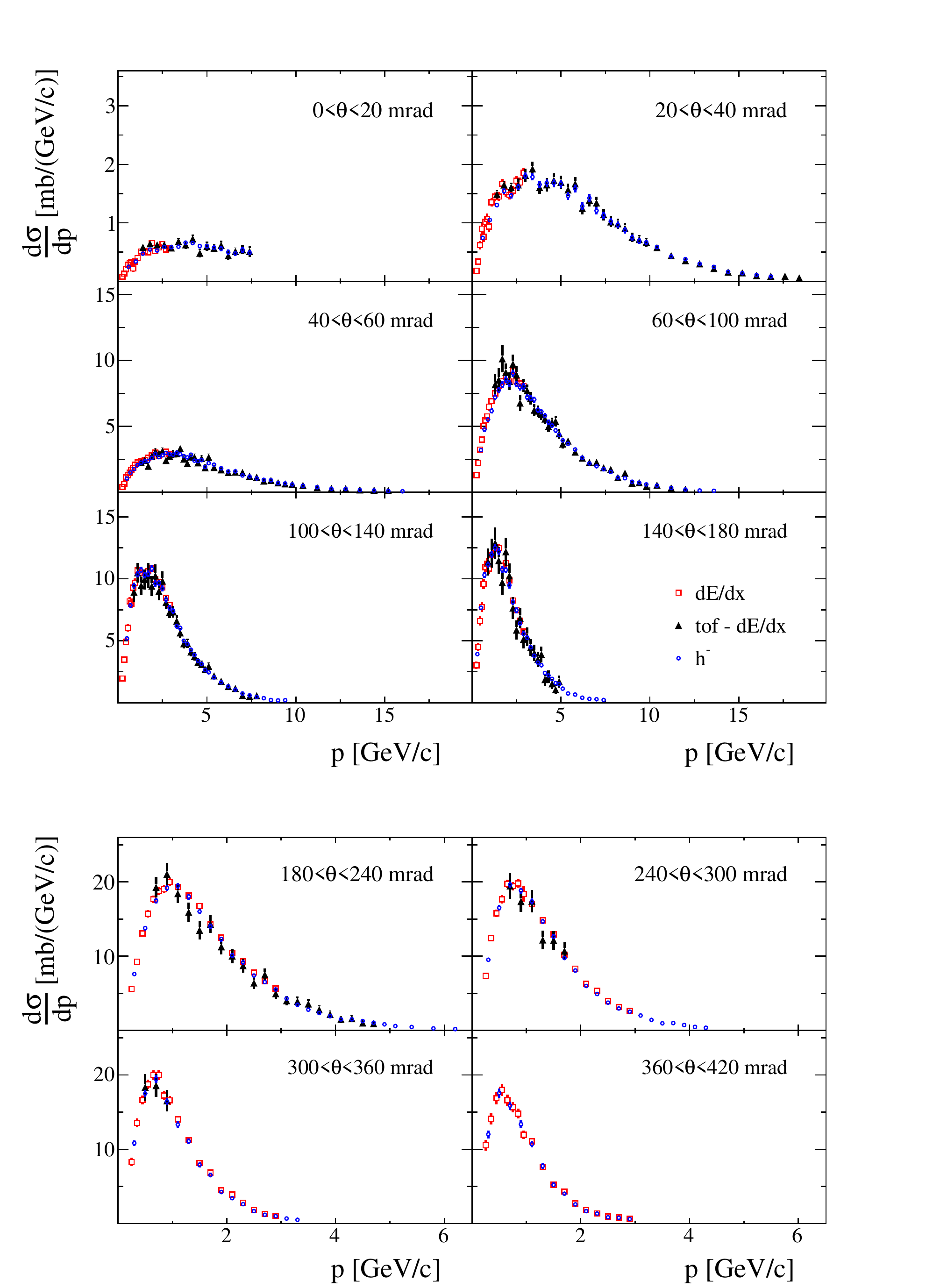}
\caption{(Color online) 
  Differential cross sections for $\pi^{-}$ meson production
  in p+C interactions at 31~GeV/\textit{c}. The spectra are presented as a
  function of laboratory  momentum ($p$)
  in different intervals of polar angle ($\theta$).
  Results obtained using three analysis methods are
  presented by different symbols: blue open circles, $h^-$ analysis;
  red open squares, $dE/dx$ analysis; and black full triangles, $tof$--$dE/dx$ analysis.
  Error bars indicate only statistical uncertainties.
}
\label{pion_minus_all_mbarn}
\end{center}
\end{figure*}

\begin{figure*}[tb]
\begin{center}
\includegraphics[width=0.9\linewidth]{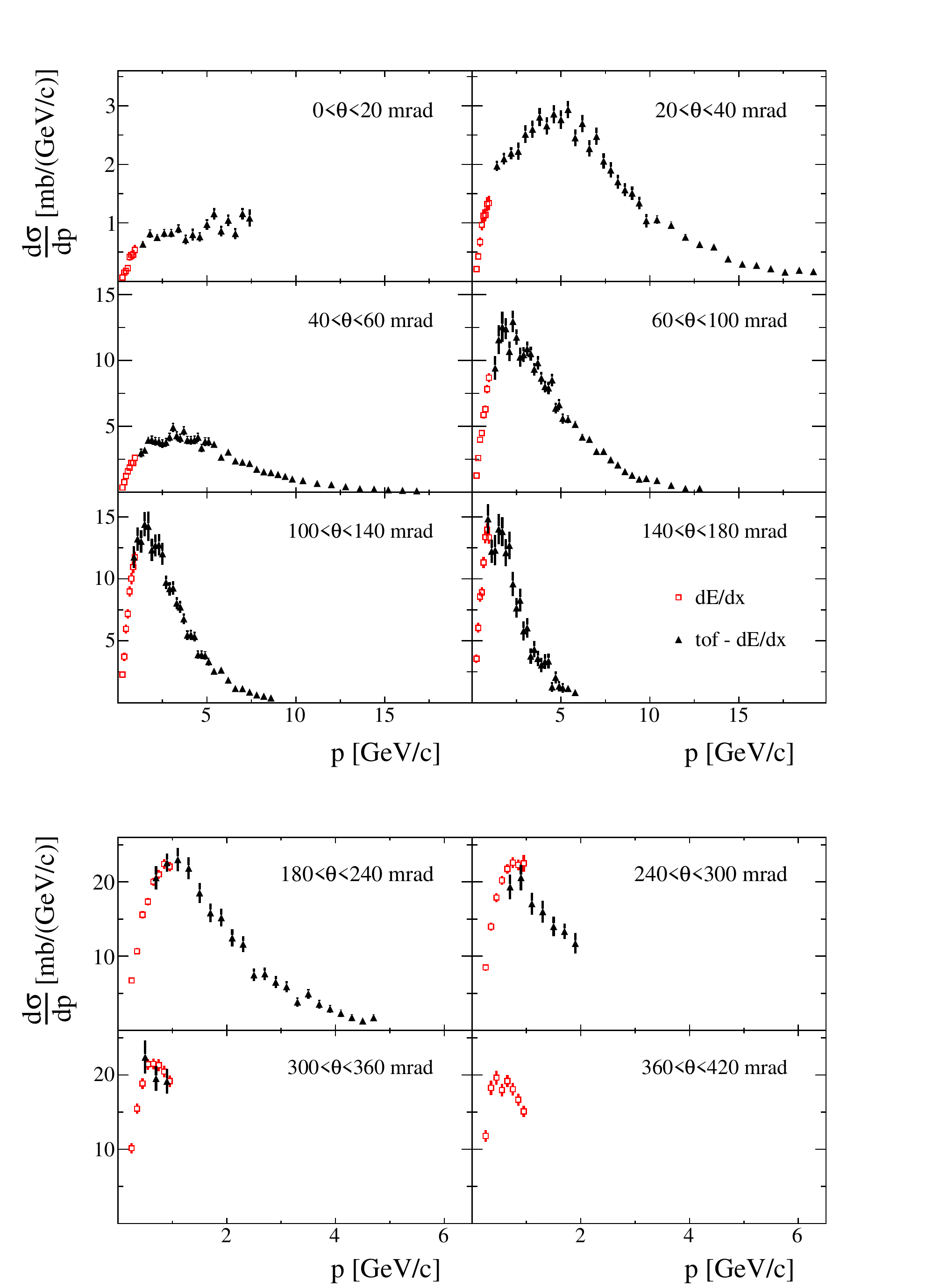}
\caption{(Color online) 
  Differential cross sections for $\pi^{+}$ meson production
  in p+C interactions at 31~GeV/\textit{c}. The spectra are presented as a
  function of laboratory  momentum ($p$)
  in different intervals of polar angle ($\theta$).
  Results obtained using two analysis methods are
  presented by different symbols:
  red open squares, $dE/dx$ analysis, and black full triangles, $tof$--$dE/dx$ analysis.
  Error bars indicate only statistical uncertainties.
}
\label{pion_plus_all_mbarn}
\end{center}
\end{figure*}

The $\pi^+$ and $\pi^-$  spectra presented in this
section refer to pions produced  in strong and
electromagnetic processes in
p+C interactions at 31~GeV/\textit{c}.

The spectra  normalized to the inclusive cross section are
shown in Figs.~\ref{pion_minus_all_mbarn} and
\ref{pion_plus_all_mbarn} for negatively and positively
charged pions, respectively.
The spectra are  presented as functions of particle momentum
in ten intervals of the polar angle.
Both quantities are calculated in the laboratory system.
The chosen binning takes into account the available statistics
of the  2007 data sample, detector acceptance, and particle production kinematics.

The spectra obtained by different methods 
(described in Sec.~\ref{sec:analysis}) are presented separately.
The agreement between them is, in general, better than 10\%.
Note that data points in the same ($p$,~$\theta$) bin
from different analysis methods
are statistically correlated as they result from the analysis of
the same data set.
In order to obtain the final spectra consisting of
statistically uncorrelated points, the measurement with the
smallest total error was selected.
The corresponding numerical values are 
presented in the Appendix,  Table~\ref{tab:xsec_results},
and are also available from~\cite{edms}, where the different 
contributions to systematic uncertainties (see Sec.~\ref{Sec:Syst}) 
are given separately.

\begin{figure*}[tb]
\begin{center}
\includegraphics[width=0.9\linewidth]{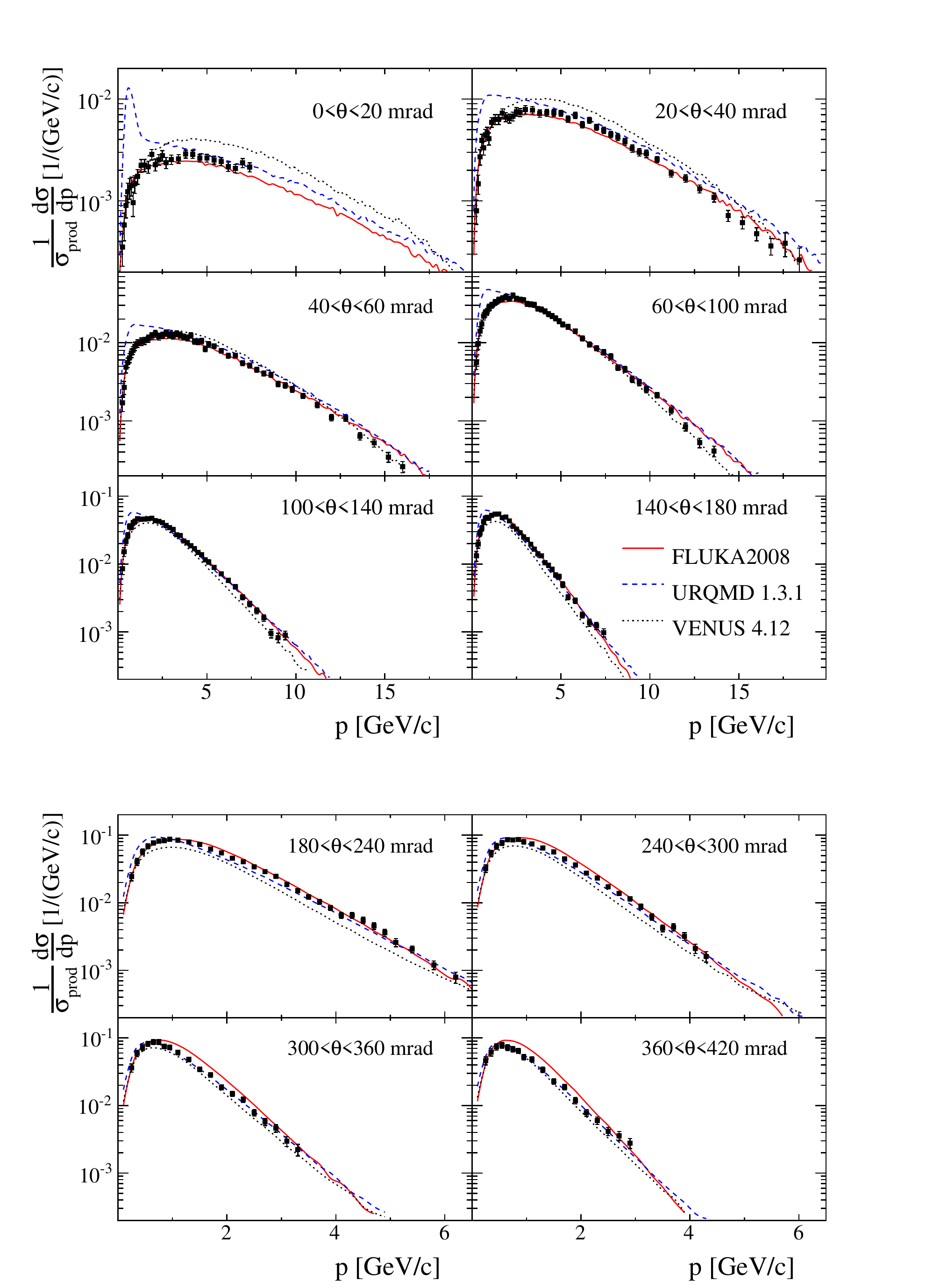}
\caption{(Color online) 
  Laboratory momentum distributions of $\pi^{-}$ mesons produced
  in production p+C interactions at 31~GeV/\textit{c}
  in different intervals of polar angle ($\theta$).
  The spectra are normalized to the mean $\pi^{-}$ multiplicity in
  all production p+C interactions.
  Error bars indicate statistical and systematic uncertainties
  added in quadrature. 
  The overall uncertainty~($2.3\%$) due to the normalization procedure is not shown.
  Predictions of hadron production models,
  \FlukaLong (solid line),
  \UrqmdLong (dashed line), and \VenusLong (dotted line), are also indicated.
}
\label{pion_minus_all_mult}
\end{center}
\end{figure*}

\begin{figure*}[tb]
\begin{center}
\includegraphics[width=0.9\linewidth]{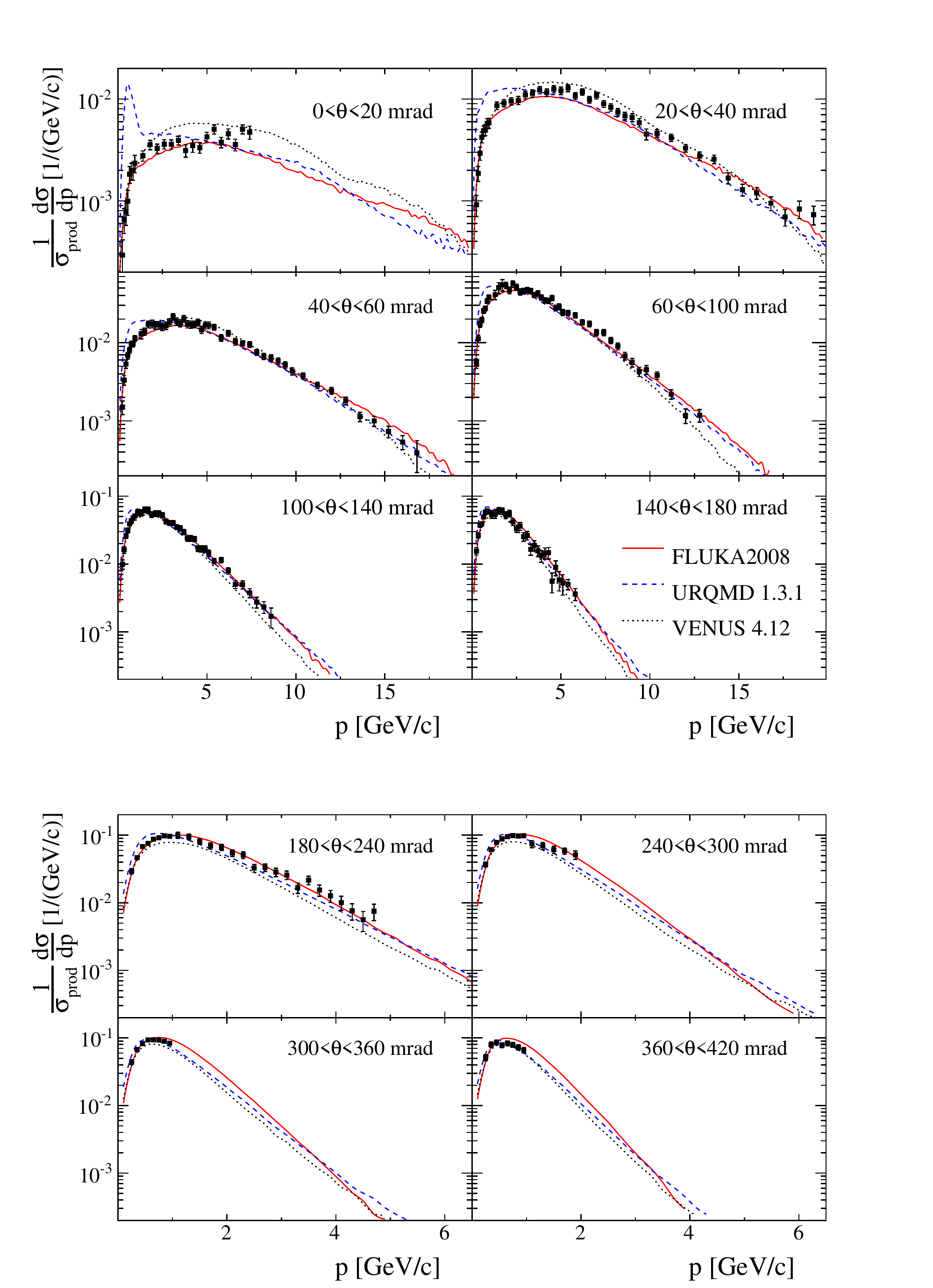}
\caption{(Color online) 
  Laboratory momentum distributions of $\pi^{+}$ mesons produced
  in production p+C interactions at 31~GeV/\textit{c}
  in different intervals of polar angle ($\theta$).
  The spectra are normalized to the mean $\pi^{+}$ multiplicity in
  all production p+C interactions.
  Error bars indicate statistical and systematic uncertainties
  added in quadrature.
  The overall uncertainty~($2.3\%$) due to the normalization procedure is not shown.
  Predictions of hadron production models,
  \FlukaLong (solid line),
  \UrqmdLong (dashed line), and \VenusLong (dotted line)  are also indicated.
}
\label{pion_plus_all_mult}
\end{center}
\end{figure*}

\begin{figure*}[tb]
\begin{center}
\includegraphics[width=0.9\linewidth]{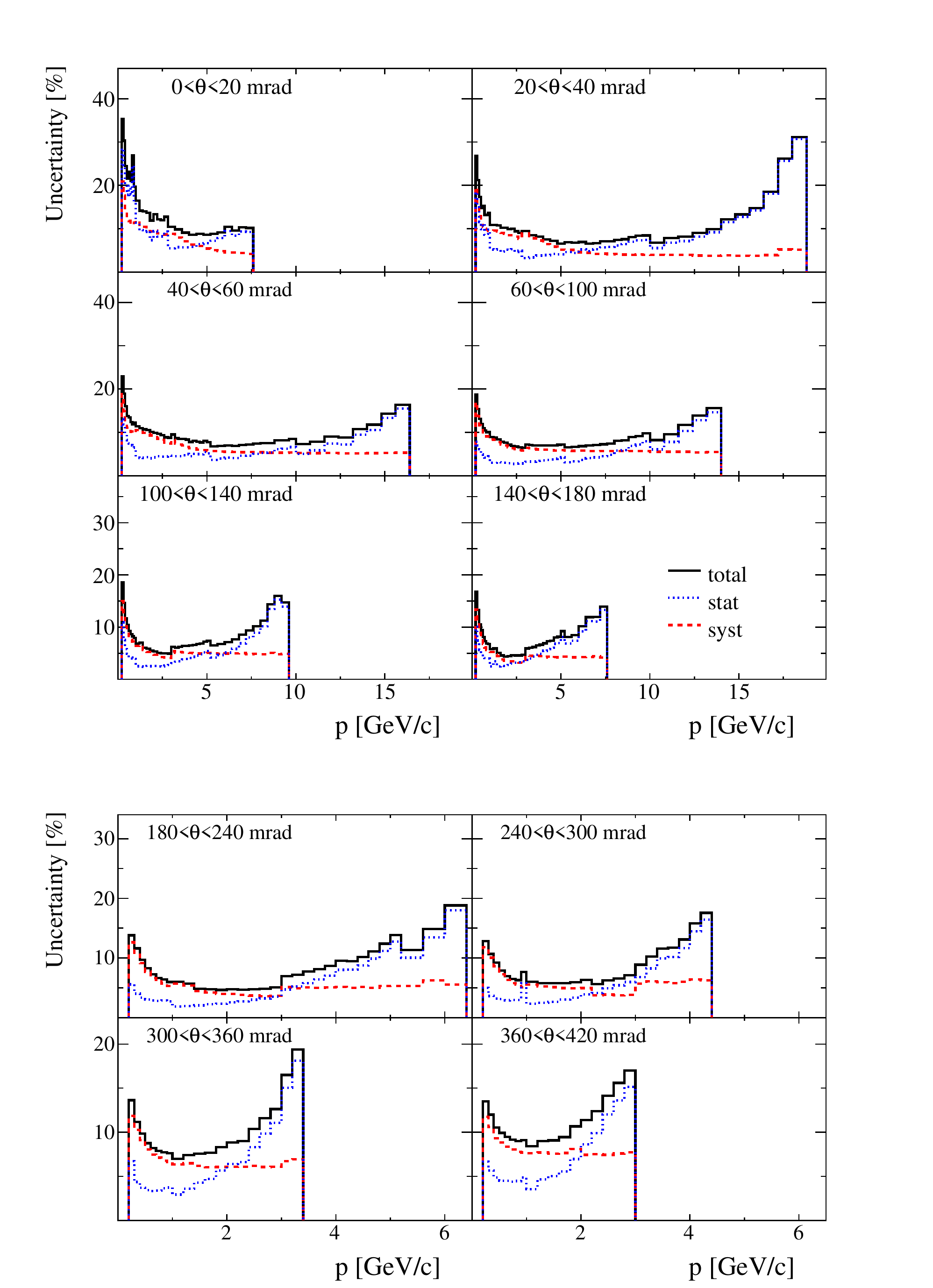}
\caption{(Color online) 
  Relative uncertainties on the $\pi^{-}$ cross sections shown in the same binning as   
  Fig.~\ref{pion_minus_all_mult}.
  Statistical (dotted line), systematic (dashed line), and total (solid line) uncertainties are   
  indicated.
  The overall uncertainty due to the normalization procedure is not shown.
}
\label{pion_minus_all_errors}
\end{center}
\end{figure*}

\begin{figure*}[tb]
\begin{center}
\includegraphics[width=0.9\linewidth]{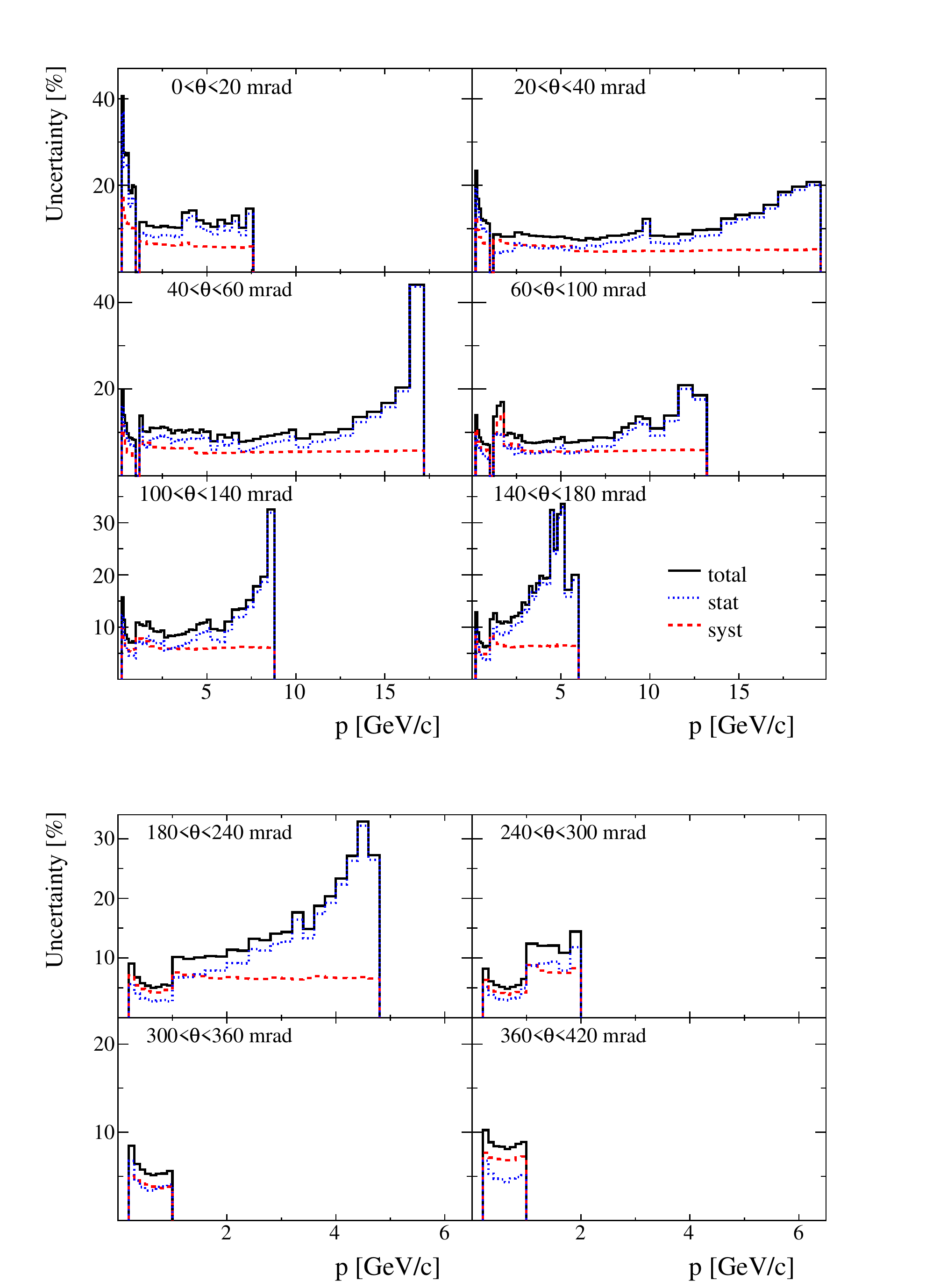}
\caption{(Color online) 
  Relative uncertainties on the $\pi^{+}$ cross sections shown in the same binning as   
  Fig.~\ref{pion_plus_all_mult}.
  Statistical (dotted line), systematic (dashed line), and total (solid line) uncertainties are   
  indicated.
  The overall uncertainty due to the normalization procedure is not shown.
}
\label{pion_plus_all_errors}
\end{center}
\end{figure*}

The final spectra are plotted in Figs.~\ref{pion_minus_all_mult}
and~\ref{pion_plus_all_mult} for $\pi^-$ and $\pi^+$, respectively,
while the corresponding fractional errors are presented 
in Figs.~\ref{pion_minus_all_errors} and~\ref{pion_plus_all_errors}.
For the purpose of a comparison of the data with model
predictions, the spectra
were normalized to the
mean $\pi^\pm$ 
multiplicity in all
production interactions.
This avoids uncertainties due
to the different treatment of quasi-elastic interactions
in models as well as problems due to the absence of predictions for
inclusive cross sections.

\begin{figure*}[!hb]
\begin{center}
\includegraphics[width=0.9\linewidth]{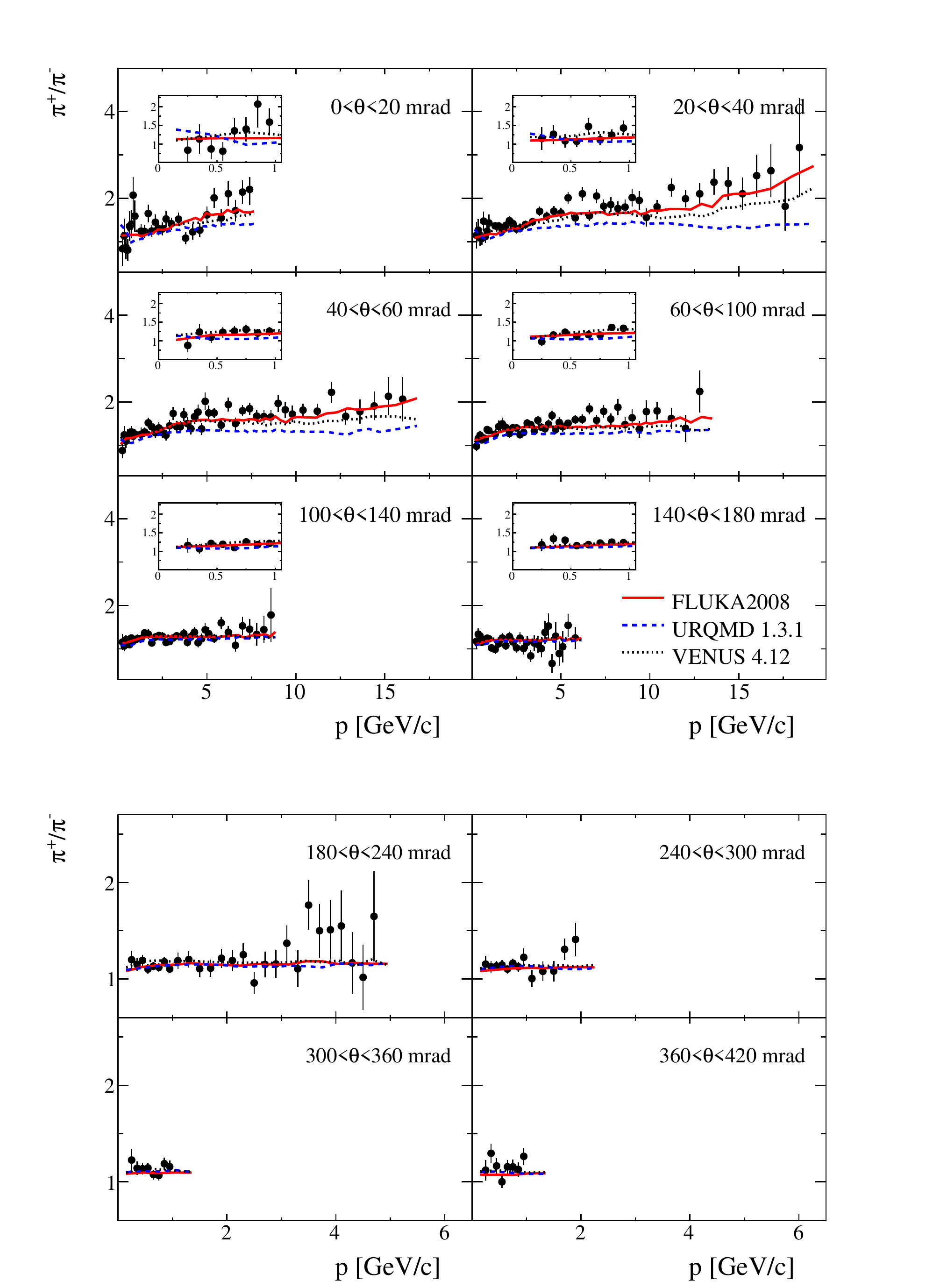}
\caption{(Color online) 
  The momentum dependence of
  the  $\pi^{+}$ to $\pi^{-}$ ratio.
  Errors are calculated taking into account only statistical uncertainties of the 
  spectra plotted in Figs.~\ref{pion_minus_all_mult} and~\ref{pion_plus_all_mult}.
  Predictions of hadron production models,
  \FlukaLong (solid line),
  \UrqmdLong (dashed line), and \VenusLong (dotted line)  are also indicated.
}
\label{fig:ratio}
\end{center}
\end{figure*}

The ratio of the final spectra of $\pi^{+}$ and $\pi^{-}$
is presented as a function of momentum
in Fig.~\ref{fig:ratio}.
The $\pi^{+}$ to $\pi^{-}$  ratio is close to 1 at
low momenta and increases with increasing momentum.

\section{Comparison to model predictions}
\label{models}

As a first application of the measurement presented in this paper, it
is interesting to compare the $\pi^-$ and $\pi^+$ spectra in p+C
interactions at 31~GeV/\textit{c} to the predictions of event generators of
hadronic interactions.  
Models that have been frequently used for the interpretation of cosmic
ray data, i.e.,\ \VenusLong~\cite{Venus}, 
\FlukaLong~\cite{Fluka}, \UrqmdLong~\cite{Urqmd}, and
\GheishaLong~\cite{Gheisha} were selected. They are part
of the \Corsika~\cite{Corsika} framework for the simulation of air
showers and are typically used to generate hadron-air interactions at
energies below 80~GeV.  In order to assure that all relevant settings of the
generators are identical to the ones used in air shower simulations,
p+C interactions at 31~GeV/\textit{c} were simulated within \Corsika in the
so-called {\itshape interaction test} mode.

The results are presented in Figs.~\ref{pion_minus_all_mult}
and~\ref{pion_plus_all_mult}
for the spectra of $\pi^-$ and $\pi^+$ respectively.
As already presented in~\cite{Unger}, \Gheisha simulations qualitatively fail
to describe
the NA61/SHINE measurements at all production angles and momenta 
(see also, e.g.,~\cite{Heck:2003br}). 
The \UrqmdLong model qualitatively disagrees with the data
only at low momenta ($p <$~3~GeV/\textit{c}) and polar angles below
about
140~mrad.
The \VenusLong and \FlukaLong models follow the data trend in
all measured polar angle intervals.

\section{Summary}

This work presents inelastic and production
cross sections as well as positively and negatively
charged pion spectra
in p+C interactions at 31~GeV/\textit{c}.
These data are essential for  precise predictions of the 
neutrino flux for the T2K 
long-baseline neutrino oscillation experiment in Japan.
Furthermore, they provide important input to improve hadron production models
needed for the interpretation of air showers initiated by ultra-high-energy cosmic particles.
The measurements were performed
with the large-acceptance NA61/SHINE spectrometer at the CERN SPS.
A set of data collected with a 4\%~$\lambda_{\mathrm{I}}$ isotropic graphite 
target during the pilot NA61/SHINE run in 2007 was used for
the analysis.
The p+C inelastic and production cross sections were found to be
\mbox{257.2 $\pm$ 1.9 $\pm$ 8.9} and \mbox{229.3 $\pm$ 1.9 $\pm$ 9.0~mb}, 
respectively.
Negatively and positively charged pion spectra as a function of laboratory
momentum  in ten intervals of the polar angle were obtained using
three different 
analysis techniques.
The final spectra were compared with predictions of hadron production models.

The data presented in this paper already provide important information
for calculating the T2K neutrino flux. Meanwhile,
a much larger data set with both the thin (4\%~$\lambda_{\mathrm{I}}$)
and the T2K replica carbon targets
was recorded in 2009 and 2010 and is presently being analyzed. This will lead
to results of higher precision for pions and extend the measurements to
other hadron species such as charged kaons, protons, $K^0_S$ and $\Lambda$. The
new data will allow a further significant reduction of the uncertainties
in the prediction of the neutrino flux in the T2K experiment.

\section{Acknowledgments}
This work was supported by the following funding agencies:
the Hungarian Scientific Research Fund (OTKA Grants No.~68506 and No.~79840),
the Polish Ministry of Science and Higher Education [Grants No.~667/N-CERN/2010/0,
No.~N N202 1267 36, No.~N N202 287838 (No.~PBP 2878/B/H03/2010/38), and No.~DWM/57/T2K/2007],
the Federal Agency of Education of the Ministry of Education and Science
of the Russian Federation (RNP Grant No.~2.2.2.2.1547), 
the Russian Academy of Sciences and
the Russian Foundation for Basic Research (Grants No.~08-02-00018 and No.~09-02-00664),
the Ministry of Education, Culture, Sports, Science and Technology,
Japan, Grant-in-Aid for Scientific Research (Grants No.~18071005, No.~19034011,
No.~19740162, No.~20740160, and No.~20039012), the Toshiko Yuasa Laboratory 
(France-Japan Particle Physics Laboratory),
the Institut National de Physique Nucl\'eaire et Physique des Particules
(IN2P3, France),
the German Research Foundation (Grant No.~GA 1480/2-1),
the Swiss National Science Foundation 
(Investigator-Driven projects and SINERGIA) and the Swiss State Secretariat 
for Education and Research (FORCE grants). \\
The authors also wish to acknowledge the support provided 
by the collaborating institutions, in particular,
the ETH Zurich (Research Grant No.~TH-01 07-3), 
the University of Bern, and the University of Geneva.

Finally, it is a pleasure to thank 
the European Organization for Nuclear Research 
for strong support and hospitality
and, in particular, the operating crews of the CERN SPS accelerator
and beam lines who made the measurements possible.

\clearpage
\appendix*
\section{}

\begin{center}
\begin {longtable*}{ cc  rr  lll lll lll  rr  lll lll lll }
  \caption{\label{tab:xsec_results}
    The NA61/SHINE results for the differential $\pi^+$ and $\pi^-$ production
    cross section in the laboratory system,
    $d\sigma^{\pi}/dp$, for p+C interactions at 31~GeV/\textit{c}.
    Each row refers to a
    different (\mbox{$p_{low} \le p<p_{up}$},
    \mbox{$\theta_{low} \le \theta<\theta_{up}$}) bin,
    where $p$ and $\theta$ are the pion momentum and polar angle
    in the laboratory frame, respectively.
    The central value as well as the statistical~($\Delta_{stat}$) and
    systematic~($\Delta_{sys}$) errors are given.
    The overall uncertainty (2.5\%) due to the normalization procedure is not included.}\\

\hline
\multicolumn{1}{c}{$\theta_{low}$} &
\multicolumn{1}{c}{$\theta_{up}$} &
\multicolumn{1}{c}{$p_{low}$} &
\multicolumn{1}{c}{$p_{up}$} &
\multicolumn{3}{c}{$\frac{d\sigma}{dp}^{\pi^+}$} &
\multicolumn{3}{c}{$\Delta_{stat}$} &
\multicolumn{3}{c}{$\Delta_{sys}$} &
\multicolumn{1}{c}{$p_{low}$} &
\multicolumn{1}{c}{$p_{up}$} &
\multicolumn{3}{c}{$\frac{d\sigma}{dp}^{\pi^-}$} &
\multicolumn{3}{c}{$\Delta_{stat}$} &
\multicolumn{3}{c}{$\Delta_{sys}$}
\\
\multicolumn{2}{c}{(mrad)} &
\multicolumn{2}{c}{(GeV/\textit{c})} &
\multicolumn{9}{c}{[mb/(GeV/\textit{c})]} &
\multicolumn{2}{c}{(GeV/\textit{c})} &
\multicolumn{9}{c}{[mb/(GeV/\textit{c})]}\\
\hline
\endfirsthead

\hline
\multicolumn{1}{c}{$\theta_{low}$} &
\multicolumn{1}{c}{$\theta_{up}$} &
\multicolumn{1}{c}{$p_{low}$} &
\multicolumn{1}{c}{$p_{up}$} &
\multicolumn{3}{c}{$\frac{d\sigma}{dp}^{\pi^+}$} &
\multicolumn{3}{c}{$\Delta_{stat}$} &
\multicolumn{3}{c}{$\Delta_{sys}$} &
\multicolumn{1}{c}{$p_{low}$} &
\multicolumn{1}{c}{$p_{up}$} &
\multicolumn{3}{c}{$\frac{d\sigma}{dp}^{\pi^-}$} &
\multicolumn{3}{c}{$\Delta_{stat}$} &
\multicolumn{3}{c}{$\Delta_{sys}$}
\\
\multicolumn{2}{c}{(mrad)} &
\multicolumn{2}{c}{(GeV/\textit{c})} &
\multicolumn{9}{c}{[mb/(GeV/\textit{c})]} &
\multicolumn{2}{c}{(GeV/\textit{c})} &
\multicolumn{9}{c}{[mb/(GeV/\textit{c})]}\\
\hline
\endhead

\hline
\multicolumn{12}{r}{{Continued on next page}}\\
\endfoot

\hline \hline
\endlastfoot

0&	20&	0.2	&	0.3	&&	0.067&	&&	0.025&	&&	0.012&	&0.2	&	0.3	&&	0.081&	&&	0.023&	&&	0.017&	\\
&	&	0.3	&	0.4	&&	0.151&	&&	0.037&	&&	0.020&	&0.3	&	0.4	&&	0.133&	&&	0.033&	&&	0.023&	\\
&	&	0.4	&	0.5	&&	0.179&	&&	0.044&	&&	0.021&	&0.4	&	0.5	&&	0.207&	&&	0.042&	&&	0.028&	\\
&	&	0.5	&	0.6	&&	0.228&	&&	0.057&	&&	0.025&	&0.5	&	0.6	&&	0.282&	&&	0.050&	&&	0.034&	\\
&	&	0.6	&	0.7	&&	0.419&	&&	0.066&	&&	0.044&	&0.6	&	0.7	&&	0.309&	&&	0.062&	&&	0.036&	\\
&	&	0.7	&	0.8	&&	0.454&	&&	0.069&	&&	0.046&	&0.7	&	0.8	&&	0.324&	&&	0.058&	&&	0.036&	\\
&	&	0.8	&	0.9	&&	0.456&	&&	0.079&	&&	0.046&	&0.8	&	0.9	&&	0.220&	&&	0.054&	&&	0.024&	\\
&	&	0.9	&	1.0	&&	0.537&	&&	0.088&	&&	0.058&	&0.9	&	1.0	&&	0.338&	&&	0.054&	&&	0.039&	\\
&	&		&		&&	&	&&	&	&&	&	&1.0	&	1.2	&&	0.399&	&&	0.045&	&&	0.047&	\\
&	&	1.2	&	1.6	&&	0.636&	&&	0.057&	&&	0.045&	&1.2	&	1.4	&&	0.512&	&&	0.049&	&&	0.053&	\\
&	&	1.6	&	2.0	&&	0.816&	&&	0.069&	&&	0.053&	&1.4	&	1.6	&&	0.516&	&&	0.048&	&&	0.054&	\\
&	&	2.0	&	2.4	&&	0.752&	&&	0.060&	&&	0.049&	&1.6	&	1.8	&&	0.495&	&&	0.047&	&&	0.049&	\\
&	&	2.4	&	2.8	&&	0.826&	&&	0.070&	&&	0.052&	&1.8	&	2.0	&&	0.655&	&&	0.048&	&&	0.061&	\\
&	&	2.8	&	3.2	&&	0.825&	&&	0.069&	&&	0.050&	&2.0	&	2.2	&&	0.519&	&&	0.049&	&&	0.049&	\\
&	&	3.2	&	3.6	&&	0.900&	&&	0.073&	&&	0.056&	&2.2	&	2.4	&&	0.581&	&&	0.048&	&&	0.052&	\\
&	&	3.6	&	4.0	&&	0.715&	&&	0.085&	&&	0.048&	&2.4	&	2.6	&&	0.639&	&&	0.053&	&&	0.055&	\\
&	&	4.0	&	4.4	&&	0.80&	&&	0.10&	&&	0.05&	&2.6	&	2.8	&&	0.543&	&&	0.051&	&&	0.047&	\\
&	&	4.4	&	4.8	&&	0.763&	&&	0.079&	&&	0.045&	&2.8	&	3.2	&&	0.584&	&&	0.032&	&&	0.052&	\\
&	&	4.8	&	5.2	&&	0.975&	&&	0.093&	&&	0.057&	&3.2	&	3.6	&&	0.593&	&&	0.034&	&&	0.047&	\\
&	&	5.2	&	5.6	&&	1.16&	&&	0.10&	&&	0.07&	&3.6	&	4.0	&&	0.662&	&&	0.038&	&&	0.047&	\\
&	&	5.6	&	6.0	&&	0.862&	&&	0.091&	&&	0.049&	&4.0	&	4.4	&&	0.655&	&&	0.038&	&&	0.041&	\\
&	&	6.0	&	6.4	&&	1.05&	&&	0.10&	&&	0.06&	&4.4	&	4.8	&&	0.604&	&&	0.039&	&&	0.036&	\\
&	&	6.4	&	6.8	&&	0.817&	&&	0.096&	&&	0.047&	&4.8	&	5.2	&&	0.604&	&&	0.040&	&&	0.033&	\\
&	&	6.8	&	7.2	&&	1.159&	&&	0.099&	&&	0.065&	&5.2	&	5.6	&&	0.577&	&&	0.042&	&&	0.029&	\\
&	&	7.2	&	7.6	&&	1.08&	&&	0.15&	&&	0.06&	&5.6	&	6.0	&&	0.558&	&&	0.044&	&&	0.027&	\\
&	&		&		&&	&	&&	&	&&	&	&6.0	&	6.4	&&	0.498&	&&	0.047&	&&	0.022&	\\
&	&		&		&&	&	&&	&	&&	&	&6.4	&	6.8	&&	0.477&	&&	0.040&	&&	0.021&	\\
&	&		&		&&	&	&&	&	&&	&	&6.8	&	7.2	&&	0.541&	&&	0.051&	&&	0.024&	\\
&	&		&		&&	&	&&	&	&&	&	&7.2	&	7.6	&&	0.492&	&&	0.046&	&&	0.021&	\\ \hline
20&	40&	0.2	&	0.3	&&	0.210&	&&	0.041&	&&	0.027&	&0.2	&	0.3	&&	0.183&	&&	0.034&	&&	0.036&	\\
&	&	0.3	&	0.4	&&	0.427&	&&	0.059&	&&	0.041&	&0.3	&	0.4	&&	0.337&	&&	0.051&	&&	0.051&	\\
&	&	0.4	&	0.5	&&	0.674&	&&	0.082&	&&	0.054&	&0.4	&	0.5	&&	0.621&	&&	0.072&	&&	0.080&	\\
&	&	0.5	&	0.6	&&	0.967&	&&	0.096&	&&	0.069&	&0.5	&	0.6	&&	0.901&	&&	0.089&	&&	0.098&	\\
&	&	0.6	&	0.7	&&	1.12&	&&	0.11&	&&	0.08&	&0.6	&	0.7	&&	0.761&	&&	0.085&	&&	0.080&	\\
&	&	0.7	&	0.8	&&	1.14&	&&	0.11&	&&	0.07&	&0.7	&	0.8	&&	1.02&	&&	0.09&	&&	0.10&	\\
&	&	0.8	&	0.9	&&	1.32&	&&	0.12&	&&	0.08&	&0.8	&	0.9	&&	1.07&	&&	0.09&	&&	0.10&	\\
&	&	0.9	&	1.0	&&	1.34&	&&	0.13&	&&	0.08&	&0.9	&	1.0	&&	0.936&	&&	0.092&	&&	0.089&	\\
&	&		&		&&	&	&&	&	&&	&	&1.0	&	1.2	&&	1.35&	&&	0.07&	&&	0.13&	\\
&	&	1.2	&	1.6	&&	1.97&	&&	0.09&	&&	0.15&	&1.2	&	1.4	&&	1.45&	&&	0.08&	&&	0.14&	\\
&	&	1.6	&	2.0	&&	2.10&	&&	0.10&	&&	0.14&	&1.4	&	1.6	&&	1.45&	&&	0.08&	&&	0.14&	\\
&	&	2.0	&	2.4	&&	2.19&	&&	0.11&	&&	0.14&	&1.6	&	1.8	&&	1.67&	&&	0.08&	&&	0.15&	\\
&	&	2.4	&	2.8	&&	2.22&	&&	0.15&	&&	0.14&	&1.8	&	2.0	&&	1.53&	&&	0.08&	&&	0.13&	\\
&	&	2.8	&	3.2	&&	2.52&	&&	0.15&	&&	0.16&	&2.0	&	2.2	&&	1.48&	&&	0.08&	&&	0.13&	\\
&	&	3.2	&	3.6	&&	2.60&	&&	0.15&	&&	0.16&	&2.2	&	2.4	&&	1.55&	&&	0.08&	&&	0.13&	\\
&	&	3.6	&	4.0	&&	2.81&	&&	0.15&	&&	0.17&	&2.4	&	2.6	&&	1.72&	&&	0.08&	&&	0.14&	\\
&	&	4.0	&	4.4	&&	2.66&	&&	0.15&	&&	0.16&	&2.6	&	2.8	&&	1.69&	&&	0.08&	&&	0.13&	\\
&	&	4.4	&	4.8	&&	2.86&	&&	0.16&	&&	0.17&	&2.8	&	3.2	&&	1.80&	&&	0.06&	&&	0.16&	\\
&	&	4.8	&	5.2	&&	2.77&	&&	0.16&	&&	0.16&	&3.2	&	3.6	&&	1.78&	&&	0.06&	&&	0.14&	\\
&	&	5.2	&	5.6	&&	2.94&	&&	0.15&	&&	0.18&	&3.6	&	4.0	&&	1.65&	&&	0.06&	&&	0.12&	\\
&	&	5.6	&	6.0	&&	2.45&	&&	0.14&	&&	0.12&	&4.0	&	4.4	&&	1.68&	&&	0.07&	&&	0.11&	\\
&	&	6.0	&	6.4	&&	2.70&	&&	0.15&	&&	0.13&	&4.4	&	4.8	&&	1.683&	&&	0.071&	&&	0.099&	\\
&	&	6.4	&	6.8	&&	2.27&	&&	0.14&	&&	0.11&	&4.8	&	5.2	&&	1.667&	&&	0.068&	&&	0.086&	\\
&	&	6.8	&	7.2	&&	2.48&	&&	0.15&	&&	0.12&	&5.2	&	5.6	&&	1.463&	&&	0.067&	&&	0.076&	\\
&	&	7.2	&	7.6	&&	2.06&	&&	0.13&	&&	0.10&	&5.6	&	6.0	&&	1.587&	&&	0.070&	&&	0.081&	\\
&	&	7.6	&	8.0	&&	1.90&	&&	0.13&	&&	0.09&	&6.0	&	6.4	&&	1.284&	&&	0.067&	&&	0.059&	\\
&	&	8.0	&	8.4	&&	1.70&	&&	0.12&	&&	0.08&	&6.4	&	6.8	&&	1.423&	&&	0.068&	&&	0.063&	\\
&	&	8.4	&	8.8	&&	1.56&	&&	0.11&	&&	0.07&	&6.8	&	7.2	&&	1.209&	&&	0.063&	&&	0.050&	\\
&	&	8.8	&	9.2	&&	1.51&	&&	0.11&	&&	0.07&	&7.2	&	7.6	&&	1.132&	&&	0.064&	&&	0.048&	\\
&	&	9.2	&	9.6	&&	1.34&	&&	0.11&	&&	0.07&	&7.6	&	8.0	&&	1.027&	&&	0.060&	&&	0.042&	\\
&	&	9.6	&	10.0	&&	1.04&	&&	0.12&	&&	0.05&	&8.0	&	8.4	&&	0.967&	&&	0.059&	&&	0.040&	\\
&	&	10.0	&	10.8	&&	1.055&	&&	0.072&	&&	0.051&	&8.4	&	8.8	&&	0.872&	&&	0.056&	&&	0.035&	\\
&	&	10.8	&	11.6	&&	0.959&	&&	0.063&	&&	0.047&	&8.8	&	9.2	&&	0.749&	&&	0.052&	&&	0.030&	\\
&	&	11.6	&	12.4	&&	0.756&	&&	0.055&	&&	0.037&	&9.2	&	9.6	&&	0.688&	&&	0.050&	&&	0.028&	\\
&	&	12.4	&	13.2	&&	0.632&	&&	0.053&	&&	0.031&	&9.6	&	10.0	&&	0.667&	&&	0.049&	&&	0.027&	\\
&	&	13.2	&	14.0	&&	0.588&	&&	0.050&	&&	0.030&	&10.0	&	10.8	&&	0.586&	&&	0.032&	&&	0.023&	\\
&	&	14.0	&	14.8	&&	0.384&	&&	0.043&	&&	0.019&	&10.8	&	11.6	&&	0.427&	&&	0.029&	&&	0.017&	\\
&	&	14.8	&	15.6	&&	0.295&	&&	0.036&	&&	0.015&	&11.6	&	12.4	&&	0.380&	&&	0.027&	&&	0.015&	\\
&	&	15.6	&	16.4	&&	0.275&	&&	0.035&	&&	0.014&	&12.4	&	13.2	&&	0.301&	&&	0.025&	&&	0.011&	\\
&	&	16.4	&	17.2	&&	0.217&	&&	0.032&	&&	0.011&	&13.2	&	14.0	&&	0.248&	&&	0.023&	&&	0.009&	\\
&	&	17.2	&	18.0	&&	0.159&	&&	0.028&	&&	0.008&	&14.0	&	14.8	&&	0.164&	&&	0.019&	&&	0.006&	\\
&	&	18.0	&	18.8	&&	0.191&	&&	0.036&	&&	0.010&	&14.8	&	15.6	&&	0.140&	&&	0.018&	&&	0.005&	\\
&	&	18.8	&	19.6	&&	0.168&	&&	0.034&	&&	0.009&	&15.6	&	16.4	&&	0.109&	&&	0.016&	&&	0.004&	\\
&	&		&		&&	&	&&	&	&&	&	&16.4	&	17.2	&&	0.083&	&&	0.015&	&&	0.003&	\\
&	&		&		&&	&	&&	&	&&	&	&17.2	&	18.0	&&	0.088&	&&	0.023&	&&	0.005&	\\
&	&		&		&&	&	&&	&	&&	&	&18.0	&	18.8	&&	0.060&	&&	0.018&	&&	0.003&	\\ \hline
40&	60&	0.2	&	0.3	&&	0.344&	&&	0.056&	&&	0.040&	&0.2	&	0.3	&&	0.391&	&&	0.051&	&&	0.074&	\\
&	&	0.3	&	0.4	&&	0.763&	&&	0.087&	&&	0.062&	&0.3	&	0.4	&&	0.616&	&&	0.073&	&&	0.091&	\\
&	&	0.4	&	0.5	&&	1.22&	&&	0.13&	&&	0.08&	&0.4	&	0.5	&&	1.12&	&&	0.11&	&&	0.14&	\\
&	&	0.5	&	0.6	&&	1.58&	&&	0.13&	&&	0.08&	&0.5	&	0.6	&&	1.28&	&&	0.10&	&&	0.14&	\\
&	&	0.6	&	0.7	&&	1.85&	&&	0.13&	&&	0.09&	&0.6	&	0.7	&&	1.47&	&&	0.11&	&&	0.16&	\\
&	&	0.7	&	0.8	&&	2.22&	&&	0.16&	&&	0.11&	&0.7	&	0.8	&&	1.69&	&&	0.12&	&&	0.17&	\\
&	&	0.8	&	0.9	&&	2.22&	&&	0.15&	&&	0.10&	&0.8	&	0.9	&&	1.83&	&&	0.12&	&&	0.18&	\\
&	&	0.9	&	1.0	&&	2.60&	&&	0.18&	&&	0.12&	&0.9	&	1.0	&&	2.07&	&&	0.13&	&&	0.21&	\\
&	&		&		&&	&	&&	&	&&	&	&1.0	&	1.2	&&	2.24&	&&	0.09&	&&	0.24&	\\
&	&	1.2	&	1.4	&&	2.98&	&&	0.33&	&&	0.24&	&1.2	&	1.4	&&	2.37&	&&	0.10&	&&	0.24&	\\
&	&	1.4	&	1.6	&&	3.18&	&&	0.23&	&&	0.23&	&1.4	&	1.6	&&	2.43&	&&	0.10&	&&	0.24&	\\
&	&	1.6	&	1.8	&&	3.94&	&&	0.26&	&&	0.30&	&1.6	&	1.8	&&	2.59&	&&	0.11&	&&	0.25&	\\
&	&	1.8	&	2.0	&&	3.97&	&&	0.34&	&&	0.28&	&1.8	&	2.0	&&	2.79&	&&	0.11&	&&	0.26&	\\
&	&	2.0	&	2.2	&&	3.85&	&&	0.34&	&&	0.25&	&2.0	&	2.2	&&	2.98&	&&	0.13&	&&	0.27&	\\
&	&	2.2	&	2.4	&&	3.84&	&&	0.34&	&&	0.25&	&2.2	&	2.4	&&	2.75&	&&	0.12&	&&	0.23&	\\
&	&	2.4	&	2.6	&&	3.69&	&&	0.34&	&&	0.24&	&2.4	&	2.6	&&	2.86&	&&	0.13&	&&	0.24&	\\
&	&	2.6	&	2.8	&&	3.78&	&&	0.34&	&&	0.24&	&2.6	&	2.8	&&	3.06&	&&	0.14&	&&	0.24&	\\
&	&	2.8	&	3.0	&&	4.18&	&&	0.35&	&&	0.27&	&2.8	&	3.0	&&	2.88&	&&	0.13&	&&	0.21&	\\
&	&	3.0	&	3.2	&&	4.91&	&&	0.37&	&&	0.31&	&3.0	&	3.2	&&	2.83&	&&	0.13&	&&	0.24&	\\
&	&	3.2	&	3.4	&&	4.28&	&&	0.35&	&&	0.27&	&3.2	&	3.4	&&	2.99&	&&	0.13&	&&	0.22&	\\
&	&	3.4	&	3.6	&&	4.09&	&&	0.34&	&&	0.26&	&3.4	&	3.6	&&	2.87&	&&	0.13&	&&	0.21&	\\
&	&	3.6	&	3.8	&&	4.65&	&&	0.36&	&&	0.29&	&3.6	&	3.8	&&	2.73&	&&	0.13&	&&	0.19&	\\
&	&	3.8	&	4.0	&&	3.95&	&&	0.33&	&&	0.25&	&3.8	&	4.0	&&	2.64&	&&	0.13&	&&	0.18&	\\
&	&	4.0	&	4.2	&&	3.92&	&&	0.34&	&&	0.25&	&4.0	&	4.2	&&	2.84&	&&	0.13&	&&	0.18&	\\
&	&	4.2	&	4.4	&&	3.96&	&&	0.34&	&&	0.21&	&4.2	&	4.4	&&	2.39&	&&	0.12&	&&	0.15&	\\
&	&	4.4	&	4.6	&&	4.16&	&&	0.35&	&&	0.22&	&4.4	&	4.6	&&	2.36&	&&	0.12&	&&	0.14&	\\
&	&	4.6	&	4.8	&&	3.35&	&&	0.31&	&&	0.17&	&4.6	&	4.8	&&	2.43&	&&	0.12&	&&	0.14&	\\
&	&	4.8	&	5.0	&&	3.84&	&&	0.33&	&&	0.20&	&4.8	&	5.0	&&	1.91&	&&	0.10&	&&	0.11&	\\
&	&	5.0	&	5.2	&&	3.83&	&&	0.33&	&&	0.20&	&5.0	&	5.2	&&	2.20&	&&	0.11&	&&	0.13&	\\
&	&	5.2	&	5.6	&&	3.63&	&&	0.22&	&&	0.19&	&5.2	&	5.6	&&	2.08&	&&	0.08&	&&	0.12&	\\
&	&	5.6	&	6.0	&&	2.64&	&&	0.21&	&&	0.14&	&5.6	&	6.0	&&	1.80&	&&	0.07&	&&	0.10&	\\
&	&	6.0	&	6.4	&&	3.04&	&&	0.21&	&&	0.16&	&6.0	&	6.4	&&	1.568&	&&	0.066&	&&	0.087&	\\
&	&	6.4	&	6.8	&&	2.36&	&&	0.20&	&&	0.13&	&6.4	&	6.8	&&	1.576&	&&	0.064&	&&	0.087&	\\
&	&	6.8	&	7.2	&&	2.27&	&&	0.13&	&&	0.12&	&6.8	&	7.2	&&	1.264&	&&	0.058&	&&	0.069&	\\
&	&	7.2	&	7.6	&&	2.18&	&&	0.13&	&&	0.12&	&7.2	&	7.6	&&	1.180&	&&	0.055&	&&	0.064&	\\
&	&	7.6	&	8.0	&&	1.74&	&&	0.11&	&&	0.09&	&7.6	&	8.0	&&	1.038&	&&	0.053&	&&	0.056&	\\
&	&	8.0	&	8.4	&&	1.54&	&&	0.11&	&&	0.08&	&8.0	&	8.4	&&	0.928&	&&	0.049&	&&	0.050&	\\
&	&	8.4	&	8.8	&&	1.48&	&&	0.11&	&&	0.08&	&8.4	&	8.8	&&	0.894&	&&	0.048&	&&	0.047&	\\
&	&	8.8	&	9.2	&&	1.34&	&&	0.10&	&&	0.07&	&8.8	&	9.2	&&	0.681&	&&	0.042&	&&	0.036&	\\
&	&	9.2	&	9.6	&&	1.189&	&&	0.097&	&&	0.065&	&9.2	&	9.6	&&	0.654&	&&	0.041&	&&	0.035&	\\
&	&	9.6	&	10.0	&&	0.995&	&&	0.090&	&&	0.055&	&9.6	&	10.0	&&	0.579&	&&	0.038&	&&	0.031&	\\
&	&	10.0	&	10.8	&&	0.870&	&&	0.057&	&&	0.048&	&10.0	&	10.8	&&	0.481&	&&	0.024&	&&	0.025&	\\
&	&	10.8	&	11.6	&&	0.657&	&&	0.052&	&&	0.036&	&10.8	&	11.6	&&	0.368&	&&	0.022&	&&	0.019&	\\
&	&	11.6	&	12.4	&&	0.562&	&&	0.046&	&&	0.031&	&11.6	&	12.4	&&	0.253&	&&	0.019&	&&	0.013&	\\
&	&	12.4	&	13.2	&&	0.414&	&&	0.038&	&&	0.023&	&12.4	&	13.2	&&	0.248&	&&	0.018&	&&	0.013&	\\
&	&	13.2	&	14.0	&&	0.260&	&&	0.032&	&&	0.015&	&13.2	&	14.0	&&	0.147&	&&	0.014&	&&	0.008&	\\
&	&	14.0	&	14.8	&&	0.230&	&&	0.031&	&&	0.013&	&14.0	&	14.8	&&	0.120&	&&	0.013&	&&	0.006&	\\
&	&	14.8	&	15.6	&&	0.168&	&&	0.027&	&&	0.009&	&14.8	&	15.6	&&	0.079&	&&	0.011&	&&	0.004&	\\
&	&	15.6	&	16.4	&&	0.123&	&&	0.024&	&&	0.007&	&15.6	&	16.4	&&	0.0598&	&&	0.0093&	&&	0.0032&	\\
&	&	16.4	&	17.2	&&	0.090&	&&	0.039&	&&	0.005&	&	&		&&	&	&&	&	&&	&	\\ \hline
60&	100&	0.2	&	0.3	&&	1.25&	&&	0.11&	&&	0.13&	&0.2	&	0.3	&&	1.28&	&&	0.10&	&&	0.22&	\\
&	&	0.3	&	0.4	&&	2.58&	&&	0.18&	&&	0.20&	&0.3	&	0.4	&&	2.23&	&&	0.15&	&&	0.31&	\\
&	&	0.4	&	0.5	&&	3.98&	&&	0.23&	&&	0.26&	&0.4	&	0.5	&&	3.23&	&&	0.20&	&&	0.38&	\\
&	&	0.5	&	0.6	&&	4.48&	&&	0.23&	&&	0.27&	&0.5	&	0.6	&&	3.99&	&&	0.20&	&&	0.43&	\\
&	&	0.6	&	0.7	&&	5.86&	&&	0.26&	&&	0.34&	&0.6	&	0.7	&&	5.02&	&&	0.23&	&&	0.50&	\\
&	&	0.7	&	0.8	&&	6.29&	&&	0.28&	&&	0.36&	&0.7	&	0.8	&&	5.45&	&&	0.23&	&&	0.52&	\\
&	&	0.8	&	0.9	&&	7.81&	&&	0.32&	&&	0.47&	&0.8	&	0.9	&&	5.74&	&&	0.24&	&&	0.52&	\\
&	&	0.9	&	1.0	&&	8.68&	&&	0.34&	&&	0.52&	&0.9	&	1.0	&&	6.49&	&&	0.29&	&&	0.59&	\\
&	&		&		&&	&	&&	&	&&	&	&1.0	&	1.2	&&	6.90&	&&	0.21&	&&	0.58&	\\
&	&	1.2	&	1.4	&&	9.42&	&&	0.91&	&&	0.91&	&1.2	&	1.4	&&	7.50&	&&	0.22&	&&	0.62&	\\
&	&	1.4	&	1.6	&&	11.6&	&&	1.1&	&&	1.5&	&1.4	&	1.6	&&	8.09&	&&	0.25&	&&	0.63&	\\
&	&	1.6	&	1.8	&&	12.6&	&&	1.1&	&&	1.8&	&1.6	&	1.8	&&	8.40&	&&	0.26&	&&	0.61&	\\
&	&	1.8	&	2.0	&&	12.41&	&&	0.79&	&&	0.85&	&1.8	&	2.0	&&	8.78&	&&	0.26&	&&	0.61&	\\
&	&	2.0	&	2.2	&&	10.68&	&&	0.75&	&&	0.73&	&2.0	&	2.2	&&	8.40&	&&	0.24&	&&	0.55&	\\
&	&	2.2	&	2.4	&&	12.96&	&&	0.81&	&&	0.96&	&2.2	&	2.4	&&	9.24&	&&	0.25&	&&	0.58&	\\
&	&	2.4	&	2.6	&&	11.76&	&&	0.58&	&&	0.85&	&2.4	&	2.6	&&	8.41&	&&	0.24&	&&	0.51&	\\
&	&	2.6	&	2.8	&&	10.25&	&&	0.73&	&&	0.62&	&2.6	&	2.8	&&	8.24&	&&	0.24&	&&	0.49&	\\
&	&	2.8	&	3.0	&&	10.43&	&&	0.57&	&&	0.61&	&2.8	&	3.0	&&	8.05&	&&	0.23&	&&	0.47&	\\
&	&	3.0	&	3.2	&&	10.88&	&&	0.56&	&&	0.61&	&3.0	&	3.2	&&	7.21&	&&	0.23&	&&	0.47&	\\
&	&	3.2	&	3.4	&&	10.51&	&&	0.55&	&&	0.59&	&3.2	&	3.4	&&	7.12&	&&	0.23&	&&	0.45&	\\
&	&	3.4	&	3.6	&&	9.34&	&&	0.48&	&&	0.51&	&3.4	&	3.6	&&	7.03&	&&	0.23&	&&	0.44&	\\
&	&	3.6	&	3.8	&&	9.82&	&&	0.52&	&&	0.54&	&3.6	&	3.8	&&	6.23&	&&	0.21&	&&	0.38&	\\
&	&	3.8	&	4.0	&&	8.66&	&&	0.48&	&&	0.47&	&3.8	&	4.0	&&	6.11&	&&	0.21&	&&	0.37&	\\
&	&	4.0	&	4.2	&&	8.01&	&&	0.45&	&&	0.44&	&4.0	&	4.2	&&	5.80&	&&	0.20&	&&	0.35&	\\
&	&	4.2	&	4.4	&&	7.90&	&&	0.46&	&&	0.43&	&4.2	&	4.4	&&	5.32&	&&	0.20&	&&	0.32&	\\
&	&	4.4	&	4.6	&&	8.52&	&&	0.48&	&&	0.47&	&4.4	&	4.6	&&	5.06&	&&	0.19&	&&	0.30&	\\
&	&	4.6	&	4.8	&&	6.37&	&&	0.42&	&&	0.35&	&4.6	&	4.8	&&	4.68&	&&	0.18&	&&	0.27&	\\
&	&	4.8	&	5.0	&&	6.64&	&&	0.43&	&&	0.37&	&4.8	&	5.0	&&	4.35&	&&	0.18&	&&	0.26&	\\
&	&	5.0	&	5.2	&&	5.60&	&&	0.39&	&&	0.31&	&5.0	&	5.2	&&	3.92&	&&	0.17&	&&	0.23&	\\
&	&	5.2	&	5.6	&&	5.53&	&&	0.29&	&&	0.31&	&5.2	&	5.6	&&	3.67&	&&	0.11&	&&	0.21&	\\
&	&	5.6	&	6.0	&&	5.14&	&&	0.27&	&&	0.29&	&5.6	&	6.0	&&	3.24&	&&	0.11&	&&	0.19&	\\
&	&	6.0	&	6.4	&&	4.17&	&&	0.24&	&&	0.23&	&6.0	&	6.4	&&	2.61&	&&	0.10&	&&	0.15&	\\
&	&	6.4	&	6.8	&&	4.00&	&&	0.24&	&&	0.22&	&6.4	&	6.8	&&	2.18&	&&	0.09&	&&	0.12&	\\
&	&	6.8	&	7.2	&&	3.09&	&&	0.21&	&&	0.17&	&6.8	&	7.2	&&	1.96&	&&	0.08&	&&	0.11&	\\
&	&	7.2	&	7.6	&&	3.09&	&&	0.21&	&&	0.17&	&7.2	&	7.6	&&	1.735&	&&	0.078&	&&	0.098&	\\
&	&	7.6	&	8.0	&&	2.46&	&&	0.16&	&&	0.14&	&7.6	&	8.0	&&	1.531&	&&	0.074&	&&	0.086&	\\
&	&	8.0	&	8.4	&&	2.06&	&&	0.17&	&&	0.12&	&8.0	&	8.4	&&	1.100&	&&	0.063&	&&	0.062&	\\
&	&	8.4	&	8.8	&&	1.55&	&&	0.15&	&&	0.09&	&8.4	&	8.8	&&	1.052&	&&	0.062&	&&	0.059&	\\
&	&	8.8	&	9.2	&&	1.28&	&&	0.14&	&&	0.07&	&8.8	&	9.2	&&	0.785&	&&	0.054&	&&	0.045&	\\
&	&	9.2	&	9.6	&&	0.98&	&&	0.12&	&&	0.06&	&9.2	&	9.6	&&	0.716&	&&	0.051&	&&	0.040&	\\
&	&	9.6	&	10.0	&&	1.04&	&&	0.12&	&&	0.06&	&9.6	&	10.0	&&	0.584&	&&	0.047&	&&	0.033&	\\
&	&	10.0	&	10.8	&&	0.878&	&&	0.081&	&&	0.052&	&10.0	&	10.8	&&	0.490&	&&	0.030&	&&	0.027&	\\
&	&	10.8	&	11.6	&&	0.505&	&&	0.063&	&&	0.030&	&10.8	&	11.6	&&	0.312&	&&	0.024&	&&	0.017&	\\
&	&	11.6	&	12.4	&&	0.270&	&&	0.054&	&&	0.016&	&11.6	&	12.4	&&	0.194&	&&	0.020&	&&	0.011&	\\
&	&	12.4	&	13.2	&&	0.271&	&&	0.048&	&&	0.016&	&12.4	&	13.2	&&	0.121&	&&	0.016&	&&	0.007&	\\
&	&		&		&&	&	&&	&	&&	&	&13.2	&	14.0	&&	0.095&	&&	0.014&	&&	0.005&	\\ \hline
100&	140&	0.2	&	0.3	&&	2.27&	&&	0.28&	&&	0.22&	&0.2	&	0.3	&&	1.96&	&&	0.21&	&&	0.29&	\\
&	&	0.3	&	0.4	&&	3.72&	&&	0.34&	&&	0.25&	&0.3	&	0.4	&&	3.47&	&&	0.26&	&&	0.44&	\\
&	&	0.4	&	0.5	&&	5.95&	&&	0.37&	&&	0.35&	&0.4	&	0.5	&&	4.92&	&&	0.29&	&&	0.50&	\\
&	&	0.5	&	0.6	&&	7.18&	&&	0.38&	&&	0.40&	&0.5	&	0.6	&&	6.04&	&&	0.30&	&&	0.56&	\\
&	&	0.6	&	0.7	&&	8.99&	&&	0.41&	&&	0.48&	&0.6	&	0.7	&&	8.18&	&&	0.34&	&&	0.68&	\\
&	&	0.7	&	0.8	&&	10.02&	&&	0.45&	&&	0.54&	&0.7	&	0.8	&&	8.00&	&&	0.34&	&&	0.61&	\\
&	&	0.8	&	0.9	&&	10.96&	&&	0.49&	&&	0.63&	&0.8	&	0.9	&&	9.28&	&&	0.37&	&&	0.67&	\\
&	&	0.9	&	1.0	&&	11.73&	&&	0.47&	&&	0.68&	&0.9	&	1.0	&&	9.67&	&&	0.38&	&&	0.67&	\\
&	&	1.0	&	1.2	&&	13.2&	&&	0.9&	&&	1.1&	&1.0	&	1.2	&&	10.68&	&&	0.27&	&&	0.68&	\\
&	&	1.2	&	1.4	&&	13.0&	&&	0.9&	&&	1.0&	&1.2	&	1.4	&&	10.50&	&&	0.26&	&&	0.69&	\\
&	&	1.4	&	1.6	&&	14.4&	&&	1.0&	&&	1.1&	&1.4	&	1.6	&&	10.52&	&&	0.25&	&&	0.59&	\\
&	&	1.6	&	1.8	&&	14.2&	&&	1.2&	&&	1.0&	&1.6	&	1.8	&&	10.62&	&&	0.27&	&&	0.56&	\\
&	&	1.8	&	2.0	&&	12.33&	&&	0.91&	&&	0.78&	&1.8	&	2.0	&&	10.83&	&&	0.27&	&&	0.53&	\\
&	&	2.0	&	2.2	&&	12.70&	&&	0.88&	&&	0.77&	&2.0	&	2.2	&&	9.96&	&&	0.26&	&&	0.47&	\\
&	&	2.2	&	2.4	&&	12.74&	&&	0.87&	&&	0.76&	&2.2	&	2.4	&&	9.74&	&&	0.24&	&&	0.43&	\\
&	&	2.4	&	2.6	&&	12.02&	&&	0.87&	&&	0.71&	&2.4	&	2.6	&&	9.32&	&&	0.23&	&&	0.40&	\\
&	&	2.6	&	2.8	&&	9.72&	&&	0.54&	&&	0.57&	&2.6	&	2.8	&&	8.46&	&&	0.22&	&&	0.35&	\\
&	&	2.8	&	3.0	&&	9.21&	&&	0.55&	&&	0.54&	&2.8	&	3.0	&&	7.87&	&&	0.22&	&&	0.32&	\\
&	&	3.0	&	3.2	&&	9.26&	&&	0.54&	&&	0.55&	&3.0	&	3.2	&&	7.39&	&&	0.23&	&&	0.40&	\\
&	&	3.2	&	3.4	&&	8.04&	&&	0.49&	&&	0.47&	&3.2	&	3.4	&&	6.18&	&&	0.21&	&&	0.31&	\\
&	&	3.4	&	3.6	&&	7.73&	&&	0.49&	&&	0.45&	&3.4	&	3.6	&&	6.02&	&&	0.21&	&&	0.31&	\\
&	&	3.6	&	3.8	&&	6.77&	&&	0.43&	&&	0.40&	&3.6	&	3.8	&&	5.00&	&&	0.19&	&&	0.26&	\\
&	&	3.8	&	4.0	&&	5.46&	&&	0.38&	&&	0.32&	&3.8	&	4.0	&&	4.74&	&&	0.18&	&&	0.24&	\\
&	&	4.0	&	4.2	&&	5.47&	&&	0.41&	&&	0.32&	&4.0	&	4.2	&&	4.27&	&&	0.18&	&&	0.22&	\\
&	&	4.2	&	4.4	&&	5.35&	&&	0.40&	&&	0.31&	&4.2	&	4.4	&&	3.86&	&&	0.17&	&&	0.19&	\\
&	&	4.4	&	4.6	&&	3.88&	&&	0.34&	&&	0.23&	&4.4	&	4.6	&&	3.39&	&&	0.16&	&&	0.17&	\\
&	&	4.6	&	4.8	&&	3.87&	&&	0.35&	&&	0.23&	&4.6	&	4.8	&&	3.13&	&&	0.15&	&&	0.15&	\\
&	&	4.8	&	5.0	&&	3.81&	&&	0.35&	&&	0.22&	&4.8	&	5.0	&&	2.65&	&&	0.14&	&&	0.13&	\\
&	&	5.0	&	5.2	&&	3.32&	&&	0.32&	&&	0.20&	&5.0	&	5.2	&&	2.46&	&&	0.13&	&&	0.12&	\\
&	&	5.2	&	5.6	&&	2.54&	&&	0.19&	&&	0.15&	&5.2	&	5.6	&&	2.05&	&&	0.09&	&&	0.10&	\\
&	&	5.6	&	6.0	&&	2.64&	&&	0.19&	&&	0.16&	&5.6	&	6.0	&&	1.645&	&&	0.077&	&&	0.081&	\\
&	&	6.0	&	6.4	&&	1.83&	&&	0.17&	&&	0.11&	&6.0	&	6.4	&&	1.321&	&&	0.068&	&&	0.065&	\\
&	&	6.4	&	6.8	&&	1.15&	&&	0.14&	&&	0.07&	&6.4	&	6.8	&&	1.061&	&&	0.062&	&&	0.053&	\\
&	&	6.8	&	7.2	&&	1.14&	&&	0.14&	&&	0.07&	&6.8	&	7.2	&&	0.746&	&&	0.052&	&&	0.037&	\\
&	&	7.2	&	7.6	&&	0.86&	&&	0.12&	&&	0.05&	&7.2	&	7.6	&&	0.593&	&&	0.048&	&&	0.029&	\\
&	&	7.6	&	8.0	&&	0.63&	&&	0.11&	&&	0.04&	&7.6	&	8.0	&&	0.472&	&&	0.042&	&&	0.023&	\\
&	&	8.0	&	8.4	&&	0.533&	&&	0.099&	&&	0.033&	&8.0	&	8.4	&&	0.369&	&&	0.038&	&&	0.018&	\\
&	&	8.4	&	8.8	&&	0.39&	&&	0.12&	&&	0.02&	&8.4	&	8.8	&&	0.218&	&&	0.029&	&&	0.011&	\\
&	&		&		&&	&	&&	&	&&	&	&8.8	&	9.2	&&	0.188&	&&	0.029&	&&	0.009&	\\
&	&		&		&&	&	&&	&	&&	&	&9.2	&	9.6	&&	0.205&	&&	0.029&	&&	0.010&	\\ \hline
140&	180&	0.2	&	0.3	&&	3.56&	&&	0.35&	&&	0.29&	&0.2	&	0.3	&&	3.02&	&&	0.30&	&&	0.41&	\\
&	&	0.3	&	0.4	&&	6.05&	&&	0.39&	&&	0.37&	&0.3	&	0.4	&&	4.51&	&&	0.34&	&&	0.50&	\\
&	&	0.4	&	0.5	&&	8.57&	&&	0.42&	&&	0.45&	&0.4	&	0.5	&&	6.59&	&&	0.37&	&&	0.58&	\\
&	&	0.5	&	0.6	&&	8.91&	&&	0.42&	&&	0.45&	&0.5	&	0.6	&&	7.72&	&&	0.39&	&&	0.62&	\\
&	&	0.6	&	0.7	&&	11.31&	&&	0.45&	&&	0.55&	&0.6	&	0.7	&&	9.58&	&&	0.42&	&&	0.65&	\\
&	&	0.7	&	0.8	&&	13.37&	&&	0.51&	&&	0.64&	&0.7	&	0.8	&&	10.94&	&&	0.44&	&&	0.67&	\\
&	&	0.8	&	0.9	&&	13.96&	&&	0.52&	&&	0.72&	&0.8	&	0.9	&&	11.20&	&&	0.45&	&&	0.63&	\\
&	&	0.9	&	1.0	&&	13.34&	&&	0.50&	&&	0.69&	&0.9	&	1.0	&&	10.82&	&&	0.43&	&&	0.59&	\\
&	&	1.0	&	1.2	&&	12.2&	&&	1.0&	&&	1.0&	&1.0	&	1.2	&&	11.95&	&&	0.29&	&&	0.63&	\\
&	&	1.2	&	1.4	&&	12.3&	&&	1.2&	&&	1.0&	&1.2	&	1.4	&&	12.44&	&&	0.43&	&&	0.60&	\\
&	&	1.4	&	1.6	&&	14.0&	&&	1.2&	&&	0.9&	&1.4	&	1.6	&&	12.49&	&&	0.31&	&&	0.55&	\\
&	&	1.6	&	1.8	&&	13.8&	&&	1.2&	&&	0.9&	&1.6	&	1.8	&&	11.04&	&&	0.28&	&&	0.42&	\\
&	&	1.8	&	2.0	&&	12.1&	&&	1.1&	&&	0.8&	&1.8	&	2.0	&&	11.27&	&&	0.29&	&&	0.40&	\\
&	&	2.0	&	2.2	&&	12.7&	&&	1.1&	&&	0.8&	&2.0	&	2.2	&&	9.89&	&&	0.27&	&&	0.34&	\\
&	&	2.2	&	2.4	&&	9.59&	&&	0.98&	&&	0.59&	&2.2	&	2.4	&&	8.25&	&&	0.25&	&&	0.29&	\\
&	&	2.4	&	2.6	&&	7.65&	&&	0.79&	&&	0.48&	&2.4	&	2.6	&&	7.46&	&&	0.24&	&&	0.25&	\\
&	&	2.6	&	2.8	&&	8.27&	&&	0.92&	&&	0.52&	&2.6	&	2.8	&&	6.62&	&&	0.22&	&&	0.21&	\\
&	&	2.8	&	3.0	&&	5.80&	&&	0.77&	&&	0.36&	&2.8	&	3.0	&&	5.74&	&&	0.20&	&&	0.18&	\\
&	&	3.0	&	3.2	&&	6.04&	&&	0.78&	&&	0.38&	&3.0	&	3.2	&&	5.24&	&&	0.19&	&&	0.24&	\\
&	&	3.2	&	3.4	&&	3.76&	&&	0.63&	&&	0.24&	&3.2	&	3.4	&&	4.48&	&&	0.18&	&&	0.20&	\\
&	&	3.4	&	3.6	&&	4.31&	&&	0.66&	&&	0.28&	&3.4	&	3.6	&&	3.80&	&&	0.16&	&&	0.17&	\\
&	&	3.6	&	3.8	&&	3.58&	&&	0.62&	&&	0.23&	&3.6	&	3.8	&&	3.24&	&&	0.15&	&&	0.15&	\\
&	&	3.8	&	4.0	&&	3.05&	&&	0.57&	&&	0.19&	&3.8	&	4.0	&&	3.02&	&&	0.15&	&&	0.14&	\\
&	&	4.0	&	4.2	&&	3.32&	&&	0.60&	&&	0.21&	&4.0	&	4.2	&&	2.40&	&&	0.13&	&&	0.10&	\\
&	&	4.2	&	4.4	&&	3.37&	&&	0.62&	&&	0.22&	&4.2	&	4.4	&&	2.20&	&&	0.13&	&&	0.10&	\\
&	&	4.4	&	4.6	&&	1.28&	&&	0.41&	&&	0.08&	&4.4	&	4.6	&&	1.92&	&&	0.12&	&&	0.08&	\\
&	&	4.6	&	4.8	&&	2.05&	&&	0.49&	&&	0.13&	&4.6	&	4.8	&&	1.58&	&&	0.11&	&&	0.07&	\\
&	&	4.8	&	5.0	&&	1.34&	&&	0.41&	&&	0.09&	&4.8	&	5.0	&&	1.50&	&&	0.10&	&&	0.06&	\\
&	&	5.0	&	5.2	&&	1.20&	&&	0.39&	&&	0.08&	&5.0	&	5.2	&&	1.141&	&&	0.093&	&&	0.050&	\\
&	&	5.2	&	5.6	&&	1.16&	&&	0.18&	&&	0.08&	&5.2	&	5.6	&&	0.748&	&&	0.051&	&&	0.033&	\\
&	&	5.6	&	6.0	&&	0.83&	&&	0.16&	&&	0.05&	&5.6	&	6.0	&&	0.661&	&&	0.049&	&&	0.028&	\\
&	&		&		&&	&	&&	&	&&	&	&6.0	&	6.4	&&	0.407&	&&	0.037&	&&	0.017&	\\
&	&		&		&&	&	&&	&	&&	&	&6.4	&	6.8	&&	0.314&	&&	0.035&	&&	0.013&	\\
&	&		&		&&	&	&&	&	&&	&	&6.8	&	7.2	&&	0.283&	&&	0.031&	&&	0.012&	\\
&	&		&		&&	&	&&	&	&&	&	&7.2	&	7.6	&&	0.223&	&&	0.030&	&&	0.009&	\\ \hline
180&	240&	0.2	&	0.3	&&	6.73&	&&	0.38&	&&	0.48&	&0.2	&	0.3	&&	5.61&	&&	0.31&	&&	0.71&	\\
&	&	0.3	&	0.4	&&	10.66&	&&	0.43&	&&	0.58&	&0.3	&	0.4	&&	9.3&	&&	0.4&	&&	1.0&	\\
&	&	0.4	&	0.5	&&	15.58&	&&	0.50&	&&	0.75&	&0.4	&	0.5	&&	13.1&	&&	0.4&	&&	1.2&	\\
&	&	0.5	&	0.6	&&	17.33&	&&	0.52&	&&	0.77&	&0.5	&	0.6	&&	15.7&	&&	0.5&	&&	1.2&	\\
&	&	0.6	&	0.7	&&	20.03&	&&	0.54&	&&	0.84&	&0.6	&	0.7	&&	17.7&	&&	0.5&	&&	1.2&	\\
&	&	0.7	&	0.8	&&	21.01&	&&	0.61&	&&	0.89&	&0.7	&	0.8	&&	18.8&	&&	0.5&	&&	1.1&	\\
&	&	0.8	&	0.9	&&	22.4&	&&	0.7&	&&	1.0&	&0.8	&	0.9	&&	19.0&	&&	0.5&	&&	1.1&	\\
&	&	0.9	&	1.0	&&	22.1&	&&	0.6&	&&	1.0&	&0.9	&	1.0	&&	20.0&	&&	0.5&	&&	1.1&	\\
&	&	1.0	&	1.2	&&	23.0&	&&	1.6&	&&	1.7&	&1.0	&	1.2	&&	19.3&	&&	0.4&	&&	1.1&	\\
&	&	1.2	&	1.4	&&	21.8&	&&	1.5&	&&	1.6&	&1.2	&	1.4	&&	18.17&	&&	0.35&	&&	0.97&	\\
&	&	1.4	&	1.6	&&	18.5&	&&	1.3&	&&	1.3&	&1.4	&	1.6	&&	16.74&	&&	0.34&	&&	0.74&	\\
&	&	1.6	&	1.8	&&	15.8&	&&	1.3&	&&	1.1&	&1.6	&	1.8	&&	14.25&	&&	0.32&	&&	0.60&	\\
&	&	1.8	&	2.0	&&	15.2&	&&	1.2&	&&	1.0&	&1.8	&	2.0	&&	12.50&	&&	0.29&	&&	0.50&	\\
&	&	2.0	&	2.2	&&	12.5&	&&	1.1&	&&	0.8&	&2.0	&	2.2	&&	10.46&	&&	0.27&	&&	0.42&	\\
&	&	2.2	&	2.4	&&	11.6&	&&	1.1&	&&	0.8&	&2.2	&	2.4	&&	9.29&	&&	0.26&	&&	0.35&	\\
&	&	2.4	&	2.6	&&	7.48&	&&	0.86&	&&	0.49&	&2.4	&	2.6	&&	7.80&	&&	0.23&	&&	0.29&	\\
&	&	2.6	&	2.8	&&	7.62&	&&	0.86&	&&	0.49&	&2.6	&	2.8	&&	6.63&	&&	0.21&	&&	0.24&	\\
&	&	2.8	&	3.0	&&	6.51&	&&	0.81&	&&	0.44&	&2.8	&	3.0	&&	5.63&	&&	0.20&	&&	0.21&	\\
&	&	3.0	&	3.2	&&	5.89&	&&	0.75&	&&	0.39&	&3.0	&	3.2	&&	4.30&	&&	0.20&	&&	0.22&	\\
&	&	3.2	&	3.4	&&	3.83&	&&	0.63&	&&	0.24&	&3.2	&	3.4	&&	3.47&	&&	0.18&	&&	0.17&	\\
&	&	3.4	&	3.6	&&	4.95&	&&	0.66&	&&	0.33&	&3.4	&	3.6	&&	2.80&	&&	0.16&	&&	0.14&	\\
&	&	3.6	&	3.8	&&	3.54&	&&	0.62&	&&	0.25&	&3.6	&	3.8	&&	2.36&	&&	0.15&	&&	0.12&	\\
&	&	3.8	&	4.0	&&	2.91&	&&	0.56&	&&	0.19&	&3.8	&	4.0	&&	1.93&	&&	0.14&	&&	0.10&	\\
&	&	4.0	&	4.2	&&	2.31&	&&	0.52&	&&	0.15&	&4.0	&	4.2	&&	1.49&	&&	0.12&	&&	0.08&	\\
&	&	4.2	&	4.4	&&	1.74&	&&	0.46&	&&	0.12&	&4.2	&	4.4	&&	1.49&	&&	0.12&	&&	0.07&	\\
&	&	4.4	&	4.6	&&	1.28&	&&	0.41&	&&	0.09&	&4.4	&	4.6	&&	1.26&	&&	0.11&	&&	0.07&	\\
&	&	4.6	&	4.8	&&	1.72&	&&	0.45&	&&	0.11&	&4.6	&	4.8	&&	1.04&	&&	0.10&	&&	0.05&	\\
&	&		&		&&	&	&&	&	&&	&	&4.8	&	5.0	&&	0.826&	&&	0.092&	&&	0.044&	\\
&	&		&		&&	&	&&	&	&&	&	&5.0	&	5.2	&&	0.593&	&&	0.076&	&&	0.032&	\\
&	&		&		&&	&	&&	&	&&	&	&5.2	&	5.6	&&	0.471&	&&	0.047&	&&	0.025&	\\
&	&		&		&&	&	&&	&	&&	&	&5.6	&	6.0	&&	0.275&	&&	0.037&	&&	0.017&	\\
&	&		&		&&	&	&&	&	&&	&	&6.0	&	6.4	&&	0.181&	&&	0.033&	&&	0.010&	\\ \hline
240&	300&	0.2	&	0.3	&&	8.49&	&&	0.44&	&&	0.54&	&0.2	&	0.3	&&	7.36&	&&	0.37&	&&	0.87&	\\
&	&	0.3	&	0.4	&&	13.98&	&&	0.53&	&&	0.66&	&0.3	&	0.4	&&	12.4&	&&	0.4&	&&	1.2&	\\
&	&	0.4	&	0.5	&&	17.89&	&&	0.59&	&&	0.78&	&0.4	&	0.5	&&	15.8&	&&	0.5&	&&	1.4&	\\
&	&	0.5	&	0.6	&&	20.21&	&&	0.61&	&&	0.84&	&0.5	&	0.6	&&	17.7&	&&	0.5&	&&	1.3&	\\
&	&	0.6	&	0.7	&&	21.75&	&&	0.64&	&&	0.84&	&0.6	&	0.7	&&	19.7&	&&	0.6&	&&	1.2&	\\
&	&	0.7	&	0.8	&&	22.60&	&&	0.71&	&&	0.91&	&0.7	&	0.8	&&	19.4&	&&	0.6&	&&	1.1&	\\
&	&	0.8	&	0.9	&&	22.28&	&&	0.74&	&&	0.96&	&0.8	&	0.9	&&	19.8&	&&	0.6&	&&	1.1&	\\
&	&	0.9	&	1.0	&&	22.5&	&&	1.1&	&&	0.9&	&0.9	&	1.0	&&	18.4&	&&	1.1&	&&	0.9&	\\
&	&	1.0	&	1.2	&&	17.1&	&&	1.5&	&&	1.5&	&1.0	&	1.2	&&	17.00&	&&	0.40&	&&	0.94&	\\
&	&	1.2	&	1.4	&&	16.0&	&&	1.5&	&&	1.3&	&1.2	&	1.4	&&	14.82&	&&	0.37&	&&	0.77&	\\
&	&	1.4	&	1.6	&&	14.0&	&&	1.3&	&&	1.1&	&1.4	&	1.6	&&	12.94&	&&	0.35&	&&	0.66&	\\
&	&	1.6	&	1.8	&&	13.3&	&&	1.1&	&&	1.0&	&1.6	&	1.8	&&	10.20&	&&	0.31&	&&	0.51&	\\
&	&	1.8	&	2.0	&&	11.7&	&&	1.4&	&&	1.0&	&1.8	&	2.0	&&	8.32&	&&	0.28&	&&	0.41&	\\
&	&		&		&&	&	&&	&	&&	&	&2.0	&	2.2	&&	6.28&	&&	0.25&	&&	0.31&	\\
&	&		&		&&	&	&&	&	&&	&	&2.2	&	2.4	&&	5.32&	&&	0.22&	&&	0.20&	\\
&	&		&		&&	&	&&	&	&&	&	&2.4	&	2.6	&&	3.96&	&&	0.19&	&&	0.15&	\\
&	&		&		&&	&	&&	&	&&	&	&2.6	&	2.8	&&	3.15&	&&	0.17&	&&	0.12&	\\
&	&		&		&&	&	&&	&	&&	&	&2.8	&	3.0	&&	2.63&	&&	0.16&	&&	0.10&	\\
&	&		&		&&	&	&&	&	&&	&	&3.0	&	3.2	&&	2.02&	&&	0.14&	&&	0.12&	\\
&	&		&		&&	&	&&	&	&&	&	&3.2	&	3.4	&&	1.42&	&&	0.12&	&&	0.09&	\\
&	&		&		&&	&	&&	&	&&	&	&3.4	&	3.6	&&	0.963&	&&	0.095&	&&	0.057&	\\
&	&		&		&&	&	&&	&	&&	&	&3.6	&	3.8	&&	1.00&	&&	0.10&	&&	0.06&	\\
&	&		&		&&	&	&&	&	&&	&	&3.8	&	4.0	&&	0.731&	&&	0.085&	&&	0.044&	\\
&	&		&		&&	&	&&	&	&&	&	&4.0	&	4.2	&&	0.484&	&&	0.070&	&&	0.031&	\\
&	&		&		&&	&	&&	&	&&	&	&4.2	&	4.4	&&	0.368&	&&	0.060&	&&	0.023&	\\ \hline
300&	360&	0.2	&	0.3	&&	10.16&	&&	0.69&	&&	0.52&	&0.2	&	0.3	&&	8.29&	&&	0.56&	&&	0.98&	\\
&	&	0.3	&	0.4	&&	15.46&	&&	0.70&	&&	0.70&	&0.3	&	0.4	&&	13.6&	&&	0.6&	&&	1.4&	\\
&	&	0.4	&	0.5	&&	18.86&	&&	0.71&	&&	0.83&	&0.4	&	0.5	&&	16.6&	&&	0.6&	&&	1.5&	\\
&	&	0.5	&	0.6	&&	21.44&	&&	0.73&	&&	0.87&	&0.5	&	0.6	&&	18.8&	&&	0.6&	&&	1.5&	\\
&	&	0.6	&	0.7	&&	21.50&	&&	0.73&	&&	0.82&	&0.6	&	0.7	&&	20.0&	&&	0.7&	&&	1.5&	\\
&	&	0.7	&	0.8	&&	21.32&	&&	0.81&	&&	0.79&	&0.7	&	0.8	&&	20.0&	&&	0.7&	&&	1.4&	\\
&	&	0.8	&	0.9	&&	20.47&	&&	0.79&	&&	0.74&	&0.8	&	0.9	&&	17.2&	&&	0.6&	&&	1.2&	\\
&	&	0.9	&	1.0	&&	19.18&	&&	0.78&	&&	0.73&	&0.9	&	1.0	&&	16.6&	&&	0.6&	&&	1.1&	\\
&	&		&		&&	&	&&	&	&&	&	&1.0	&	1.2	&&	14.02&	&&	0.41&	&&	0.89&	\\
&	&		&		&&	&	&&	&	&&	&	&1.2	&	1.4	&&	11.09&	&&	0.40&	&&	0.72&	\\
&	&		&		&&	&	&&	&	&&	&	&1.4	&	1.6	&&	7.89&	&&	0.34&	&&	0.49&	\\
&	&		&		&&	&	&&	&	&&	&	&1.6	&	1.8	&&	6.53&	&&	0.31&	&&	0.39&	\\
&	&		&		&&	&	&&	&	&&	&	&1.8	&	2.0	&&	4.25&	&&	0.24&	&&	0.26&	\\
&	&		&		&&	&	&&	&	&&	&	&2.0	&	2.2	&&	3.43&	&&	0.22&	&&	0.21&	\\
&	&		&		&&	&	&&	&	&&	&	&2.2	&	2.4	&&	2.81&	&&	0.19&	&&	0.17&	\\
&	&		&		&&	&	&&	&	&&	&	&2.4	&	2.6	&&	1.81&	&&	0.15&	&&	0.11&	\\
&	&		&		&&	&	&&	&	&&	&	&2.6	&	2.8	&&	1.31&	&&	0.13&	&&	0.08&	\\
&	&		&		&&	&	&&	&	&&	&	&2.8	&	3.0	&&	1.06&	&&	0.12&	&&	0.06&	\\
&	&		&		&&	&	&&	&	&&	&	&3.0	&	3.2	&&	0.68&	&&	0.10&	&&	0.05&	\\
&	&		&		&&	&	&&	&	&&	&	&3.2	&	3.4	&&	0.512&	&&	0.093&	&&	0.036&	\\ \hline
360&	420&	0.2	&	0.3	&&	11.81&	&&	0.80&	&&	0.91&	&0.2	&	0.3	&&	10.5&	&&	0.7&	&&	1.2&	\\
&	&	0.3	&	0.4	&&	18.3&	&&	1.0&	&&	1.3&	&0.3	&	0.4	&&	14.1&	&&	0.8&	&&	1.5&	\\
&	&	0.4	&	0.5	&&	19.6&	&&	0.9&	&&	1.4&	&0.4	&	0.5	&&	16.9&	&&	0.8&	&&	1.6&	\\
&	&	0.5	&	0.6	&&	18.0&	&&	0.8&	&&	1.2&	&0.5	&	0.6	&&	18.0&	&&	0.8&	&&	1.6&	\\
&	&	0.6	&	0.7	&&	19.2&	&&	0.8&	&&	1.3&	&0.6	&	0.7	&&	16.6&	&&	0.7&	&&	1.4&	\\
&	&	0.7	&	0.8	&&	18.1&	&&	0.9&	&&	1.2&	&0.7	&	0.8	&&	15.7&	&&	0.7&	&&	1.3&	\\
&	&	0.8	&	0.9	&&	16.7&	&&	0.8&	&&	1.2&	&0.8	&	0.9	&&	14.8&	&&	0.7&	&&	1.2&	\\
&	&	0.9	&	1.0	&&	15.1&	&&	0.8&	&&	1.1&	&0.9	&	1.0	&&	11.97&	&&	0.58&	&&	0.92&	\\
&	&		&		&&	&	&&	&	&&	&	&1.0	&	1.2	&&	11.07&	&&	0.39&	&&	0.84&	\\
&	&		&		&&	&	&&	&	&&	&	&1.2	&	1.4	&&	7.76&	&&	0.36&	&&	0.60&	\\
&	&		&		&&	&	&&	&	&&	&	&1.4	&	1.6	&&	5.22&	&&	0.26&	&&	0.39&	\\
&	&		&		&&	&	&&	&	&&	&	&1.6	&	1.8	&&	4.30&	&&	0.24&	&&	0.33&	\\
&	&		&		&&	&	&&	&	&&	&	&1.8	&	2.0	&&	2.71&	&&	0.19&	&&	0.22&	\\
&	&		&		&&	&	&&	&	&&	&	&2.0	&	2.2	&&	1.80&	&&	0.16&	&&	0.13&	\\
&	&		&		&&	&	&&	&	&&	&	&2.2	&	2.4	&&	1.38&	&&	0.14&	&&	0.10&	\\
&	&		&		&&	&	&&	&	&&	&	&2.4	&	2.6	&&	0.96&	&&	0.11&	&&	0.07&	\\
&	&		&		&&	&	&&	&	&&	&	&2.6	&	2.8	&&	0.82&	&&	0.11&	&&	0.06&	\\
&	&		&		&&	&	&&	&	&&	&	&2.8	&	3.0	&&	0.636&	&&	0.096&	&&	0.049&	\\
\hline

\end {longtable*}
\end{center}

\end{document}